\pdfoutput=1
\documentclass[10pt,preprint]{aastex}
\usepackage{color}

\newcommand{\av}{$A_V$}

\newcommand{\etal}{et~al.}

\newcommand{\msun}{M$_{\sun}$}

\newcommand{\mum}{$\mu$m}

\begin{document}

\title{Young Stellar Object Variability (YSOVAR): Long Timescale Variations in the Mid-Infrared}

\slugcomment{Version from \today}

\author{L.~M.~Rebull\altaffilmark{1},  
A.~M.~Cody\altaffilmark{1},              
K.~R.~Covey\altaffilmark{2},             
H.~M.~G\"unther\altaffilmark{3},         
L.~A.~Hillenbrand\altaffilmark{4},       
P.~Plavchan\altaffilmark{5},             
K.~Poppenhaeger\altaffilmark{3,6},       
J.~R.~Stauffer\altaffilmark{1},          
S.~J.~Wolk\altaffilmark{3},              
R.~Gutermuth\altaffilmark{7},           
M.~Morales-Calder\'on\altaffilmark{1,8},
I.~Song\altaffilmark{9},
D.~Barrado\altaffilmark{8},           
A.~Bayo\altaffilmark{10,11},          
D.~James\altaffilmark{12}, 
J.~L.~Hora\altaffilmark{3},           
F.~J.~Vrba\altaffilmark{13},          
C.~Alves de Oliveira\altaffilmark{14},
J.~Bouvier\altaffilmark{15,16}, 
S.~J.~Carey\altaffilmark{1},       
J.~M.~Carpenter\altaffilmark{4},   
F.~Favata\altaffilmark{17},
K.~Flaherty\altaffilmark{18},      
J.~Forbrich\altaffilmark{19,3},   
J.~Hernandez\altaffilmark{20},     
M.~J.~McCaughrean\altaffilmark{17},
S.~T.~Megeath\altaffilmark{21},
G.~Micela\altaffilmark{22},        
H.~A.~Smith\altaffilmark{3},
S.~Terebey\altaffilmark{23},       
N.~Turner\altaffilmark{24},
L.~Allen\altaffilmark{25}, 
D.~Ardila\altaffilmark{26},        
H.~Bouy\altaffilmark{8},           
S.~Guieu\altaffilmark{27}        
}

\altaffiltext{1}{Spitzer Science Center (SSC), Infrared Processing and
Analysis Center (IPAC), 1200 E.\ California Blvd., California
Institute of Technology, Pasadena, CA 91125 USA; rebull@ipac.caltech.edu}
\altaffiltext{2}{Lowell Observatory, 1400 West Mars Hill Road, Flagstaff,
AZ 86001 USA}
\altaffiltext{3}{Harvard-Smithsonian Center for Astrophysics, 60 Garden St.,
Cambridge, MA 02138 USA}
\altaffiltext{4}{Department of Astronomy, California Institute of
Technology, Pasadena, CA 91125 USA}
\altaffiltext{5}{NASA Exoplanet Science Institute (NExScI), Infrared Processing and
Analysis Center (IPAC), 1200 E.\ California Blvd., California
Institute of Technology, Pasadena, CA 91125 USA, and Missouri State
University, 901 S National Ave, Springfield, MO 65897 USA}
\altaffiltext{6}{NASA Sagan Fellow}
\altaffiltext{7}{Dept.\ of Astronomy, University of Massachusetts, Amherst, MA  01003 USA}
\altaffiltext{8}{Depto. Astrof\'{\i}sica, Centro de Astrobiolog\'{\i}a
(INTA-CSIC), ESAC campus, P.O. Box 78, E-28691 Villanueva de la Ca\~nada, Spain}
\altaffiltext{9}{Physics and Astronomy Department, University of
Georgia, Athens, GA 30602-2451 USA}
\altaffiltext{10}{Max Planck Institut f\"ur Astronomie, K\"onigstuhl
17, 69117, Heidelberg, Germany}
\altaffiltext{11}{Departamento de F\'isica y
Astronom\'ia, Facultad de Ciencias, Universidad de Valpara\'iso, Av.
Gran Breta\~na 1111, 5030 Casilla, Valpara\'iso, Chile} 
\altaffiltext{12}{Cerro Tololo InterAmerican Observatory (CTIO), Casilla 603,
La Serena, Chile}
\altaffiltext{13}{US Naval Observatory, Flagstaff Station
10391 W. Naval Observatory Road, Flagstaff, AZ 86005 USA} 
\altaffiltext{14}{European Space Agency (ESA/ESAC), P.O. Box 78, 28691
Villanueva de la Ca\~ada, Madrid, Spain}
\altaffiltext{15}{Univ. Grenoble Alpes, IPAG, F-38000 Grenoble, France}
\altaffiltext{16}{CNRS, IPAG, F-38000 Grenoble, France}
\altaffiltext{17}{European Space Agency, ESTEC, Postbus 299, 2200 AG
Noordwijk, The Netherlands}
\altaffiltext{18}{Astronomy Department, Wesleyan University, 96 Foss
Hill Dr., Middletown, CT 06459 USA}
\altaffiltext{19}{University of Vienna, Department of Astrophysics,
T\"urkenschanzstr. 17, 1180 Vienna, Austria}
\altaffiltext{20}{Centro de Investigaciones de Astronom\'ia, Apdo.
Postal 264, M\'erida 5101-A, Venezuela}
\altaffiltext{21}{Ritter Astrophysical Observatory, Department of
Physics and Astronomy, University of Toledo, Toledo, OH 43606 USA}
\altaffiltext{22}{INAF -- Osservatorio Astronomico di Palermo, Piazza
del Parlamento 1, 90134 Palermo, Italy}
\altaffiltext{23}{Department of Physics and Astronomy, 5151
State University Drive, California State University at Los Angeles,
Los Angeles, CA 90032 USA}
\altaffiltext{24}{Jet Propulsion Laboratory, 4800 Oak Grove Drive, Pasadena, CA
91109 USA}
\altaffiltext{25}{NOAO, 950 N. Cherry Ave., Tucson, AZ USA}
\altaffiltext{26}{NASA Herschel Science Center (NHSC), Infrared Processing and
Analysis Center (IPAC), 1200 E.\ California Blvd., California
Institute of Technology, Pasadena, CA 91125 USA}
\altaffiltext{27}{ESO, Chile}

\begin{abstract}

The YSOVAR (Young Stellar Object VARiability) Spitzer Space Telescope
observing program obtained the first extensive mid-infrared (3.6 and
4.5 \mum) time-series photometry of the Orion Nebula Cluster plus
smaller footprints in eleven other star-forming cores (AFGL 490, NGC
1333, Mon R2, GGD 12-15, NGC 2264, L1688, Serpens Main, Serpens South,
IRAS 20050+2720, IC 1396A, and Ceph C).  There are $\sim$29,000 unique
objects with light curves in either or both IRAC channels in the
YSOVAR data set.  We present the data collection and reduction for the
Spitzer and ancillary data, and define the ``standard sample" on which
we calculate statistics, consisting of fast cadence data, with
epochs very roughly twice per day for $\sim$40d. We also define a
``standard sample of members", consisting of all the IR-selected
members and X-ray selected members. We characterize the standard
sample in terms of other properties, such as spectral energy
distribution shape. We use three mechanisms to identify variables in
the fast cadence data -- the Stetson index, a $\chi^2$ fit to a flat
light curve, and significant periodicity.  We also identified
variables on the longest timescales possible of 6-7 years, by
comparing measurements taken early in the Spitzer mission with the
mean from our YSOVAR campaign. The fraction of members in each cluster
that are variable on these longest timescales is a function of the
ratio of Class I/total members in each cluster, such that clusters
with a higher fraction of Class I objects also have a higher fraction
of long-term variables. For objects with a YSOVAR-determined period
and a [3.6]$-$[8] color, we find that a star with a longer period is
more likely than those with shorter periods to have an IR excess. We
do not find any evidence for variability that causes [3.6]$-$[4.5]
excesses to appear or vanish within our dataset; out of members and
field objects combined, at most 0.02\% may have transient IR excesses.

\end{abstract}

\keywords{circumstellar matter -- stars: pre-main sequence --
stars:protostars -- stars: variables: general}

\section{Introduction}
\label{sec:introduction}

Optical variability was one of the original, defining characteristics
of the class of object later determined to be stars in the process of
formation (Joy 1945; Herbig 1952), or young stellar objects (YSOs).
Optical and near-infrared (NIR) monitoring over timescales of weeks to
months of the nearest star-forming regions (SFRs) have shown that the
surfaces of YSOs are often mottled, with both hot spots (where gas
accretion columns from the inner disk impact the stellar surface) and
cool spots (starspots analogous to sunspots; Rydgren \& Vrba 1983;
Vrba \etal\ 1986; Bouvier 1993). Because the stars are also rotating,
the presence of spots causes their apparent luminosities and colors to
vary with the stellar rotation period. As summarized in Herbst \etal\
(1994), cool spots are found on YSOs without disks, or at least
without substantial accretion disks (weak-lined T Tauri stars, or
WTTs), which is expected since those stars do not generally have other
signatures of active accretion; however, both cool spots and hot spots
have been identified on YSOs with substantial disks (classical T Tauri
stars, or CTTs). The largest amplitude, most variable optical light
curves are generally attributed to hot spots (Vrba \etal\ 1993).

The periodicities found in spot-dominated light curves have been taken
to be the rotation period of the star, and the derivation of periods
has long been the most common analysis of time series data of young
stars. For solar mass YSOs and ages $\sim$ few Myr, the distribution
of rotational velocities is bimodal, with one set of stars having
periods of order 2-4 days and the other with characteristic periods of
8-12 days (e.g., Cieza \& Baliber 2007 and references therein). This
period distribution has been interpreted in terms of a model where the
rotational periods of the accreting stars are magnetically locked to
the Keplerian rotation period of their inner disks (with periods of
order 10 days), whereas stars that are no longer accreting spin up as
they contract, thus associating the short-period peak in the rotation
period distribution with stars that have lost their disks at young
ages (e.g., Bouvier \etal\ 1997). This correlation at young ages
appears to persist to later ages, with the slow rotators on the
Zero-Age Main Sequence (the stars with long-lived accretion disks)
being more likely to have debris disks, which could suggest that these
slow rotators are more likely to have formed planets (see, e.g.,
Bouvier 2008, McQuillan \etal\ 2013ab).

In the past, periodicities have been most frequently determined using
ground-based optical time-series observations. Optical observations
are primarily sensitive to phenomena associated with the stellar
photosphere or with other energetically ``hot" regions (hot spots,
accretion columns, chromospheres), and are limited in regions of high
extinction. In contrast, observations at longer wavelengths penetrate
extinction, and also offer a new perspective by being sensitive to
variability associated with ``warm" or ``cool" regions -- the disks
and envelopes of YSOs. The dominant contributions to YSO photometric
variability in the IR include the Rayleigh-Jeans tail of the hotter
processes, as well as dust reprocessing of emission from these hotter
processes, along with phenomena uniquely associated with the disk.
Relevant disk processes might involve thermal emission from an
over-dense (or over-warmed) region of the inner disk, variable disk
accretion, structure in the disk rotating into and out of view causing
changes in the measured \av\ towards the star, or disk instabilities
(e.g., Fedele \etal\ 2007; Plavchan \etal\ 2008a, 2013; Herbst \etal\
2010). Finally, standard geometric effects due to orbiting companions
can also be probed in the mid-infrared, uniquely so for more embedded
sources.  Because many more physical processes can affect the
variability of YSOs in the infrared, relatively few infrared light
curves are periodic and thus straightforward to analyze (see, e.g.,
Cody \etal\ 2014, Morales-Calder\'on \etal\ 2011).

One of the first monitoring programs of YSOs at NIR wavelengths  was
Skrutskie \etal\ (1996), which monitored 15 YSOs in Taurus-Auriga.
They found periodic variability, as well as variability due to
accretion and extinction.

The first large program of time series photometry of YSOs at
wavelengths longward of 1 micron was by Carpenter, Hillenbrand \&
Skrutskie (2001, CHS01). CHS01 obtained $JHK_s$ monitoring of $\sim$3
square degrees of the Orion Nebula Cluster (ONC) over a $\sim$1 month
time period as part of the Two-Micron All Sky Survey (2MASS; Skrutskie
\etal\ 2006). About 1000 Orion members showed NIR photometric
variability in their
data\footnote{http://www.astro.caltech.edu/~jmc/variables/orion}.
Typical light curve amplitudes were of order 0.2 mag, but with some
stars exhibiting amplitudes up to 2 mag; periods were determined for
about a quarter of of their stars. CHS01 attributed the variability
for somewhat more than half of the stars to cool spots; they suspected
that most of the others could be explained by hot spots, variable
extinction or variable accretion. However, they could not make a
definitive determination and suggested multiple mechanisms could be
involved.  2MASS monitored other star-forming regions as well,
including Chamaeleon and Rho Oph.  Carpenter \etal\ (2002) reported on
the more limited 2MASS study of the Chamaeleon star forming region and
similarly characterized variability amplitudes and behaviors, along
with identifying new candidate young star members via their infrared
variability.  Plavchan \etal\ (2008b) and Parks \etal\ (2014) report
on the 2MASS observations of a small region in Rho Oph, finding about
100 variables, with roughly similar variability properties as Orion in
that the amplitude variations were found to be between a few tenths
and 2 mag, with periods obtained for about a third of the sample.
Subsequent to 2MASS, more recently, there have been a number of NIR
monitoring programs studying other star-forming regions, such as Wolk
\etal\ (2013), which monitored Cyg OB7, finding several classes of YSO
NIR variability.

In the mid-infrared (Cohen \& Schwarz 1976), as for the near-infrared
(e.g. Elias \etal\ 1978; Rydgren \& Vrba 1983), previous literature suggested
at least small variations on timescales of months to years, likely
attributable to circumstellar disk processes. However, in the same way
that charge-coupled devices (CCDs) revolutionized our ability to
discern precisely optical variability trends and 2MASS did the same
for near-infrared variability, the Spitzer Space Telescope (Werner
\etal\ 2004) has allowed us to probe even small variations in the
mid-infrared; Spitzer is a photometrically stable (better than 1\%),
sensitive, wide-field ($5\arcmin\times5\arcmin$), Earth-trailing
(avoiding orbital day/night aliasing) platform. Spitzer, specifically
the Infrared Array Camera (IRAC; Fazio \etal\ 2004), observes at bands
sensitive to both YSO photospheres and circumstellar dust.

Cycle 6 was the first post-cryogen Spitzer cycle, using just IRAC's
first two channels (3.6 and 4.5 \mum, often abbreviated IRAC-1 and
IRAC-2, or I1 and I2; when reporting measurements in magnitudes, the
bands are written with brackets, e.g., [3.6]=16.38 mag). The YSOVAR
(Young Stellar Object VARiability)  Spitzer Space Telescope Cycle-6
Exploration Science (ES) Program was approved for 550 hours of
observations, with the goal of obtaining the first extensive
mid-infrared time-series photometry of the central $\sim1\arcdeg$ of
the Orion Nebula Cluster plus smaller footprints in eleven other
star-forming cores; see Table~\ref{tab:clusterproperties} for a list
of the clusters.   There are several other ES and smaller programs
exploring YSOs in the time domain with Spitzer, many of which are
affiliated (to varying degrees) with the larger YSOVAR effort. We have
incorporated under the YSOVAR umbrella some additional strongly
related programs pre-dating and arising from YSOVAR, resulting in a
total of 786 hours of Spitzer time. About 130 hours of that is
dedicated observations of NGC 2264 (Coordinated Synoptic
Investigation: NGC 2264, or CSI 2264), which is discussed by Cody
\etal\ (2014), among other papers (e.g., Stauffer \etal\ 2014, 2015).
CSI 2264 is not discussed in the same way as the rest of the data here,
in no small part because the observations are generally different and
because it involves coordination of more telescopes.  A list of core
YSOVAR programs and affiliated programs is presented in
Table~\ref{tab:programs} and discussed in
Section~\ref{sec:newspitzeroverview}. We sometimes refer to the
components of the original YSOVAR program as `YSOVAR-classic' to
distinguish them from the smaller affiliated observations obtained
over the same time period.  There are $\sim$29,000 unique objects with
light curves from either (or both) of the IRAC channels in the YSOVAR
data set, matched to $\sim$39,000 individual light curves. These light
curve counts include light curves from both cluster members and a
significant number of non-member stars, and also likely include
extragalactic objects. There are more light curves than objects
because most objects have light curves at just I1 or I2, but many have
light curves at both I1 and I2.

YSOVAR data were obtained to help reveal the structure of the inner
disk region of YSOs, provide new constraints on accretion and
extinction variability, assess timescales of mid-IR variability from
seconds to years, identify new young eclipsing binaries, help identify
new very low mass substellar members of the surveyed clusters,
constrain the short and long-term stability of hot spots on the
surfaces of YSOs, and determine rotational periods for objects too
embedded for such monitoring in the optical.

In this paper, in addition to presenting an overview of the data set,
one of our goals is to specifically address the longest timescale
variations that we can quantify in these clusters, 6-7 years. We look
for large changes between the earliest Spitzer observations (from
observations obtained early in the cryogenic era) and the YSOVAR
monitoring observations. We discuss many of the technical details
associated with the data reduction across the YSOVAR effort. The Orion
data were first described by Morales-Calder\'on \etal\ (2011;
hereafter MC11). The other 11 smaller-field clusters are introduced in
this paper, but will be discussed in detail in other papers (the first
of which, on L1688, is G\"unther \etal\ 2014).  Here, we first provide
a summary of the observations and data reduction
(Sec.~\ref{sec:obsanddatared}), followed by a definition of the
samples we use (Sec.~\ref{sec:sampledefinition}). We present some
global statistics on all the clusters in
Sec.~\ref{sec:ensembleanalysis}. We delve into variable selection in
Sec.~\ref{sec:findingvars}, discussing different tests for selecting
variables and simulating the sensitivity of the techniques given the
actual data set. We present some analysis that is best done with all
the clusters together in Sec.~\ref{sec:discussion}, such as fractions
of long-term variables across clusters, correlations between rotation
rate and IR excess, and absence of transient disks. We summarize in
Sec.~\ref{sec:conclusions}.

For completeness, we note here that we refer to the 12 regions of
recent star formation that we observed for YSOVAR as `clusters',
knowing that others may prefer `associations' or other nomenclature.
Our targets resemble small condensations within a region; Gutermuth
\etal\ (2009), among others, addresses formal clustering in these
regions.

\section{Observations and Data Reduction}
\label{sec:obsanddatared}

In this section, we review the target selection and some general
properties of the clusters, and describe the Warm Spitzer
observations: observing strategy, data reduction, cadence, and the
noise floor. We also describe the data reduction for the cryogenic-era
data, other archival data, and the Chandra X-ray data. 

\begin{deluxetable}{lcccccp{9.5cm}}
\tabletypesize{\scriptsize}
\rotate
\tablecaption{Summary of Cluster Properties\label{tab:clusterproperties}}
\tablewidth{0pt}
\tablehead{
\colhead{Cluster\tablenotemark{a}} & \colhead{Dist.} &
\colhead{Gal.~Lat.\tablenotemark{b}}& \colhead{Class II/I} &
\colhead{Class II/I} & \colhead{Class I/tot} & \colhead{Notes} \\
 & \colhead{(pc)} & & \colhead{(G09)\tablenotemark{c}} &
\colhead{(obj.~w/ L.C.)\tablenotemark{d}} & \colhead{(mem.~w/
L.C.)\tablenotemark{e}} & }
\startdata
AFGL 490       & 900 & +1.8$\arcdeg$   & $\sim$3.2 & 4.5$\pm$0.9& 0.20$\pm$0.04& distance: Testi \etal\ (1998); ratio: Masiunas \etal\ (2012), using the same approach as G09 but deeper data, report a ratio of $\sim$5\\ 
NGC 1333       & 235 & $-$20.5$\arcdeg$& $\sim$2.7 & 4.7$\pm$1.3& 0.16$\pm$0.04& distance: Hirota \etal\ (2008); see also Hirota \etal\ (2011) \\ 
Orion          & 414 & $-$19.0$\arcdeg$& \nodata   & 9.7$\pm$0.9& 0.07$\pm$0.01& distance: Menten \etal\ (2007) \\ 
Mon R2         & 830 & $-$12.6$\arcdeg$& $\sim$4.7 & 6.0$\pm$1.4& 0.15$\pm$0.03& distance: Herbst \& Racine (1976); see discussion in Carpenter \& Hodapp (2008)\\ 
GGD 12-15      & 830 & $-$11.9$\arcdeg$& $\sim$4.2 & 5.8$\pm$1.6& 0.13$\pm$0.03& distance: Herbst \& Racine (1976); see discussion in Carpenter \& Hodapp (2008)\\ 
NGC 2264       & 760 & +2.1$\arcdeg$   & \nodata   & 3.0$\pm$0.6& 0.15$\pm$0.03& distance: Sung \etal\ (1997)  \\ 
L1688          & 120 & +16.6$\arcdeg$  & $\sim$3.0 & 1.9$\pm$0.6& 0.28$\pm$0.08& distance: Wilking \etal\ (2008), Loinard \etal\ (2012), Loinard \etal\ (2008); G09 cites Wilking \etal\ (2005) for 150 pc\\ 
Serpens Main   & 415 & +16.5$\arcdeg$  & $\sim$1.4 & 2.2$\pm$0.6& 0.21$\pm$0.05& distance: Dzib \etal\ (2010), Loinard \etal\ (2012) (see also Eiroa \etal\ 2008); G09, as do many other authors, cites 260 pc from Strai\v{z}ys \etal\ (1996) \\ 
Serpens South  & 415 & +3.8$\arcdeg$   & $\sim$0.7 & 0.9$\pm$0.2& 0.41$\pm$0.08& G09 ratio calculated from numbers in Gutermuth \etal\ (2008b); assumed to be same distance as Serpens Main \\ 
IRAS 20050+2720& 700 & $-$2.6$\arcdeg$ & $\sim$1.9 & 2.2$\pm$0.4& 0.30$\pm$0.05& distance: Wilking \etal\ (1989), G\"unther \etal\ (2012)\\ 
IC 1396A       & 900 & +3.9$\arcdeg$   & \nodata   & 7.9$\pm$2.8& 0.04$\pm$0.01& distance: Contreras \etal\ (2002); II/I ratio given is in region with both I1 and I2 light curves -- II/I ratio calculated in exactly the same way as other clusters is 11.6$\pm$4.0 \\ 
Ceph C         & 700 & +2.1$\arcdeg$   & $\sim$2.3 & 3.2$\pm$1.0& 0.17$\pm$0.05& distance: Moscadelli \etal\ (2009); II/I ratio given is in region with both I1 and I2 light curves -- II/I ratio calculated in exactly the same way as other clusters is 3.8$\pm$1.1\\ 
\enddata
\tablenotetext{a}{Clusters appear in RA order, in this and subsequent
tables. RA for our specific observations appears in
Table~\ref{tab:programs}.  All of these clusters are thought to be
between 1 and 5 Myr old.}
\tablenotetext{b}{Approximate Galactic latitude, provided as a rough
proxy of the total number of background/foreground stars (from
Galactic contamination) expected in the region.}
\tablenotetext{c}{Ratio of Class II to Class I sources, as presented
in G09. For a brief definition of SED classes, see
App.~\ref{sec:sedsection}.}
\tablenotetext{d}{Ratio of Class II to Class I sources, using the same
classification scheme as G09 (where the classes are only available for
IR-selected members),  but using the reprocessed cryogenic data, and
the ratio is calculated just for objects with light curves in the
YSOVAR data. Error bars are derived assuming independent Poisson
uncertainties on the numbers used to compute the ratio. See
Section~\ref{sec:clusterparameterization} for more discussion. }
\tablenotetext{e}{Ratio of Class I to total number of member
sources with viable light curves in the YSOVAR data, using the
standard set of members (Sec.~\ref{sec:standardsetofmembers}) and our
SED classification (App.~\ref{sec:sedsection}). Error bars are derived
assuming independent Poisson uncertainties on the numbers used to
compute the ratio. See Section~\ref{sec:clusterparameterization} for
more discussion on this ratio and how it compares to the Class
II/Class I ratio.}
\end{deluxetable}

\subsection{Target Selection}

In this section, we discuss first our criteria for picking targets,
and then review basic properties of each cluster.

\subsubsection{Overview of target selection} 
\label{sec:targetseloverview}

The Orion star forming region has been the subject of variability
studies more than any other star forming region. Because Orion is so
well-studied, particularly for optical and NIR variability, we elected
to include it as a very significant part of our mid-infrared observing
effort. 

Besides Orion, we selected the cores of 11 additional young clusters
for monitoring; see Table~\ref{tab:clusterproperties}. Our
smaller-field, more embedded cluster sample (``smaller field
clusters'') was selected based on detailed examination of all the
star-forming regions surveyed with Spitzer by mid-2007, when YSOVAR
targets were selected. We chose regions which satisfy the following
criteria: (a) a relatively high fraction of Class I sources (for a
brief definition of spectral energy distribution -- SED -- classes,
such as Class I, see App.~\ref{sec:sedsection}), such that we would
obtain monitoring for some of the most heavily embedded objects; (b) a
high density of YSOs within one or a few IRAC fields of view (FOVs) --
typically $>$40 YSOs per field; (c) moderate cirrus backgrounds; and
(d) minimal problems with crowding or very bright nearby sources.
Several of the regions we monitored are very difficult to observe from
the ground due to their high level of obscuration.  We added IC~1396A
even though it is not as embedded, because there has already been a
significant investment of Spitzer time in monitoring this region (more
on this below).  NGC 2264, like Orion, has been intensively surveyed
for variability, from the ground (see, e.g., Makidon \etal\  2004,
Lamm \etal\ 2005, Cieza \& Baliber 2007) and from space (see, e.g.,
Alencar \etal\ 2010 for CoRoT, Zwintz \etal\ 2009 for MOST). We
focused on the most embedded region of NGC 2264 for the
`YSOVAR-classic' part of the program, and monitored a much larger
region for CSI 2264.

\subsubsection{Cluster properties}
\label{sec:clusterproperties}

We now briefly discuss general properties of each of these clusters,
in RA order (see Tables~\ref{tab:clusterproperties} and
\ref{tab:programs}).  We tabulate several characteristics of these
clusters in Table~\ref{tab:clusterproperties}, including some values
from Gutermuth \etal\ (2009, 2010; hereafter G09).  All of these
clusters are thought to be between 1 and 5 Myr old.  Ages more
accurate than that for clusters as young and embedded as these are
difficult to obtain, even in a relative sense. Cryogenic Spitzer
observations for many of these clusters were discussed in G09, who
calculated the Class II to Class I ratio for various regions and
sub-clusters within the cryo Spitzer maps. This ratio is easier to
obtain than an age, and helps place the clusters' evolutionary state
in context with each other; it is provided here (in
Table~\ref{tab:clusterproperties} and the text below) in part as a
link back to existing literature. In general, our observations enclose
just the most embedded parts of these clusters; the full cluster
membership generally includes objects beyond the regions we monitored.
(Orion is different in that the map is much larger than the other
smaller-field cluster maps, but, even then, the map does not include
all objects thought to be part of Orion.) Thus, we have recalculated
the Class II to Class I ratio for just objects with YSOVAR light
curves, in the same way as was done in G09 (e.g., just for the
IR-selected members); this value appears in
Table~\ref{tab:clusterproperties}, again to place our observations in
context with the prior literature. We discuss this parameterization in
more detail below in Section~\ref{sec:clusterparameterization}, and
consider a different parameterization of the clusters' relative
evolutionary states, the ratio of the number of Class I sources to
total YSOs; anticipating this discussion, these values are included in
Table~\ref{tab:clusterproperties}. The total cluster membership we use
for this metric includes objects with light curves that we selected
using both the IR data from Spitzer and X-ray data from the Chandra
X-ray Observatory (Sec.~\ref{sec:sampledefinition}). The X-ray data
are discussed further in Sec.~\ref{sec:chandrareduction}. 

Table~\ref{tab:clusterproperties} also includes an approximate
Galactic latitude as a very rough guide to the overall Galactic
background/foreground source density expected near each location.
Targets close to the Galactic plane (e.g., Serpens Main) have a higher
surface density of objects in our fields of view than targets further
from the plane.

\paragraph{AFGL 490:} The cluster associated with AFGL 490 is a
portion of the larger Cam OB1 Association, centered on a massive
object (8-10 \msun; sometimes the name AFGL 490 is used to refer just
to this massive protostar). The cluster is thought to lie between
$\sim$900 pc (Testi \etal\ 1998) and $\sim$1010 pc (Strai\v{z}ys \&
Laugalys 2008). We adopt a distance of 900 pc (as does G09).  G09,
based on the cryogenic Spitzer data for this region, found at least
100 young stars or candidates in this vicinity, and found the overall
Class II to Class I ratio to be about 3. Masiunas \etal\ (2012) also
report on the cryogenic Spitzer observations and, incorporating
additional data, find many new YSO candidates and a Class II
to Class I ratio overall of $\sim$5. Among the objects with YSOVAR
light curves, using the G09 color cuts to identify Class I and II
sources (see Sec.~\ref{sec:cryodatareduction} and
App.~\ref{sec:sedsection}), the ratio is $\sim$4.5. This region is the
only cluster in our set of clusters that does not have archival
Chandra X-ray observations (see Sec.~\ref{sec:chandrareduction}).

\paragraph{NGC~1333:} NGC~1333 is on the western edge of the Perseus
molecular cloud, and, at only $\sim$235 pc (Hirota \etal\ 2008, 2011),
it is one of the youngest and most well-studied star forming regions,
with $>$200 refereed publications. The region is riddled with outflows
from young stars (e.g., Plunkett \etal\ 2013), which can clearly be
seen in the Spitzer 4.5 \mum\ image of this region.   Gutermuth \etal\
(2008a, 2009) analyzed the cryogenic Spitzer data for this region, and
found more than 130 young stars or candidates; the Class II to Class I
ratio they found in the core is $\sim$2.7.  For the region where we
have light curves, the ratio is $\sim$4.7. Despite the large number of
prior studies in the literature, NGC 1333 has few prior studies
specifically investigating time variability. The YSOVAR data
will be discussed in detail in Rebull \etal\ (in prep); Raga \etal\
(2013) report on proper motions of the outflows in this region using
the YSOVAR data.

\paragraph{Orion:} Orion has been the subject of
extensive variability studies. Haro (1969), Herbig \& Kameswara Rao
(1972), and Walker (1978) all conducted pioneering studies of the
variability of young stars in Orion, leading to the modern era using
two-dimensional imaging cameras as initiated by Herbst and his team
(Attridge \& Herbst 1992; Choi \& Herbst 1996). They obtained
multi-year, optical time series photometry of several regions of the
ONC, eventually obtaining light curves for hundreds of YSOs, often
spanning several years.  Many other groups subsequently obtained time
series photometry of other portions of the ONC, using optical (Stassun
\etal\ 1999; Rebull 2001; Herbst \etal\ 2002; Stassun \etal\ 2006,
2007; Irwin \etal\ 2007; Rodr\'iguez-Ledesma, Mundt, \& Eisl\"offel
2009) and near-IR (CHS01) imaging data. More recently, some far-IR
monitoring has been conducted in Orion as well (Billot \etal\ 2012). In
many cases, the light curve shapes are well-fit by models of
rotational modulation via hot or cold spots, normally at moderately
high latitudes since the light curves are seldom ``flat-bottomed.''
However, for a significant fraction of the light curves, particularly
those in the near-IR, spots do not seem to provide a good explanation
for the observed variability (CHS01).  We included the ONC as a YSOVAR
target because of the substantial amount of extant monitoring
available in the literature. Note that it is likely slightly older
than most of the other embedded regions studied here. For the
cryogenic-era data (see Sec.~\ref{sec:cryodatareduction}), we used the
data reduction, YSO identification, and YSO classification from
Megeath \etal\ (2012), which are very similar to that from G09.
Therefore, while there is no ratio of Class II to Class I objects from
G09 itself to report, we calculated this ratio using the Megeath
\etal\ (2012) classifications, for just the objects for which we have
light curves, obtaining $\sim$9.7. While this ratio is affected by the
much larger region monitored by YSOVAR (just the cores are monitored
in most of the other regions), this value is consistent with Orion
being slightly older than our other more embedded targets. We adopt a
distance of 414 pc from Menten \etal\ (2007).

\paragraph{Mon R2:}
Mon R2 appears in G09, with a Class II to Class I ratio in the central
region of $\sim$4.7. This region, part of the Monoceros R2 molecular
cloud, is near vdB 67 and 69 (see, e.g., Carpenter \& Hodapp 2008). It
is typically thought to be at $\sim$830 pc (Herbst \& Racine 1976,
Carpenter \& Hodapp 2008).  There are several sources that are very
bright in the infrared in this location, 
which affects the completeness of the catalog extracted from the
Spitzer data. The Class II to Class I ratio in the region with YSOVAR
monitoring is $\sim$6. These data will be discussed in more detail by
Hillenbrand \etal\ (in prep).

\paragraph{GGD 12-15:}
GGD 12-15 is a dense core also located in the Monoceros R2 molecular
cloud (see, e.g., Carpenter \& Hodapp 2008). As such, we assume it to
also be at $\sim$830 pc. G09 find a Class II to Class I ratio of
$\sim$4.2; we calculate $\sim$5.8 for the region with YSOVAR light
curves. Several time-variable radio sources are located here
(Carpenter \& Hodapp 2008 and references therein). We also monitored
this region with Chandra during portions of our YSOVAR Spitzer
campaign. These data will be discussed in depth by Wolk \etal\ (in prep). 

\paragraph{NGC 2264:} NGC 2264 is thought to be comparable in age to
or slightly older than Orion (see e.g., Ramirez \etal\ 2004a). This
region does not appear in G09 so we do not have a similarly-obtained
Class II to Class I ratio for comparison.  However, the region we
monitored as part of the YSOVAR-classic data (original YSOVAR program
61027; see Section~\ref{sec:newspitzeroverview} and
Table~\ref{tab:programs}), which is the region discussed here, is
centered on the Spokes Cluster (Teixeira \etal\ 2006), the most
embedded portion of NGC~2264 (analogous to the BN region in Orion). We
calculate a Class II to Class I ratio in this monitored region of
$\sim$3.0, which is comparable to the ratio for many of the other
embedded clusters here. We work just with the YSOVAR-classic  data in
the present paper; Cody \etal\ (2014) discuss the much larger CSI 2264
dataset. We have adopted a distance to this cluster of 760 pc (Park
\etal\ 2000).

\paragraph{L1688:} Lynds 1688 (L1688) is located within the $\rho$
Ophiuchi molecular cloud. It is also one of the best-studied
star-forming regions in this YSOVAR data set, with more than 500
refereed articles.  The distance to this region is a subject of some
debate.  There is recent evidence for 130 pc (Wilking \etal\ 2008 and
references therein),  131 pc (Mamajek 2008), and 120 pc (Loinard
\etal\ 2008; Loinard 2012; Lombardi \etal\ 2008).  Lombardi \etal\
(2008) offer a plausible explanation for the ``discrepancy" in the
VLBI results noted by Wilking \etal\ (2008): maybe there are other
subregions in Oph at distinct distances, but the evidence is unclear
at this point. Our results are not particularly dependent on distance;
we have adopted 120 pc as one of the most recent determinations. Lynds
1688 has a high surface density of embedded objects. This region
appears in G09 with a Class II to Class I ratio of $\sim$3.0; we
calculate a value of $\sim$1.9 for the region we monitored. Barsony
\etal\ (2005) first reported YSO variability in the MIR in the objects
in this core. By comparison to Infrared Space Observatory (ISO) data,
they found significant variability in 18 out of 85 objects detected,
on timescales of years. They found such variability in all SED classes
with optically thick disks, and suggest that this might be due to
time-variable accretion.   The Multiband Imaging Photometer for
Spitzer (MIPS; Rieke \etal\ 2004) data for this region were presented
in Padgett \etal\ (2008); no variability in sources at 24 \mum\ was
found to a level of 10\% on timescales of hours. Alves de Oliviera \&
Casali (2008) recently reported on deep NIR monitoring of this region,
with 14 epochs over 2 years, finding that 41\% of the known YSOs are
variable.   The $\rho$ Oph core region was included in one of the
``calibration'' 2MASS fields, and as such has monitoring data in the
NIR (Parks \etal\ 2014) with a cadence of $\sim$1 day over 3 observing
seasons spanning $\sim$2.5 years. Parks \etal\ (2014) found 101
variables, 72 of which are identified with known YSOs in the region.
Plavchan \etal\ (2013) report on YLW 16A, finding a 93 day
periodicity. The YSOVAR data are discussed in detail by G\"unther
\etal\ (2014).

\paragraph{Serpens Main:} The Serpens core has been studied for
decades, but has become known as `Serpens Main' to distinguish it from
the relatively recently discovered embedded star-forming core known as
Serpens South (Gutermuth \etal\ 2008b; see below). The original
Spitzer cryo-era data for Serpens Main were presented in Harvey \etal\
(2007); they found no clear evidence at the $\sim$25\% level for
IRAC-band variability in any sources in the field over the $\sim$6 hr
timescale of their observations. The Spitzer cryogenic data were also
used in G09, who determined the ratio of Class II to Class I objects
at $\sim$1.4, the lowest of all of the clusters from YSOVAR that also
appear in G09. It has a high surface density of embedded objects, and
it is another very well-studied star forming region, with more than
500 refereed articles. While the Strai\v{z}ys \etal\ (1996) distance
of 260 pc to Serpens Main is well-cited in the literature, more recent
studies (Dzib \etal\ 2010; Loinard 2012; see also Eiroa \etal\ 2008)
using, e.g., Very Long Baseline Interferometry (VLBI) instead suggest
that the distance of 260 pc may be portions of clouds associated with
Aquila, and that a better distance for Serpens itself is actually 415
pc, which we adopt here. Hodapp (1999) reports on NIR variability of
knots, jets, and young stars; in terms of point sources, Hodapp (1999)
primarily discusses one particular source (OO Ser), in this region.
There was a brief Spitzer monitoring program of this region conducted
during Cycle-3 as part of guaranteed time (see
Table~\ref{tab:programs}); the observations conducted as part of
YSOVAR were designed to be well-matched to these observations. We
obtain a Class II to Class I ratio of $\sim$2.2 for the region with
light curves. 

\paragraph{Serpens South:}
Serpens South was discovered by Gutermuth \etal\ (2008b) as a dense,
embedded cluster in the Serpens-Aquila Rift. It is thought that this
cluster is at about the same distance as Serpens Main, which we have
taken to be 415 pc. From the numbers in Gutermuth \etal\ (2008b), it
has a Class II to Class I ratio of only $\sim$0.7. Considering just
the region we monitored, the ratio is comparable at $\sim$0.9.

\paragraph{IRAS 20050+2720:} Observations from Spitzer and Chandra of
the cluster associated with IRAS 20050+2720 (abbreviated IRAS 20050)
have been discussed by G\"unther \etal\ (2012). This cluster
is part of the Cygnus Rift and is likely at $\sim$700 pc (G\"unther
\etal\ 2012 and references therein).  G09 determined the ratio of
Class II to Class I objects to be $\sim$1.9; in the region with light
curves, we obtain $\sim$2.2. The YSOVAR data for this cluster will be
discussed in detail by Poppenhaeger \etal\ (in prep).

\paragraph{IC 1396A:} IC 1396A is the most prominent globule in the IC
1396 complex; this cluster is sometimes called the Elephant Trunk
Nebula. Morales-Calder\'on \etal\ (2009) presented the first Spitzer
monitoring of young stars, centered on this target. This region is
also likely to be very young, but it is not particularly embedded, at
least compared to other YSOVAR clusters. It is not included in G09,
but over the entire region within which light curves were obtained, we
obtain a Class II to Class I ratio of $\sim$11.6. Because IC 1396A is
at a high ecliptic latitude (see Sec.~\ref{sec:footprints} below for
discussion of ecliptic latitude dependencies), there are light curves
with only a few single-band points for many objects. For the much
smaller number of objects that have light curves in both IRAC-1 and 2,
we obtain a much lower Class II to Class I ratio of $\sim$7.9, though
within Poisson errors calculated assuming independent errors in the
numerator and denominator, these values are consistent (11.6$\pm$4.0
and 7.9$\pm$2.8).  We assume it is at a distance of $\sim$900 pc
(Contreras \etal\ 2002).

\paragraph{Ceph C:} Ceph C is part of the Cep OB 3 molecular cloud,
and is included in the G09 study, with a Class II to Class I ratio
obtained there of $\sim$2.3. G09 used a distance of 730 pc (Blauw
1964); we adopt a distance of 700 pc based on maser parallax from
Moscadelli \etal\ (2009).  Ceph C is one of the less well-studied
clusters in our set. For the region with any light curves (see
Section~\ref{sec:footprints} below), we obtain a Class II to Class I
ratio of $\sim$3.8. Because it, like IC 1396A, is at a high ecliptic
latitude, we repeated this calculation for the much smaller number of
objects that have light curves in both IRAC-1 and 2, obtaining a ratio
of $\sim$3.2; again assuming Poisson counting statistics, these ratios
are comparable (3.8$\pm$1.1 and 3.2$\pm$1.0). This cluster was
monitored during our YSOVAR campaign with Chandra as well; these data
will be discussed in depth by Covey \etal\ (in prep).

\begin{deluxetable}{llllcp{2cm}llccp{5cm}}
\tabletypesize{\scriptsize}
\rotate
\tablecaption{Summary of YSOVAR (and closely related) Spitzer observations\label{tab:programs}}
\tablewidth{0pt}
\tablehead{
\colhead{Cluster} & \colhead{PID\tablenotemark{a}} &
\colhead{PI\tablenotemark{b}} & \colhead{Approx.\ Obs.} &
\colhead{Approx.\ } & \colhead{Date Observed} &  \colhead{Epochs} &
\colhead{Cadence\tablenotemark{c}} & \colhead{\# obj} & \colhead{Total
exp.} & \colhead{Notes} \\
 & & & \colhead{Center (J2000)} & \colhead{Ecl.Lat.} & &  & & \colhead{w/LCs\tablenotemark{d}} & \colhead{time (hr)}}
\startdata
AFGL 490       &  60014 & J.~R.~Stauffer & 03:27:24 +58:44:00 & +38 & 2011 Oct - Nov & 46 & fast & $\sim$1970 & 9.1 & 2 IRAC FOVs \\
NGC 1333       &  61026 & J.~R.~Stauffer & 03:29:06 +31:19:30 & +12 & 2011 Oct - Nov & 73 & fast & $\sim$690 & 18.5 & 2$\times$2 IRAC FOVs\\
Orion          &  61028 & J.~R.~Stauffer & 05:35:15 -05:21:00 & $-$28 & 2009 Oct - Dec; 2010 Oct - Dec & 80 & fast & $\sim$7500 & 365.2 & very large IRAC map ($\sim$0.9 sq.~deg.)\\
Orion          &  61028 & J.~R.~Stauffer & 05:35:27 -04:47:31 & $-$28 & 2011 Nov & $\sim$270 & staring & $\sim$180 & 10 & staring eclipsing binary follow-up \\
Orion          &  61028 & J.~R.~Stauffer & 05:35:27 -04:47:31 & $-$28 & 2011 Nov & 94 & EB & $\sim$480 & 14.7 & 2 FOVs, mapping; EB follow-up \\
Orion          &  70025 & J.~R.~Stauffer & 05:35:02 -05:18:30 & $-$28 & 2010 Nov - Dec & 74 & $\sim$hourly & \tablenotemark{e} &30.3 & Dipper follow-up \\
Mon R2         &  61025 & J.~R.~Stauffer & 06:07:48 -06:25:00 & $-$30 & 2010 Nov - Dec & 46 & fast & $\sim$710 & 5.6 & 1 IRAC FOV\\
GGD 12-15      &  61021 & J.~R.~Stauffer & 06:10:48 -06:12:30 & $-$30 & 2010 Nov - Dec & 77 & fast & $\sim$1010 & 14.8 & 2 IRAC FOVs\\
GGD 12-15      &  70172 & J.~Forbrich    & 06:10:48 -06:12:30 & $-$30 & 2010 Dec & $\sim$530 & staring & $\sim$370 & 20.0 & 1 staring FOV; simul.~w/ CXO, \& YSOVAR-like 2-field obs.~at start/end\\
NGC 2264       &  61027 & J.~R.~Stauffer & 06:41:04 +09:35:10 & $-$13 & 2010 Nov - Dec & 39 & fast &  $\sim$780 & 4.7 & one IRAC FOV\\
NGC 2264       &  80040 & J.~R.~Stauffer & 06:40:48 +09:42:00 & $-$13 & 2011 Dec & \tablenotemark{f} & CSI & $\sim$16,500 & 99.1 & Cy8 -- CSI 2264, staring \& mapping \\
NGC 2264       &  90098 & J.~R.~Stauffer & 06:40:48 +09:42:00 & $-$13 & 2013 Dec - 2014 Jan & 80 & CSI & $\sim$4700 & 30.3 & Cy9 -- CSI 2264 followup on $\sim$15 targets (cluster targets) \\
L1688          &  61024 & J.~R.~Stauffer & 16:27:10 -24:37:30 & $-$3  & 2010 Apr - May; 2010 Sep - Oct; 2011 Apr - May; 2011 Oct - Nov & 108  & fast/slow & $\sim$840 & 30.7 & 3 IRAC FOVs, fast cadence 2010 Apr-May\\
L1688          &  60109 & P.~Plavchan    & 16:27:31 -24:40:45 & $-$3  & 2010 Apr & $\sim$2500 & staring & $\sim$90 & 24.0 & reverberation mapping \\
L1688          &  90128 & H.~M.~G\"unther& 16:27:31 -24:40:45 & $-$3  & 2013 May - June & 10 & 3-4 days & $\sim$840 & 2.8 & Cy9 follow-up \\
Serpens Main   &  30319 & G.~Fazio       & 18:29:59 +01:13:53 & +24 & 2006 Sep & 29 & cryo & $\sim$2800 & 6.5 & cryo obs; same map as PC; 9 more hrs staring obtained on subset\\ 
Serpens Main   &  61029 & J.~R.~Stauffer & 18:29:59 +01:13:53 & +24 & 2011 May - Jun & 82 & fast & $\sim$3400 & 16.2 & 2 IRAC FOVs\\
Serpens South  &  61030 & J.~R.~Stauffer & 18:30:04 -02:02:05 & +21 & 2011 May - Jun & 82 & fast & $\sim$1540 & 10.0 & 1 IRAC FOV\\
IRAS 20050+2720&  61023 & J.~R.~Stauffer & 20:07:04 +27:29:14 & +46 & 2010 Jun - Aug & 102 & fast & $\sim$3030 &  13.4 & 1 IRAC FOV\\
IC 1396A       &   470  & B.~T.~Soifer   & 21:36:30 +57:29:48 & +64 & 2008 Jan - Feb & 29 & cryo & $\sim$4500 & 15.0 & 2$\times$3 IRAC FOVs; DDT; see Morales-Calder\'on \etal\ (2009) \\ 
IC 1396A       &   497  & J.~R.~Stauffer & 21:36:45 +57:30:20 & +64 & 2008 Dec &  10& cryo & $\sim$1800 & 1.5 & 1 IRAC FOV; DDT; 16 $\sim$contemporaneous MIPS 24 \mum\ epochs obtained, 2.1 more hrs. \\
IC 1396A       &  61022 & J.~R.~Stauffer & 21:36:30 +57:29:48 & +64 & 2009 Aug - 2010 Mar; 2010 Aug - 2011 Feb & 143\tablenotemark{g} & fast/slow & $\sim$5100 & 38.9 & 2$\times$3 IRAC FOVs; fast cadence 2009 Sep-Nov\\
Ceph C         &  61020 & J.~R.~Stauffer & 23:05:51 +62:30:55 & +59 & 2009 Dec - 2010 Mar; 2011 Jan - Mar; 2011 Sep - Nov & 39 & fast/slow & $\sim$1950 & 4.8 & 1 IRAC FOV; fast cadence 2010 Aug-Nov \\
(Ceph C)       & (61020)& (K.~Covey)     & 23:05:51 +62:30:55 & +59 & 2010 Aug - Nov  & 105 & fast & (as above)& 13.5 & 1 FOV; simul.~w/ CXO (CXO Cy11 P.I.:K.~Covey); AORs entirely included within program 61020 \\
\enddata
\tablenotetext{a}{PID=Program IDentification number; data from each
set of observations of each cluster was grouped within the same
program. The data associated with each PID can be retrieved all at
once from the Spitzer Heritage Archive using this number.
YSOVAR-classic programs are 60014 and 61020-61030, inclusive.}
\tablenotetext{b}{PI= Program's Primary Investigator.}
\tablenotetext{c}{Cadence, meaning rate at which observations were
obtained. For the YSOVAR-classic clusters, the fast/slow cadence is 
described in Sec.~\ref{sec:cadence}. Additional values found in this
column include staring (meaning that the IRAC observations were
staring rather than mapping mode), EB (observations designed for
follow up of specific eclipsing binary candidates), hourly, CSI
(cadence described in Cody \etal\ 2014), cryo  (observations conducted
in cryogenic era and cadence was different for each program).}
\tablenotetext{d}{Approximate number of unique objects with a light
curve in either or both IRAC channels. Counting just the original
YSOVAR programs, there are  $\sim$29,000 unique objects with a light
curve in either or both IRAC channels. Counting IRAC-1 and IRAC-2 for
the same object as two distinct light curves, there are $\sim$11,000
objects with light curves in both channels, and a total of
$\sim$39,000 light curves.}
\tablenotetext{e}{Counted as part of light curves listed for program 61028.}
\tablenotetext{f}{344 AORs used in this program.}
\tablenotetext{g}{One more epoch was also obtained during
the IRAC Warm Instrument Calibration (IWIC), which was before IRAC was
properly calibrated during the initial days of the warm mission, and
so that epoch is not included in our final dataset.}
\end{deluxetable}

\begin{deluxetable}{lcccccc}
\tabletypesize{\scriptsize}
\rotate
\tablecaption{Summary of YSOVAR Fast Cadence Observations\label{tab:programs2}}
\tablewidth{0pt}
\tablehead{
\colhead{Cluster} & \colhead{Dates Observed} & \colhead{total
$\Delta$t\tablenotemark{c}} & \colhead{Typical \#
epochs} & \colhead{Typical min($\Delta$t)\tablenotemark{d}} & \colhead{Typical
max($\Delta$t)\tablenotemark{d}} & \colhead{\# obj with $\geq$5 epochs,} \\
 & & \colhead{(d)}  & & \colhead{(d)}& \colhead{(d)} & \colhead{both
 channels}}
\startdata
AFGL 490                 & 2011 Oct 19 - Nov 25 & 37.7 & 46 & 0.34 & 1.95 &  $\sim$600 \\ 
NGC 1333                 & 2011 Oct 12 - Nov 14 & 33.5 & 72 & 0.04 & 1.95 &  $\sim$240 \\ 
Orion\tablenotemark{a}   & 2009 Oct 23 - Dec 1  & 39.0 & 80 & 0.21 & 0.80 &  $\sim$5200\\ 
Mon R2                   & 2010 Nov 8 - Dec 23  & 44.4 & 46 & 0.26 & 2.67 &  $\sim$240 \\ 
GGD 12-15                & 2010 Nov 16 - Dec 24 & 38.5 & 78 & 0.01 & 1.82 &  $\sim$340 \\ 
NGC 2264\tablenotemark{b}& 2010 Nov 18 - Dec 24 & 35.7 & 39 & 0.18 & 2.51 &  $\sim$260 \\ 
L1688\tablenotemark{a}   & 2010 Apr 12 - May 17 & 35.4 & 80 & 0.11 & 0.75 &  $\sim$200 \\ 
Serpens Main             & 2011 May 20 - Jun 24 & 34.7 & 81 & 0.09 & 0.75 &  $\sim$700 \\ 
Serpens South            & 2011 May 20 - Jun 24 & 34.7 & 81 & 0.09 & 0.75 &  $\sim$370 \\ 
IRAS 20050+2720          & 2010 Jun 12 - Aug 9  & 57.8 & 99 & 0.06 & 3.68 &  $\sim$690 \\ 
IC 1396A\tablenotemark{a}& 2009 Sep 13 - Nov 9  & 56.1 & 105& 0.13 & 2.52 &  $\sim$1480\\ 
Ceph C\tablenotemark{a}  & 2010 Sep 18 - Oct 30 & 42.0 & 106& 0.03 & 3.68 &  $\sim$400 \\ 
\enddata
\tablenotetext{a}{Additional observations exist in slower cadence.}
\tablenotetext{b}{Additional observations are part of CSI 2264.}
\tablenotetext{c}{Total change in time, in days, between first and last fast
cadence point.}
\tablenotetext{d}{Typical minimum or maximum change in time, in days,
between adjacent points in the time series. Also see
Fig.~\ref{fig:dtpercluster} below, which has histograms of the time
step distributions.}
\end{deluxetable}

\begin{deluxetable}{lp{3.4cm}p{4.4cm}llp{5cm}}
\tabletypesize{\scriptsize}
\rotate
\tablecaption{Summary of Cryo-Epoch Spitzer observations\label{tab:cryoprograms}}
\tablewidth{0pt}
\tablehead{
\colhead{Cluster} & \colhead{AORKEY\tablenotemark{a}}
& \colhead{Obs.~Date} & \colhead{Adop.~Date\tablenotemark{b}} & $\Delta(t)$\tablenotemark{c}& \colhead{Notes}  }
\startdata
AFGL 490       & 3654656, 3663104 & 2004-10-08, 2004-09-24 & 2004.9 & 6.9 & Reduction from G09\\
NGC 1333       & 3652864, 5793280, 4316672 & 2004-02-10, 2004-09-08, 2004-02-03 & 2004.5 & 7.3 & Reduction from G09, Gutermuth \etal\ (2008b); some data from c2d included where necessary \\
Orion          & (many) & 2004-03 -- 2004-10 & 2004.5 & 5.9 & Reduction from Megeath \etal\ (2012) \\
Mon R2         & 3659008, 3668224 & 2004-03-13, 2005-04-07 & 2004.3 & 6.6 & Reduction from G09\\
GGD 12-15      & 3658752, 3667968 & 2004-03-31, 2004-03-15 & 2004.3 & 6.6 &Reduction from G09\\
NGC 2264       & 3956480, 3956736, 3956992, 3957248 &  2004-03-06, 2004-10-08, 2004-03-06, 2004-10-08 & 2004.5 & 6.4 &re-reduced in G09 style \\
L1688          & 3652096, 4321280 & 2004-03-07, 2004-08-26 & 2004.3 & 6.7 & Reduction from G09; some data from c2d included where necessary\\
Serpens Main   & 3652352, 3661568, 12649216 & 2004-04-01, 2004-04-06, 2005-04-12& 2004.3 & 7.2& Reduction from G09; some data from c2d included where necessary\\
Serpens South  & 19957248, 19958016, 19998464, 19998720, 20002304, 20002560, 20018688, 20018944& 2006-10-27, 2006-10-27, 2006-10-28, 2006-10-28, 2006-10-27, 2006-10-27, 2006-10-28, 2006-10-28& 2006.9 & 4.6 & Reduction from Gutermuth \etal\ (2008b) \\
IRAS 20050+2720& 3656448, 3665152 & 2004-06-09, 2004-09-25 & 2004.5 & 6.1 & Reduction from G09\\
IC 1396A       & 3959040, 4316416 & 2003-12-20, 2004-06-23 & 2004.0 & 6.4 & Re-reduced in G09 style \\
Ceph C         & 3655936, 3664384 & 2003-12-19, 2003-12-19 & 2004.0 & 6.4 & Reduction from G09\\
\enddata
\tablenotetext{a}{An AOR is an Astronomical Observation Request, the
fundamental unit of Spitzer observing. An AORKEY is the unique 8-digit
integer identifier for the AOR, which can be used to retrieve these
specific data from the Spitzer Heritage Archive.}
\tablenotetext{b}{This is the adopted date of the IRAC observations
used in the present paper for the cryo epoch (e.g., the time that was
used to represent the date of those observations taken over more than
one epoch). Note that two clusters have monitoring observations taken
during the cryo era, but this $\sim$single-epoch earliest cryo
observation is what is used in the present paper as the `cryo epoch'
for comparison to the `YSOVAR epoch.' }
\tablenotetext{c}{$\Delta(t)$, in years, between adopted cryo epoch 
and mean YSOVAR fast-cadence time.}
\end{deluxetable}

\subsection{Warm Spitzer Observations: General properties}
\label{sec:newspitzeroverview}

To better manage the observation planning and data downloading for the
12 clusters we observed, we separated each cluster into an individual
observing program.  The individual observing programs and clusters are
listed in Table~\ref{tab:programs}; one can download the data
from the Spitzer Heritage Archive (SHA) using these program numbers. 
We will deliver all of our extracted photometry to the Spitzer Science
Center (SSC) and Infrared Science Archive
(IRSA\footnote{http://irsa.ipac.caltech.edu/}) for public access via
IRSA tools, and through those tools, the Virtual Observatory.

Most of the observations were IRAC mapping mode Astronomical
Observation Requests (AORs) using full-array 12 sec high-dynamic-range
(HDR) mode (which is defined such that a 0.4 and 10.4 sec exposure is
obtained at each pointing -- see the IRAC Instrument Handbook,
available at the SSC/IRSA
website\footnote{http://irsa.ipac.caltech.edu/data/SPITZER/docs/}). 
In some cases, as noted in Table~\ref{tab:programs}, the AORs were
staring AORs, meaning that they were continuous or semi-continuous
observations without dithering or mapping.  These staring data will be
discussed in other YSOVAR papers. The present paper is limited to the
HDR mapping observations (and is largely further restricted to the
`fast cadence' observations; see Sec.~\ref{sec:cadence}). 

Roughly half of the original YSOVAR time allocation was devoted to
observations of Orion.  The Orion Spitzer campaign included a
$\sim$0.9 square degree region centered on the Trapezium cluster in
Orion; this is far larger than can be
obtained in a single AOR at a single epoch. The observed area was thus
broken into five segments with a central region of $\sim$20$\arcmin
\times$25$\arcmin$ and four flanking fields. The central part was
observed in full array mode with 1.2 seconds exposure time and 20
dither positions to avoid saturation by the bright nebulosity around
the Trapezium stars. The remaining four segments of the map were
observed in 12 sec HDR mode, and 4 dither positions. More details on
these original YSOVAR Orion observations appear in MC11.  We
reallocated time within the YSOVAR time budget to follow up some
eclipsing binaries in Orion (Morales-Calder\'on \etal\ 2012). We also
obtained time in Cycle 7 to follow up some AA~Tau analogues in Orion
presented in MC11. These follow-up observations did not map the entire
Orion region.

The rest of the original YSOVAR time allocation was spread among the
rest of the star-forming regions. In contrast to Orion, for the 11
other clusters, the monitoring observations are one or at most a few
IRAC FOVs; see the last column of Table~\ref{tab:programs}.  As noted
above, we sometimes refer to these ``11 other clusters'' as ``the
smaller field clusters.''

The observations are typically spread over weeks to months, usually at
a cadence of about twice per day, but with an interval between
observations that varied both by design (to reduce aliasing problems)
and due to constraints imposed by other Spitzer programs or
infrastructure operations that were executed during the same campaigns
(more on the cadence below in Sec.~\ref{sec:cadence}).  

NGC~2264 was included as a smaller-field cluster to be monitored as part of
the original YSOVAR program, and while this original program was still
executing, CSI 2264 was approved. We continued to execute the original
YSOVAR small-field observations in this region (program 61027 in
Table~\ref{tab:programs}), and these small-field regions are discussed
here, since they resemble the rest of the original YSOVAR progams more
closely than they do CSI 2264. 

Two of our clusters, IC~1396A and Serpens Main, were monitored in the
cryogenic era with Spitzer with the primary intention of monitoring
changes in these objects in the Spitzer bands (as opposed to removing
artifacts). In YSOVAR, we re-observed these clusters in the same way
such that the same objects continue to be monitored, and we re-reduced
all of these data in the same fashion as discussed here. In the
context of this paper, we are not particularly focused on these
cryogenic-era monitoring data; however, these data will be included in
the YSOVAR cluster-specific papers.

\begin{figure}[ht]
\epsscale{0.5}
\plotone{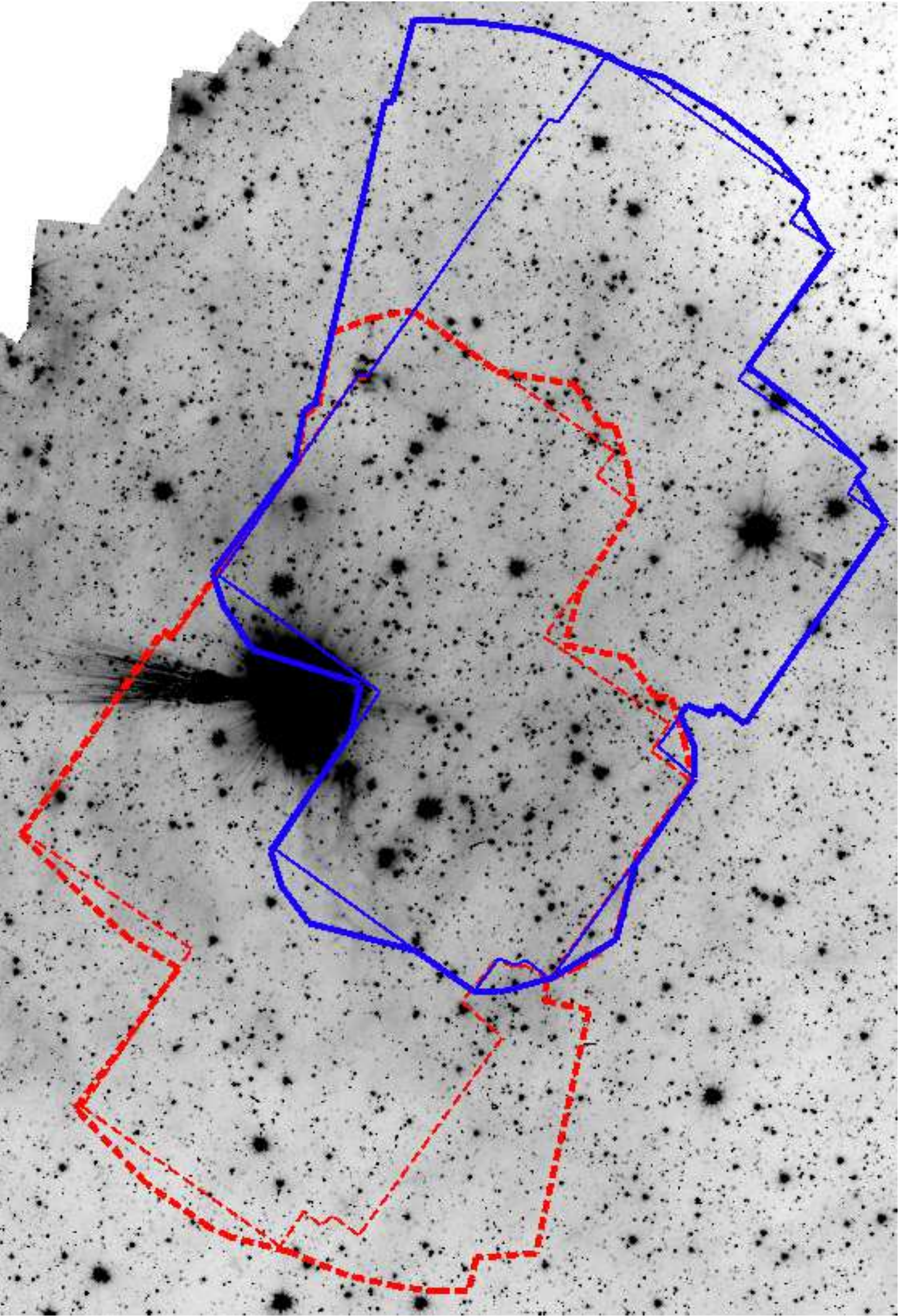}
\caption{The approximate sky coverage for a summed-up image consisting
of all epochs of YSOVAR AFGL 490 observations, superimposed on a
reverse greyscale image of AFGL 490 at 4.5 \mum\ obtained during
the cryogenic mission. The thicker blue solid line is 3.6 \mum\ 
and the thicker red dashed line is 4.5 \mum. A single epoch of
observation is also indicated by thinner blue solid and red dashed
lines, with the difference between the single epoch and the larger
polygon due to (ecliptic latitude dependent) field rotation effects.
North is up and east is to the left. The distance between the farthest
north and farthest south coverage here is $\sim$27$\arcmin$. The field
rotation here is not as significant as in those with ecliptic
latitudes $\sim$60$\arcdeg$. There is no Chandra coverage for this
cluster. Similar Figures for the remaining 11 clusters are included in
Appendix~\ref{sec:footprintfigs}.}
\label{fig:afglfootprints}
\end{figure}

\clearpage

\subsection{Warm Spitzer Observations: Footprints and Operational Constraints}
\label{sec:footprints}

Because of the nature of Spitzer and IRAC, the footprints of our
observations and how they vary with time are both complicated issues.
We now discuss these issues as they pertain to YSOVAR observations.

Figure~\ref{fig:afglfootprints} (and the analogous
Figures~\ref{fig:n1333footprints}--\ref{fig:cephcfootprints} in
Appendix~\ref{sec:footprintfigs} for the remaining 11 clusters)
present the outline (footprint) of the YSOVAR observations. The
original YSOVAR Orion observations (Fig.~\ref{fig:orionfootprints})
and the CSI 2264 observations (Fig.~\ref{fig:csi2264footprints}) both
cover a substantially larger area than those for the other,
smaller-field clusters. Chandra footprints are included in these
figures for reference, and discussed below in
Sec~\ref{sec:chandrareduction}.

The focal plane of the IRAC camera is such that the 3.6 and 4.5 \mum\
fields of view (FOVs) are not the same; data are obtained
in both FOVs at once, with one field placed on the target of interest,
and the other obtaining serendipitous data in a non-overlapping
$\sim5\arcmin\times5\arcmin$ field with a center offset
$\sim6.5\arcmin$ from the target field; see the IRAC Instrument
Handbook for more details. The placement of these FOVs also changes
with time. For all Spitzer observations, as discussed in the Spitzer
Space Telescope Handbook (also available at the SSC/IRSA
website\footnote{http://irsa.ipac.caltech.edu/data/SPITZER/docs/}),
the ecliptic latitude of the target defines when one can observe the
target and for how long. At any one time, Spitzer can observe in an
annulus defined by the operational pointing zone, which can be 
conceptually summarized as ``neither too close nor too far away from
the Sun." It is about 40$\arcdeg$ wide, and rotates with the Sun at a
rate of about a degree a day. An object near the ecliptic plane thus
can only be observed for a period of about 40 days, twice a year;
objects near the ecliptic pole can be observed at any time during the
year, for as long as needed.  Related to this, the IRAC FOV, as
projected onto the sky for any given object on the ecliptic equator,
is at an essentially constant angle with respect to North for the
duration of the $\sim$40 day observing window, and, $\sim$6 months
later, is flipped by 180$\arcdeg$ but then also essentially constant
for that $\sim$40 day observing window. However, the FOV for an object
at the ecliptic pole rotates by about a degree a day.

For each of the Figures showing the IRAC footprints
(Fig.~\ref{fig:afglfootprints};
Figs.~\ref{fig:n1333footprints}--\ref{fig:cephcfootprints}), the
projected outline of the observation covered by the entire
YSOVAR data set is indicated on top of a 4.5 \mum\ observation
obtained (in most cases) during the cryogenic era. The different
regions observed by the 3.6 and 4.5 \mum\ cameras are identified; the
nominal target of the observation is covered in both FOVs. In addition
to the target of the observations, there are serendipitous data
obtained offset from the target area, as seen in the Figures. The
footprint of the serendipitously obtained data in each band can change
considerably depending on the time of observation and the ecliptic
latitude of the target. For targets at higher ecliptic latitudes,
where field variation with time is important, a single epoch of
observation is indicated in the Figure in addition to the area covered
by the entire YSOVAR data set.  

The approximate ecliptic latitude for each target is included in
Table~\ref{tab:programs} as a rough indication of how long the
observing window was and how much field rotation one can expect to see
in these Figures.  For Orion, we designed our observations to be less
sensitive to the rotation; that plus the fact that the Orion map
contains many FOVs means that the rotation is much less of a factor in
terms of how much of the sky may be included in both channels (as
opposed to just one). For the smaller-field cluster observations
containing only 1-4 IRAC FOVs, a primary target field is monitored in
both IRAC bands, but substantial serendipitous monitoring data have
also been obtained, typically in just one band. For most of the
smaller-field YSOVAR cluster targets, such as NGC 1333
(Fig.~\ref{fig:n1333footprints}), the central region has complete
2-band coverage, and most of the objects in the region of the
single-band coverage north and south of the target field have data
over most if not all of the campaign. For L1688
(Fig.~\ref{fig:l1688footprints}), with an ecliptic latitude of
$-$3$\arcdeg$, the field rotation is essentially zero during
one $\sim$40 day window. However, it was monitored over more than one
window, so the observations flip by 180$\arcdeg$ for the next $\sim$40
day window. (The targets of observation in
Fig.~\ref{fig:l1688footprints} are the three regions with brighest
stars and nebulosity in the centers of the three `stripes' of
coverage.) In contrast, for IC 1396A or Ceph C (ecliptic latitudes of
$\sim$60$\arcdeg$), the central $5\arcmin\times5\arcmin$ region is
covered throughout the monitoring window, but the rapid field rotation
creates a ``fan'' of coverage such that the serendipitous coverage in
the outer periphery of those targeted regions have data in just one
band, and only for a fraction of the campaign.  Some objects are
monitored in 3.6 \mum\ at the beginning of the campaign, and in 4.5
\mum\ at the end of the campaign (see
Figs.~\ref{fig:ic1396footprints} and \ref{fig:cephcfootprints}).  

If one considers only objects with light curves in both channels for
the smaller-field clusters, even for the fields that do not rotate,
one loses a significant fraction of the available data; while the
nominal targets of our observations are the regions with two-band
coverage, there are still good light curves for cluster members in the
regions with coverage in only one band. Generally, for the
smaller-field clusters, we did not require data in both channels, and
instead retained all light curves for our analysis, even if obtained
in only one channel (Sec.~\ref{sec:standardsetforstatistics}). Note
also that the highest ecliptic latitude among our targets is
$\sim$65$\arcdeg$; no higher ecliptic latitude targets were included,
despite the possibility of much longer observing windows, at least in
part because we could find no suitable star-forming regions at these
latitudes at the time we planned our observations. The field rotation
for these targets at $\sim$65$\arcdeg$ is already substantial.

In general, because of the field rotation, as any given object moves
in to (or out from) the mapped region, the first (or last) few epochs
may be unusable as the object may appear in only a fraction of the
dithers at that position and the photometry may thus be compromised by
edge effects. Substantial field rotation with time can have a
significant effect on faint stars near bright stars. Diffraction
spikes of bright stars rotate with time within the IRAC FOV, so if a
target star falls near a diffraction spike, this can introduce a
source of false longer timescale variability; also see
Section~\ref{sec:spitzernewdata}. 

We note here that the low ecliptic latitude of L1688 (and specifically
the 180$\arcdeg$ flip between observing windows) enables G\"uenther
\etal\ (2014) to use these observations to characterize possible
artifacts in the light curve related to the position in each frame.
These tests confirm that the statistical criteria for selecting
variable sources (Sec.~\ref{sec:findingvars} below) is restrictive
enough that differences between positions within the frame cannot be a
primary contributor for the sources identified as variable.

\subsection{Warm Spitzer Observations: Observing Cadence}
\label{sec:cadence}

More than 95\% (524 hours) of our original YSOVAR Spitzer Cycle-6
program observing time was devoted to a ``fast cadence'' mode,
designed to be sensitive to timescales from $\sim$0.15 to $\sim$40
days, consistent with the known timescales of YSOs due to
accretion-related flickering, rotational modulation of star spots and
other effects.  The upper limit to this range of timescales was set by
the typical duration of Spitzer visibility windows near the ecliptic
plane, as discussed above in Sec.~\ref{sec:footprints}.

Sampling a light curve at evenly spaced intervals introduces period
aliasing, and a bias towards the detection of false periods at integer
fraction multiples of the sampling interval (see, e.g., Plavchan
\etal\ 2008b or Dawson \& Fabrycky 2010).  This period aliasing is
common at integer fraction multiples of 1 day in ground-based
photometric surveys due to the natural day-night cycle of the Earth's
rotation.   Given typical YSO rotation periods of one to several days,
we therefore chose to execute our program with a cadence designed to
be compatible with the Spitzer scheduling, and minimize
period-aliasing while retaining sensitivity to variability on a broad
range of timescales.  

The fast cadence mode for each region had a total time baseline of
$\sim$40 days, varying depending on the actual duration of the
visibility window, with additional days on either end of the
visibility window for scheduling flexibility. The actual total
duration of the fast cadence observations for each cluster is listed
in Table~\ref{tab:programs2}.  In particular, for AFGL 490, Mon R2,
GGD12-15, IRAS 20050+2720, IC 1396A, and Ceph C, we originally planned
a 42 day timespan; for Orion, Serpens Main, and Serpens South, we
planned a 38.5 day timespan; and for NGC 1333, NGC 2264 and L1688, we
planned a 35 day timespan. We divided the time baseline into a
repeating set of 3.5 day sub-cadences to ease scheduling.  Within
each 3.5 day period, we made 8 visits, with time steps between visits
of 4, 6, 8, 10, 12, 14, and 16 hours respectively. By using a linearly
increasing time-step, we were able to evenly sample in Fourier space a
range of higher frequency (shorter timescale) photometric variability
than we would have been able to otherwise.  Additionally, we were also
able to minimize the total amount of necessary observing time to
sample these high frequencies and to minimize the period aliasing. 
For comparison, splitting 8 visits evenly across a 3.5 day cadence
would have removed our sensitivity to variability timescales shorter
than $\sim$0.5 days.

To further accommodate scheduling flexibility, we also included a
window of $\pm$2 hours in which to obtain a given epoch of observation
for our fast cadence observations.  This $\pm$2 hour window
effectively resulted in a randomization of the cadence about our
desired times of observation, which had the additional
benefit of further reducing period aliasing, in particular aliases
with a period of 3.5 days.  We also allowed for interruptions
due to data downlink transfers and higher priority observations such
as staring mode observations of transiting exoplanets. 
We are grateful to the SSC scheduling staff for accommodating this
fairly complex cadence over the two years of the survey program.

\begin{figure}[ht]
\epsscale{0.8}
\plotone{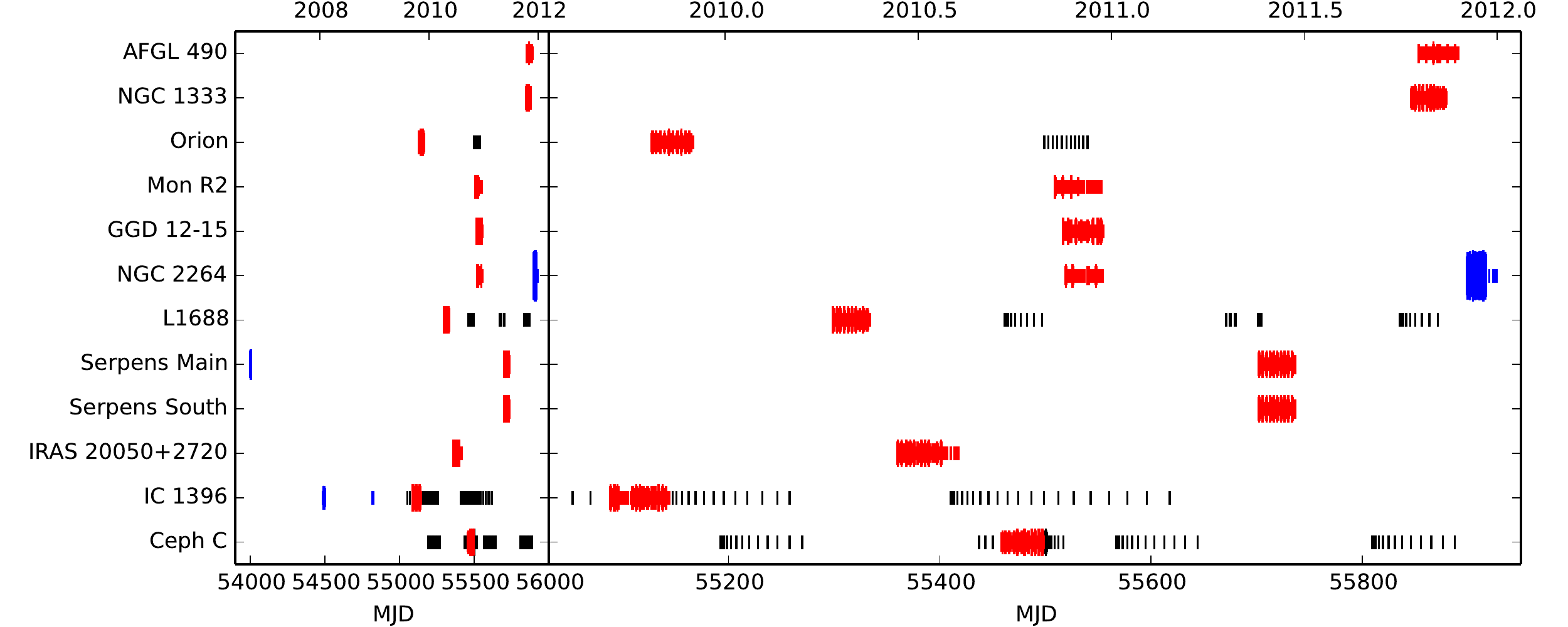}
\caption{Visual overview of the observing seasons and the cadence for
all clusters in the YSOVAR project. On the left side, all of the
YSOVAR data are depicted  (where Spitzer's cryogen ran out 15 May
2009, MJD 54966.925). On the right side is an expanded view of the
data highlighted here and used for the `standard set for statistics'
(see Sec.~\ref{sec:standardsetforstatistics}). The designated ``fast
cadence" observations are shown in red, with other  YSOVAR-classic
observations shown in black. Other YSOVAR-affiliated programs are
shown in blue; the large post-cryo monitoring of NGC 2264 is CSI 2264,
and the cryogenic observations of IC 1396A and Serpens are also shown.
Each individual observation is marked by a ``$|$'' symbol in which 
the length of the line is proportional to the number of observations
in a window 12 hours before and after that particular observation.
When the observations are so dense that the symbols overlap, the
frequency of observations can thus be judged from the thickness of the
line. An enlargement of the red, fast-cadence sequences can be found
in Figure~\ref{fig:fastcadence} below.}
\label{fig:cadence}
\end{figure}

Overall, we obtained $\sim$40-100 epochs per cluster, which are
visualized in Figure~\ref{fig:cadence}. Table~\ref{tab:programs} lists
all the clusters, with information about the observations (including
the total number of epochs); Table~\ref{tab:programs2} summarizes just
the fast cadence observations, including typical values of the minimum
and maximum size of the timestep between epochs, and the total time
baseline.  Our unevenly spaced epochs enables a characterization of
different types of observed variability on a variety of timescales;
see further discussion below (Sec.~\ref{sec:varlimits}).   While
periodic analysis tools take advantage of such unevenly spaced data,
Findeisen \etal\ (2014) point out that for auto-correlation functions
of non-periodic light curves, timescale sensitivity may be limited by
the largest (not the smallest) adjacent timestep in the time series.

For three star-forming regions (IC 1396A, L1688, and Ceph C) with
longer visibility windows, or that we observed in multiple visibility
windows, we also observed in a ``slow cadence." Aside from the 42 day
fast cadence, we executed visits with time-steps between visits of 1,
2, 3, 4, 5, etc.\ days to cover the remaining duration of the
visibility window. These observations constituted $\sim$5\% of our
Cycle-6 observing program (25 hours), and these observations allowed
us to be sensitive to variability timescales of $\sim$40 days to
$\sim$2 years, e.g., to objects such as KH15D (e.g., Winn \etal\ 2006)
and WL4 (Plavchan \etal\ 2008a), as well as long term trends on
timescales of a year identified in the NIR (e.g., Parks \etal\ 2014).  For
L1688 and Ceph C, we also executed this slow cadence observing mode
during the three other visibility windows during our survey. 
Additionally, Orion had some slower cadence follow-up, and NGC 2264
had the CSI 2264 program at a different cadence.

For the primary statistical analysis discussed in this paper, we use
the ``standard statistical sample,'' e.g., only the fast cadence data
(where there are at least 5 epochs in the lightcurve; see
Sec.~\ref{sec:sampledefinition} below). For about half of the
clusters, this is the entirety of the data set; see
Tables~\ref{tab:programs} and \ref{tab:programs2}, and the red points
in Figure~\ref{fig:cadence}. A detailed analysis of the ensemble of
all of these observations for each cluster will be presented in papers
in preparation, customized to each cluster.

For completeness, we note here the following specifics about those
clusters with data beyond the YSOVAR fast cadence. L1688, IC 1396A,
and Ceph C all have slow cadence data. Orion has slower cadence data
over a second year. Data for Ceph C over its fast cadence window
include those originally obtained as part of the original YSOVAR
project, as well as data taken using a very similar distribution of
time steps as part of a YSOVAR-related Chandra program; see
Table~\ref{tab:programs}. For consistency with the maximum timescales
sampled in other clusters, we define the Ceph C fast cadence light
curve to be a window spanning 09-19-2010 through 42 days later, which
is more representative of the fast cadence window in the other
clusters.  We note, however, that the sampling within this window is
nonetheless enhanced relative to other smaller-field clusters, due to
the inclusion of extra Spitzer observations obtained to support the
coordinated Chandra observations. Two other clusters (IC 1396A and
IRAS 20050) also sample longer timescales than the more typical
$\sim$40 day window (see Figure~\ref{fig:fastcadence} below for an
enlargement of the fast cadence). We did not truncate the last few
slower cadence points from the IRAS 20050 time series because there
are no other slower cadence observations, and truncating it would
effectively `orphan' the last few points. The IC 1396A cadence doesn't
provide a clean breakpoint, so it also has points covering a slightly
longer window than `typical' included in its fast cadence. (See
Table~\ref{tab:programs2} for the total time in the fast cadence for
all clusters.) IC 1396A was one of the first star-forming regions to
be monitored intensively with Spitzer (Morales-Calder\'on \etal\
2009), so has additional cryo-era monitoring not shown in
Fig.~\ref{fig:cadence}. Serpens Main also has some cryo-era
monitoring.  GGD 12-15 and L1688 also have staring data, not included
here.

\subsection{Warm Spitzer Observations: Data Reduction}
\label{sec:spitzernewdata}
\label{sec:spitzernewdatadatared}

We started with the IDL package Cluster Grinder (G09), which has been
used by several other projects (e.g., Megeath \etal\ 2012), and then
made custom modifications to make it suitable for this time-series
data set.  We started with the basic calibrated data (BCD) images
released by the SSC. The BCD images used correspond largely to
software version S18.18 (with about 20\% S19.0 and S19.1); earlier
reductions such as those in MC11 used a mix of S18.12, 18.14 and
18.18.  Each BCD frame was processed for standard bright source
artifacts and combined into mosaics (at 0.86$\arcsec$ per pixel grid
resolution) by exposure time (long or short), epoch, and bandpass.
Cosmic ray hits and other transient artifacts are flagged during
mosaic construction using redundant data at each pixel position in the
final mosaic grid as a reference for the nominal value and allowable
internal variation.  

Since the HDR mode obtains a long and a short frame at each pointing,
we need to combine data from the two exposures. HDR mosaics are merged
together on a pixel-by-pixel basis, with appropriately scaled
short-frame HDR values above a set threshold (individual pixel values
corresponding to sources $\sim$10th mag, dependent on background
level) replacing significantly compromised pixels in the long-frame
HDR mosaic.  

Point source detection and aperture photometry are then performed for
each channel and each AOR/epoch.  The aperture radius was
2.4$\arcsec$, with a sky annulus having inner and outer radii of
2.4$\arcsec$ and 7.2$\arcsec$. We adopted Vega standard magnitude zero
points of 19.30 and 18.64 for 1 DN s$^{-1}$ total flux at 3.6 and 4.5
\mum, respectively. These values include standard corrections for our
chosen aperture and sky annulus sizes (Reach \etal\ 2005; Carey \etal\
2010).   

Extensive tests were run on aperture photometry obtained in three
different ways.  Our initial approach was to obtain photometry from
the mosaic created for each AOR, with the assigned time for that point
being the average time for the AOR. We also obtained photometry on
each BCD frame individually, with the individual, specific time of
that observation being assigned to that point. Finally, we took the
set of all individual BCD measurements for a given source within a
given AOR, and took the mean brightness and the mean time for that set
of observations.  This latter approach proved to have the best quality
photometry. It facilitated the treatment of two well-known systematic
effects (residual gain and pixel phase effects) while keeping our
signal-to-noise ratios (SNRs) close to ideal. It provided the highest
SNR light curves, even if the time resolution is slightly lower than
what might be theoretically possible. Effectively, this choice means
that a SNR $\sim$5 is required for a viable source in a given BCD,
such that the net SNR of the source over the whole AOR is at least
$\sim$10. If a source fell on the edge of the BCD such that the two
native pixel radius was not entirely within a given BCD, no
measurement is obtained.  Measurements were also dropped as non-viable
if the measurement (in the aperture or annulus) included a masked IRAC
pixel such that obtaining a finite measurement was not possible, or if
the source was faint enough (or the sky variation high enough) that
the net measured flux was $<$0.  If there were three or more valid
detections of a given source in a given AOR, we could perform outlier
rejection, and outliers were flagged; the rest of the measurements
were combined by uniform weight arithmetic mean.  The uncertainties
were added in quadrature and divided by the included measurement
count. 

Source matching between epochs and wavelengths was performed by
position, taking the nearest source within 1$\arcsec$. Final positions
for each source were the mean of the individual measurements. Our
astrometric residuals as compared to 2MASS suggest an uncertainty of 
$<$200 mas root-mean-square (RMS), after a single iteration refinement
of the astrometry using a first set of mosaics constructed blindly
with the BCD-delivered world coordinate system (WCS).  Objects with
fewer than five viable detections over separate epochs in a single
band were not retained as valid light curves. 

We note here that there are some residual instrumental column pulldown
effects (see the IRAC Instrument Handbook for more information) in
some areas near bright sources. This is a more significant problem for
clusters with a large number of very bright sources (e.g., Mon R2; see
brightnesses in  Fig.~\ref{fig:afglfootprints},
Figs.~\ref{fig:n1333footprints}--\ref{fig:cephcfootprints}).  The
location of these artifacts can change with time, particularly in
fields at high ecliptic latitude and thus with significant field
rotation during a given visibility window (see
Sec.~\ref{sec:footprints}), and can have particularly significant
effects on sources fainter than $\sim$12th mag. Objects
falling in such regions that are faint and that had non-repeating
structure in their lightcurves should be particularly scrutinized
(including visual inspection of the input frames) for the validity of
their lightcurves. Points obviously compromised by these effects
should be additionally identified as non-viable points and rejected.
Discussion of this process for individual objects will appear in the
clusters papers.

Even after all of the processing to this point, some occasional
outliers in the light curves remain. To identify outliers within the
fast cadence data, we initially identified points several $\sigma$
away from the mean for each light curve. We used this approach in
MC11, with a different (much earlier) processing of the original
Spitzer data. However, our more recent, improved processing described
here reduced the number of outliers. We experimented with algorithms
for rejecting outliers from each light curve, but were unable to
identify a technique that rejected visually spurious points while
preserving apparent bona fide variability.  An approach as we used in
MC11 would, for example, drop points at the beginning and end of a
light curve with a long trend. Our estimates suggest that implementing
an outlier rejection strategy affects $\lesssim$10\% of the light
curves tagged as intrinsically variable (as per methods described
further below), so we retained all points in the light curves. 
Outliers may still be identified in a few specific individual cases;
those outliers are then effectively dropped from the light curve, and
are discussed where relevant.

\subsection{Warm Spitzer Observations: Noise Floor}
\label{sec:noisefloor}

\begin{figure}[ht]
\epsscale{0.8}
\plotone{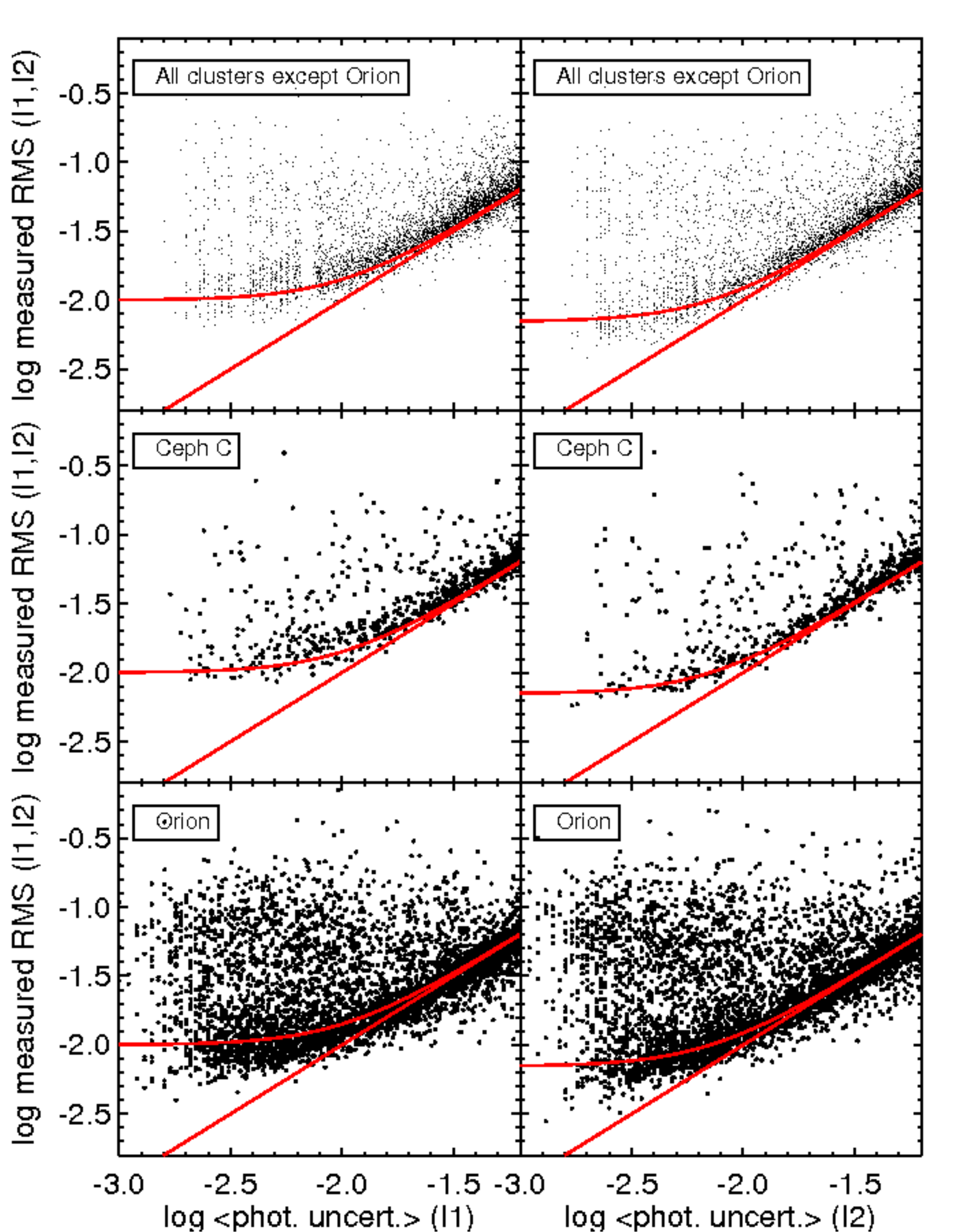}
\caption{Plots of measured RMS ($\sigma$) vs.\ median photometric
uncertainty for I1 (left) and I2 (right). The top row is for all
clusters except Orion, the middle row is just Ceph C, and the bottom
row is just Orion. The straight red line is the 1-to-1 relationship
between these parameters, e.g., the expected relationship if there was
no `noise floor'; the curved red line is the empirically derived curve
which we used to determine a noise `floor' of 0.01 mags for I1 and
0.007 mags for I2, which we then added in quadrature to the individual
errors obtained for each point. To appear in this plot, objects must
have more than 20 epochs. Ceph C appears separately to give an
indication of what these plots look like for individual clusters. We
combined all the clusters together to better determine the empirical
floor, even though there are some variations among the clusters; see
the text. }
\label{fig:fixerrors}
\end{figure}

\begin{figure}[ht]
\epsscale{0.8}
\plotone{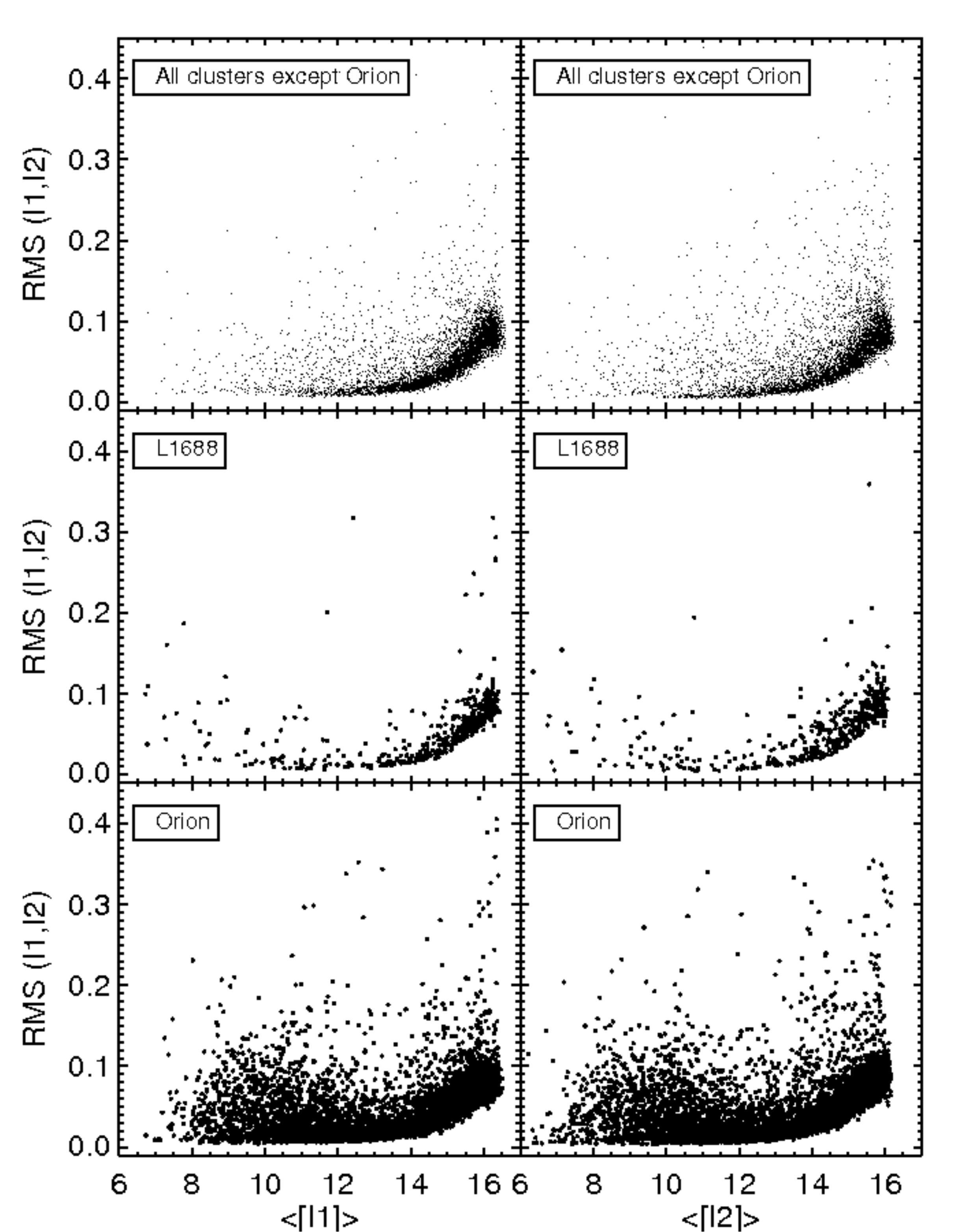}
\caption{Plots of measured RMS ($\sigma$) vs.\ mean magnitude for I1
(left) and I2 (right). The top row is for all clusters except Orion,
the middle row is just L1688, and the bottom row is just Orion. There
is more, and more significant, variation among the brighter sources.
(Some very large RMS values fall above the range shown in this plot;
the range is limited here for clarity.)
\label{fig:rmsmag}}
\end{figure}

Many tests of variability (e.g., a $\chi^2$ test; see
Sec.~\ref{sec:findingchi2}) depend on  accurate error estimates for
each data point in the time series. Initial estimates of uncertainties
were derived by combining three terms in quadrature (see also
discussion in Megeath \etal\ 2004): shot noise in the aperture, shot
noise in the mean background flux per pixel integrated over the
aperture, and the standard deviation of the sky annulus pixels to
account for the influence of non-uniform nebulous background.
Such error values proved to be a slight underestimate of
the true uncertainty, and do not take into account a `floor' in the
errors from calibration errors, primarily the intrapixel gain
variation. 

We wanted to find a value for this floor that worked for all clusters
(since all of the mapping observations discussed here were obtained in
the same way). We also wanted to be conservative in that we wanted the
set of objects we selected as variable at the end of this process to
be reliably variable (with very few if any non-variables selected as
variable), even if it meant that our sample would not be complete in
that some legitimately (lower-level) variable objects would not be
selected. In other words, we wished to err slightly on the side of
overestimating errors. We investigated the RMS error for each light
curve for each object vs.\ magnitude and vs.\ the median photometric
uncertainty for all sources. In MC11, we estimated the true error as a
function of magnitude to be single-valued as a function of magnitude
bin. In general, this means that the errors will be correct on
average, but there will be specific objects (e.g., those in high
background regions) for which they are too low.  

Here, we aimed to improve on this estimate by finding a value for the
noise floor that can be added in quadrature to the error estimates for
each point, preserving the errors obtained for each point, localized
in time and space. For each channel, for every object in each cluster
having at least 20 epochs in its light curve so as to ensure
well-determined values, we plotted the measured RMS ($\sigma$) of the
light curve against its median photometric uncertainty; see
Figure~\ref{fig:fixerrors}.  Orion alone represents about a quarter of
the light curves shown in Fig.~\ref{fig:fixerrors}, so it is useful in
some contexts to separate Orion from the analysis of the rest of the
ensemble. Data for an individual smaller-field cluster are thus
shown in the Figure for reference.

The largest amplitude sources in Fig.~\ref{fig:fixerrors} are true
variables.  To determine the noise floor, we are interested in how the
observed RMS for non-variables over many epochs compares to the formal
errors. The bulk of the distribution of points in
Fig.~\ref{fig:fixerrors}, e.g., the non-variable and less noisy
sources in the ensemble of points, fall in the same region in each
plot; the distributions run smoothly down to an asymptotic value.
However, there are three components contributing to the overall
scatter of the regions with more objects seen in
Fig.~\ref{fig:fixerrors}, not necessarily seen as distinct features
there. As we move to brighter objects (generally those with smaller
median photometric uncertainty), we have a statistically larger chance
that the objects will be cluster members, e.g., legitimately young and
therefore expected to have a greater likelihood of being highly
variable. There is a transition between the short and long HDR frames
(at $\sim$10th mag) which affects the measured error for sources that
are faint in the short frames but not in the long frames. There are
variations in depth ($\sim$16th-17th mag) reached across clusters due
to variable and nebulous backgrounds (the variations in depth can also
be seen in the faint limits of
Fig.~\ref{fig:jdistpart1}-\ref{fig:jdistpart2}, discussed below) such
that a given median photometric uncertainty need not necessarily refer
to stars of roughly the same brightness among the clusters.
Fig.~\ref{fig:fixerrors} includes an individual cluster (Ceph C) to
give an indication of this cluster-dependent variation in the floor;
it can be seen that fewer points fall below the red line on the left
side of each panel in Ceph C than they do in the Orion plots on the
bottom row.

Close inspection of Fig.~\ref{fig:fixerrors} reveals that the
asymptotic noise floor value is on average slightly higher for the 3.6
\mum\ channel and slightly lower for the 4.5 \mum\ channel.
This is consistent with expectations that the dominant source of error
in the asymptotic noise floor is from residual intrapixel gain
variations, since the intrapixel gain effect is lower in I2 than I1. 
There are also more points with significantly larger $\sigma$ in the
4.5 \mum\ channel. The 4.5 \mum\ channel has more contributed noise in
any given light curve -- shocks and scattered light, which are very
common in our target regions, contribute to the overall scatter in
$\sigma$ at 4.5 \mum. There is also some lower-level contribution from
PAH features (also common in these regions) in 3.6 \mum\ (see, e.g.,
Flagey \etal\ 2006). These contributions to both channels mean that
there is some cluster-to-cluster variation in the uncertainties
associated with individual sources at a given magnitude.

We determined that a systematic error of 0.01 mags in I1 and 0.007
mags in I2 (both indicated in Fig.~\ref{fig:fixerrors}) provides the
best fit to the photometric RMS seen in each channel in that it is a
conservative representation of the distribution as the distributions
approach the asymptote.  These values are also quite comparable to the
magnitude of the intrapixel gain effect in each of these channels, and
it is not expected that the errors in IRAC photometry from mapping
observations would be less than 0.1\%. Very similar values are
obtained via a slightly different approach for the CSI 2264 mapping in
Cody \etal\ (2014), where the populations of members and non-members
are better understood.  We added our empirically-derived systematic
values in quadrature to each of the individual errors obtained by our
pipeline. 

Figure~\ref{fig:rmsmag} plots the light curve RMS ($\sigma$) derived
from the corrected light curves against the mean magnitude for that
light curve.  Here, again, one can see more, and more significant,
variation among the brighter sources. Relatively few sources have much
larger uncertainties because they are close to bright neighbors, chip
edges, or affected by other instrumental artifacts; many of the
sources with large RMS are legitimately variable. For sources brighter
than 13th mag, the total uncertainty is dominated by the error floor. 
The overall noise increases significantly fainter than about 14th
magnitude, and the rapid falloff of the number of objects detected in
the survey beyond 16th magnitude is immediately apparent. (Also see
Fig.~\ref{fig:jdistpart1}-\ref{fig:jdistpart2} below.) There are small
variations between clusters, which are caused by different levels of
diffuse emission and a different degree of instrumental artifacts
from, e.g., residual pull-down. A sample individual cluster is
included in Fig.~\ref{fig:rmsmag} as an indication of what is seen in
the individual smaller-field clusters.

\subsection{Cryogenic-era Spitzer Data}
\label{sec:cryodatareduction}

Early in the Spitzer mission, each of our 12 regions were observed,
often as part of the guaranteed time observations or the
original Cores-to-Disks (c2d) Legacy program (Evans \etal\ 2003,
2009a); these observations are summarized in
Table~\ref{tab:cryoprograms}.  In most cases, the data were obtained
at two epochs. For clusters near the ecliptic plane, asteroids that
happen to be passing through the region at the time of observation
appear as long wavelength sources that can resemble embedded low-mass
YSOs. Thus, the two epochs were often planned to be separated by a
time of order hours (e.g., the earliest epochs of Serpens, much of
L1688) to allow the moving Solar System objects enough time to move,
and the corresponding pixels to be rejected as outliers when the
individual frames were combined.  For some observations (e.g., Orion),
to facilitate bright source artifact mitigation, the epochs were
separated by $\sim$6 months such that the orientation of the arrays in
the later observation is 180$\arcdeg$ from that in the early
observation (see Sec.~\ref{sec:footprints} above). For the most robust
detections, the cryogenic data can be combined into a single early
epoch, or one can compare the two epochs to constrain variability on
timescales of months (e.g., Megeath \etal\ 2012 for Orion at IRAC
bands) or hours (e.g., Rebull \etal\ 2007 for Perseus at 24 \mum).  

To retain information from the individual epochs, the cryogenic data
for all of our clusters were reduced identically to the YSOVAR
monitoring data, except using cryogenic calibrations (e.g., the zero
point and threshold between the short and long HDR frames).  Source
detection was performed independently in each band for each epoch.
However, for reasons we now describe, measurements derived from
individual epochs from the cryogenic era cannot generally be used over
the entire region, and the identical approach adopted for the rest of
the YSOVAR data cannot generally be used for the cyrogenic data. 

For the YSOVAR data, each epoch covered the region of interest
with sufficient redundancy, so that data reduction can be performed on
a per-epoch basis as described in Sec.~\ref{sec:spitzernewdata} above.
For some of the cryogenic era data, the {\em observations were
designed to have sufficient redundancy in the region of interest only
once all of the cryogenic era observations were combined.}
Specifically, for some of those clusters that were observed in two
epochs, for  moving object identification, the observations from a
single epoch were not designed to robustly measure objects at that
epoch; it was envisioned that the two epochs would be combined to
create a net mosaic of the inertial targets.  For example, one epoch
might have two frames, with another two frames at a later epoch.
Combining those after the fact provides 4 frames of dithered
observations to remove artifacts and moving objects. However, a single
epoch, having only 2 frames, does not necessarily have enough data for
robust photometry.  A complication comes from the combination of
individual frames to compose a map at a given epoch. The maps at any
one epoch are constructed from multiple pointings, with overlap
between each pointing. Therefore, in those thin regions between
adjacent pointings, there are more than 2 frames at a given epoch.
Because our pipeline requires at least 3 frames per position, it only
derives (and archives) high-quality photometry at either of those two
epochs for that thin region between adjacent pointings in the map, not
the whole map at that epoch. Therefore, if one just downloads the
earliest epoch data from our archive for those clusters, one obtains a
``grid-shaped web" of viable data, not complete coverage of the
region. We have retained all of these individual epoch data in the
data we  deliver to IRSA because it may be useful to future
researchers to have individual measurements of sources serendipitously
falling in those regions. But, for users not aware of this `feature,'
it will seem strange to only have cryogenic data over a web of
coverage in those early epochs.

In the context of the present paper, therefore, all of the earliest
cryogenic-era Spitzer observations were combined into one mosaic per
epoch per channel, which was then run through Cluster Grinder. The
time assigned to this early epoch of observation is the average time
of all of the observations that went into the mosaic, summarized in
Table~\ref{tab:cryoprograms}. 

As previously stated, IC~1396A and Serpens Main were explicitly
monitored in the cryogenic era with Spitzer, and these data were also
reprocessed, retaining individual epochs of observation. 

The data were bandmerged across Spitzer bands by position, within a
search radius of $1\arcsec$.  (Recall that the IRAC photometry
apertures we used were 2.4$\arcsec$.)  

Gutermuth \etal\ (2008a, 2009, 2010) present methodology for
identifying YSOs from the cryogenic catalog. The details of the
selection process appear in those papers, but in summary, multiple
cuts in multiple color-color and color-magnitude diagrams are used to
identify YSO candidates, as distinct from, e.g., extragalactic and
nebular contamination. This color selection is part of Cluster
Grinder, and thus we have this classification, based on the cryogenic
data, for all of our clusters; we used it in
Sec.~\ref{sec:clusterproperties} and in part of
Table~\ref{tab:clusterproperties}. We have adopted this YSO selection
mechanism as part of one of the primary YSOVAR sample definitions; see
Section~\ref{sec:sampledefinition}. 

\clearpage

\subsection{New Data at Optical and NIR Wavelengths}

For completeness, we note here that, in addition to the new IRAC data,
we often also obtained contemporaneous optical and NIR observations
from the ground using a variety of telescopes. The details of each
telescope/camera and set of accompanying observations is beyond the
scope of this paper, and is (or will be) provided in each paper
discussing the results from each cluster. For example, MC11 discussed
observations obtained from four other ground-based telescopes obtained
in 2009-2010, contemporaneous with the first epoch of Orion
observations; Cody \etal\ (2014) discussed observations obtained
contemporaneously with the CSI 2264 campaign.  Because these ancillary
monitoring data are different for each cluster, we omit them here for
clarity.

\subsection{Other IR Archival Data}
\label{sec:otherir}

We also included 2MASS $JHK_s$ data for all of our clusters. For two
clusters (NGC~1333, L1688), deeper than survey 2MASS data (the
6$\times$ data) are available from the 2MASS archive, and are included
where available. Such detections were also bandmerged to the rest of the
catalog by position, within a search radius of $1\arcsec$. 

For three of our clusters (L1688, Mon R2, Serpens Main), deeper NIR
data from the UKIRT Infrared Deep Sky Survey (UKIDSS; Lawrence \etal\
2007) broadband data were publicly available, but not necessarily at
all or even most of the UKIDSS bands ($Z$, $Y$, $J$, $H$, $K$); in all
three cases, there were at least $K$-band data, which is important for
our calculation of SED class (see App.~\ref{sec:sedsection}). Where
available, these data are also bandmerged to the rest of the catalog
by position, within a search radius of $1\arcsec$. Details of which
bands are available and to what depth will be included in the
individual papers associated with the relevant clusters.

For two of our clusters, AFGL 490 and Ceph C, data in 2
Spitzer bands are available from one of the warm portions of the
Spitzer program called the  Galactic Legacy Infrared Mid-Plane Survey
Extraordinaire (GLIMPSE; Benjamin \etal\ 2003). Where available, those
single-epoch, shallow data were also bandmerged to the rest of the
catalog by position, within a search radius of $1\arcsec$. Since those
measurements are independent measures of these objects at [3.6] and
[4.5], these measurements were retained as distinct points from the
cryo [3.6] and [4.5], or the means of our light curves. 

For three other clusters, NGC 1333, L1688 and Serpens Main, data are
available from the Cores to Disks (c2d; Evans \etal\ 2003, 2009a)
program data deliveries. The data used for these deliveries are
typically the same BCDs as were used in the cryogenic data
(Sec.~\ref{sec:cryodatareduction}). As such, then, they are not
independent measurements, and these data were only used to supplement
our cryogenic-era catalog if a band was missing, e.g., because G09 had
identified it as having insufficient SNR in that reduction.

Naturally, each cluster has different amounts of additional 
data in the literature, and the details of exactly which data are
included (and how it was merged to the rest of the catalog) will
appear in the corresponding YSOVAR cluster paper.

\subsection{Chandra data}
\label{sec:chandrareduction}

\begin{deluxetable}{lllllp{9.5cm}}
\tabletypesize{\scriptsize}
\rotate
\tablecaption{Summary of X-ray Observations\label{tab:xrayobs}}
\tablewidth{0pt}
\tablehead{
\colhead{Cluster} & \colhead{Aimpoint\tablenotemark{a} (J2000)} & \colhead{Obsid} &
\colhead{Exp.\ time (ks)} & \colhead{obs date} &  \colhead{notes} }
\startdata
AFGL 490 &  \nodata &\nodata &\nodata &\nodata & No X-ray observations \\
NGC 1333 & 03:29:05.60, +31:19:19.00 &   642 & 43.2 & 2000-07-12 & see also analysis by Getman \etal\ (2002), Winston \etal\ (2010) \\
NGC 1333 & 03:29:02.00, +31:20:54.00 &  6436 & 39.5 & 2006-07-05 & (as above) \\
NGC 1333 & 03:29:02.00, +31:20:54.00 &  6437 & 36.6 & 2006-07-11 & (as above) \\
Orion    & \nodata&\nodata &\nodata &\nodata & reanalysis beyond the scope of this paper; values taken from Getman \etal\ (2005) and Ramirez \etal\ (2004)  \\
Mon R2   & 06:07:49.50, -06:22:54.70 &  1882 & 98.1 & 2000-12-02 & see also analysis by Kohno \etal\ (2002), Nakajima \etal\ (2003) \\
GGD12-15 & 06 10 50.00, -06 12 00.00 & 12392 & 67.3 & 2011-12-23  & X-ray catalog not yet in literature; X-ray properties to be discussed in detail in the corresponding YSOVAR cluster paper \\
NGC 2264 & \nodata& \nodata& \nodata& \nodata& reanalysis beyond the scope of this paper; values taken from Flaccomio \etal\ (2006) and Ramirez \etal\ (2004a) \\
L1688    & 16 27 17.18, -24 34 39.00 &  635 & 100.7 & 2001-05-16 & see also analysis by Imanishi \etal\ (2001)\\
Serpens Main    &18 29 50.00, +01 15 30.00 & 4479 & 88.5 & 2005-06-28 &  see also analysis by  Winston \etal\ (2007), Giardino \etal\ (2007)\\
Serpens South   &18 30 03.00, -02 01 58.20 & 11013 & 99.5& 2011-06-10& X-ray catalog not yet in literature; X-ray properties to be discussed in detail in the corresponding YSOVAR cluster paper \\
IRAS 20050+2720 &20 07 13.60, +27 28 48.80 & 6438 & 22.6 & 2006-12-10 & see also analysis by G{\"u}nther \etal\ (2012) \\
IRAS 20050+2720 &20 07 13.60, +27 28 48.80 & 7254 & 20.9 & 2007-06-07 & (as above)\\  
IRAS 20050+2720 &20 07 13.60, +27 28 48.80 & 8492 & 50.5 & 2007-01-29 & (as above)\\
IC1396A   & 21 36 50.30, +57 30 24.00 & 10990 & 29.7 & 2010-06-09 & see also analysis by in Getman \etal\ (2012) \\
IC1396A   & 21 36 50.30, +57 30 24.00 & 11807 & 29.8 & 2010-03-31 & (as above) \\
Ceph C    & 23 05 51.00, +62 30 55.00 & 10934 & 44   & 2010-09-21 & X-ray catalog not yet in literature; X-ray properties to be discussed in detail in the corresponding YSOVAR cluster paper \\
\enddata
\tablenotetext{a}{Formally, ``aimpoint'' is the location in chip
coordinates on the Chandra detector where the target lands. This is
different from the focal point, which is the location of the sharpest
(narrowest) point-spread function. In practice, this can be thought of
as the target of the observation. } 
\end{deluxetable}

X-ray observations are an excellent complement to mid- and near-IR
observations in regions of star formation. Thanks to the strong
correlation between X-ray luminosity and age, X-ray emission is
particularly effective in identifying YSOs which are free of IR excess
(Class IIIs), such that X-ray surveys provide samples of young stars
unbiased by their IR characteristics. 

X-ray observations can penetrate up to \av$\sim$500 mag into a star
forming cloud with very deep integrations (Grosso \etal\ 2005). 
However, even with shallow integrations, one can reach as deep or
deeper in the X-rays than many NIR surveys in the $JHK$ bands. The
Chandra X-ray Observatory, with its high angular resolution mirrors
and low-noise detectors, is particularly effective in resolving
crowded fields down to 0.5$\arcsec$ scales. Furthermore, in X-rays, OB
stars are often not much brighter than pre-main-sequence stars, so
close companions, even when associated with OB stars, can be
identified (Stelzer \etal\ 2005). Chandra data (or any other X-ray
data, e.g., from the European Space Agency's XMM-Newton observatory)
and Spitzer data combined have been shown to yield a more complete
survey of members; there are many examples of this in the literature.
For some of the clusters in our target list, an incomplete listing of
such work could include Winston \etal\ (2007, 2010) for NGC 1333 and
Serpens Main, respectively, G\"unther \etal\ (2012) for IRAS 20050,
Getman \etal\ (2006) for IC 1396A, or Imanishi \etal\ (2001) for
L1688.

Identification of additional cluster members is the primary purpose
for which we include X-ray data, where available, for the YSOVAR
clusters. We can then compare variability characteristics for the
X-ray detected sample with that from, e.g., the IR-selected sample. 
With the exception of AFGL 490\footnote{As of early 2014, AFGL 490 has
not been observed by XMM-Newton either.}, all of the YSOVAR clusters
have been observed by Chandra using the Advanced CCD Imaging
Spectrometer for wide-field imaging (ACIS-I), and nearly all of them
have already been published. The date, duration, and aimpoints of
these observations are listed in Table~\ref{tab:xrayobs}, along with
citations to the literature where possible; footprints of the
observations are included in
Figures~\ref{fig:n1333footprints}-\ref{fig:cephcfootprints}.  

In order to create a more unified set of detection criteria, and in
order to reach fainter sources, X-ray data for the 9 smaller-field
clusters with X-ray data were reprocessed to achieve internal
consistency and maximize source detection. We used Chandra pipeline DS
8.4.5.  Detailed X-ray data and analysis have already been published
for the very deep observation of the Orion Nebula Cluster (Feigelson
\etal\ 2005, Getman \etal\ 2005), and for two shallower fields north
and south of the Orion Nebula Cluster (Ramirez \etal\ 2004b);
similarly, for NGC 2264, Flaccomio \etal\ (2006) and Ramirez \etal\
(2004a) have published deep Chandra data (our region here is primarily
covered by Flaccomio \etal\ 2006). Because those two regions include
X-ray data substantially deeper than those for the rest of the
clusters, we have not reprocessed these X-ray data here, but instead
taken those source lists from the literature. (A primary reason we
reprocessed the 9 smaller-field clusters is to reach fainter sources,
which the deeper integrations in Orion and NGC 2264 already
accomplish.) Additional X-ray data obtained contemporaneously with
Spitzer monitoring will be discussed in the appropriate
cluster-specific papers.

The region with ACIS-I data is usually smaller than the full region
observed with Spitzer during the cryogenic era, but often covers the
region monitored for YSOVAR; see
Figures~\ref{fig:n1333footprints}--\ref{fig:cephcfootprints}. Aside
from AFGL 490 (which has no X-ray data), the region with the least
spatial Chandra fractional coverage is Orion; the Orion region with IR
light curves is by far the largest of our sample. The other 10
clusters have complete or nearly complete X-ray coverage of the region
with 2-band IRAC light curves; X-ray coverage of the regions with
light curves in only 1-band varies. The Chandra sensitivity varies
across the field of view, with higher sensitivity in the center of the
pointing, decreasing towards the edges.

Source detection from the archival Chandra data is performed with the
wavdetect algorithm in CIAO (Chandra Interactive Analysis of
Observations; Fruscione \etal\ 2006), versions 4.5 and 4.6. The
details for the detection process used here are identical to those
used in G\"unther \etal\ (2012). Our chief goal is identification of
all X-ray sources, since even a weak detection -- provided it is
coincident with an IRAC source -- has a high likelihood of being a
legitimate source, though this does not necessarily prove cluster
membership.  Table~\ref{tab:xrayprops} lists the number of X-ray
sources detected above the 2$\sigma$ significance level in the entire
ACIS-I field (a single ACIS-I FOV is
$\sim16\arcmin\times\sim16\arcmin$).  For comparison,
Table~\ref{tab:xrayprops}  also indicates the number of detections
reported in the literature, often for exactly the same data. We
recover essentially all the previously reported sources, and add a
significant number of weaker ones due to the lower threshold employed
here. Note that without the additional criterion imposed here of
requiring a match between an X-ray source and an IRAC source, the
2$\sigma$ significance level used here would be too low and would
result in a high number (about 1/3) of spurious sources. The
requirement of a positional match with an IRAC source allows us to use
such a low threshhold. We found that below 2 sigma, we do not find
matches between X-ray and IR sourcess in excess of matches expected
due to random chance. 

To determine if a given X-ray source is coincident with an IRAC
source, we matched X-ray sources to IRAC source lists generated for
Spitzer cryogenic-era observations, as described in
Sec.~\ref{sec:cryodatareduction} above. Due to the highly
spatially-dependent Chandra point spread function, three matching
radii were used. For sources within 3\arcmin\ of the aimpoint, the
matching radius was 1\arcsec. For sources more than 6\arcmin\ off-axis
from the aimpoint, the matching radius was 2\arcsec. In between, the
matching radius was 1.5\arcsec. We note that IRAS 20050+2720 and IC
1396A are both exceptions since the fields were each observed multiple
times, at different roll angles. Since IRAS 20050+2720 covered a wide
range of rotation angles, in that case, the source positions are
composites of different point spread functions; matches were made more
carefully in these observations. For IC 1396A, the range of roll
angles was smaller, so the same approach was used as for the rest of
the clusters. For each cluster, in Table~\ref{tab:xrayprops}, we list
the total number of X-ray sources with a match to an IRAC source
within the appropriate radius. We note here for completeness that
there are some very bright X-ray sources without IR counterparts; for
our purposes within YSOVAR, we drop these sources.

If an X-ray source is bright enough for spectral fitting, one can
determine four key characteristics: X-ray temperature, gas absorption,
mean flux ($F_x$), and variability.  We fit a single Astrophysical
Plasma Emission Code (APEC) thermal emission model (Smith \etal\ 2001)
for each source detected above 30 net counts, using C-statistics and
leaving the absorbing column density ($N_H$), the temperature ($T$),
and the volume emission measure as free parameters to determine the
first three of the above characteristics.  The metallicity is fixed at
30\% of the Solar abundances (Anders \& Grevesse 1989), in keeping
with typical values from coronal emission from late-type stars.
Details of the extraction and fitting processes are similar to those
discussed by Winston \etal\ (2010), which follows the procedure for
automated processing laid out for the ANCHORS (AN archive of CHandra
Observations of Regions of Star formation) pipeline (Spitzbart \etal\
2005). The key point is that all fits are done assuming that the
source is a star with a thermal one-temperature spectrum. We calculate
luminosities ($L_x$) for each source using the fluxes ($F_x$) from 0.3
keV to 8.0 keV and line-of-sight absorptions, assuming the distance to
each cluster given in Table~\ref{tab:clusterproperties}.  Detailed
flux errors due to the fit were not calculated separately for each
source, since systematics are likely to dominate; instead, global flux
errors were determined using the CIAO tool dmextract. This process
calculates a simple error estimate based on photon statistics and the
mean value of the exposure map in the source region. These errors are
about 4\% at 2000 counts, about 35\% at 100 counts, and the errors
reach 100\% below about 50 counts. The error budget is dominated by
photon counts and uncertainty in $N_H$. There is no evidence that such
errors are markedly biased by the C-statistic.

Since errors on $L_x$ can be dominated by systematic effects, we have
reprocessed all sources (in those 9 smaller-field clusters with
Chandra data), even those with previously published fluxes, to ensure
uniform spectral fitting methodology.  For faint sources with fewer
than 30 counts, no fit can be performed. In this case, we determine a
median photon energy, which, when combined with the count rate, leads
to an approximate flux determination. All spectral properties
presented here are effectively time averaged over all observations. 

To determine the fourth of the characteristics above (variability), we
tested the light curves for variability using the Gregory-Loredo
method (GL-vary; Gregory \& Loredo 1992). This method uses
maximum-likelihood statistics and evaluates a large number of possible
break points from the prediction of constancy. It assigns an index to
each lightcurve -- the higher the value of the index, the greater the
variability. Index values greater than 7 indicate $>99$\% variability
probability. Values of the GL-vary index $>$9 usually indicate flares.
GL-vary is not reliable below about 30 raw counts, the same limit we
used for performing spectral fits. The value of this index will be
provided where relevant on a source-by-source basis in the individual
cluster papers.

There are five broad classes of X-ray sources in the field: 
\begin{itemize}
\item First, background active galactic nuclei (AGN) -- these will be
numerous; up to 50 of these per 16\arcmin$\times$16\arcmin\ Chandra
field are expected, depending on the depth of exposure (Getman \etal\
2006). However, they tend to be faint in both X-rays and mid-IR, or
are not matched in the IRAC bands at all. 
\item Second, background starburst galaxies -- while much
less common than AGN, they are brighter than AGN, and have colors
similar to Class II YSOs. They tend to be fainter in the IR than
typical Class II YSOs, and may be tentatively identified via the
faintness of the IR counterpart. 
\item Third, compact objects such as white
dwarfs -- these tend to be very faint and usually undetected in the
IRAC bands. 
\item Fourth, YSOs -- those with disks have already been
identified via their IR excesses. We identify the probable disk-free
objects that are detected in X-rays by their star-like IR colors and
magnitudes consistent with membership.  In addition to matching an
IRAC source, the star-like colors are required to ensure that any
newly revealed sources are probable Class III objects and not distant
starburst galaxies.  
\item Fifth and finally, active late-type field stars -- both
foreground and background stars can appear with X-ray fluxes
comparable to our targets. Based on a study of IC 1396, Getman \etal\
(2006) estimate that there could be $\lesssim 10$  of these per
16\arcmin$\times$16\arcmin\ Chandra field. Since the contaminants are
a mixture of foreground and background objects of comparable fluxes to
our targets (with less foreground and more background contamination
likely for closer clusters), without optical spectra, they are very
hard to discern from Class III objects. Necessarily, then, these
remain in our sample of candidate members selected via X-rays and
represent a source of contamination. 
\end{itemize} 
We note that Getman \etal\ (2012) estimate (again for IC 1396, a
region in the Galactic plane) a 22\% probability that any X-ray source
with any IRAC counterpart is not a member of the cluster. That rate
drops to about 10\% in IC 1396 if one eliminates sources without 2MASS
$JHK_s$ detections.   Because the contamination rate is affected by
absorption, exposure time, and the depth to which one extracts
sources, it may be different for the other clusters.

We can improve our inventory of YSOs in these clusters by identifying
objects with X-ray detections, IRAC counterparts, and SEDs that are
consistent with those of stars. We have adopted this YSO selection
mechanism as the other main component of the primary sample
definitions; see Section~\ref{sec:sampledefinition}.  Note that we
define `SEDs consistent with stars' to be those with a fitted SED
Class III (see App.~\ref{sec:sedsection}) but that there is room to
create an augmented membership list (Sec.~\ref{sec:augmentedsample})
to include objects that have an X-ray detection, an IRAC counterpart,
were not identified as a YSO from the IR alone, but that have an SED
consistent with a YSO. With regard to foreground or background
stellar contamination, because any appropriate brightness cutoff is a
function of cluster distance (and \av), we defer any detailed
exclusion of likely foreground or background stars from the member
sample to the individual cluster papers.

\begin{deluxetable}{lcccccc}
\tabletypesize{\scriptsize}
\rotate
\tablecaption{X-ray Source detection characteristics\label{tab:xrayprops}}
\tablewidth{0pt}
\tablehead{
\colhead{Cluster} & \colhead{\# $> 2\sigma$\tablenotemark{a} } &
\colhead{\# cryo match\tablenotemark{b}} &
\colhead{Detections in literature } &
\colhead{min(log $L_x$)\tablenotemark{c}} &
\colhead{max(log $L_x$)\tablenotemark{c}} &
\colhead{med(log $L_x$)\tablenotemark{c}} 
}
\startdata
AFGL 490 & \nodata & \nodata& (no X-ray data) & \nodata&\nodata&\nodata\\
NGC 1333        & 192 & 119 & 109 (Getman \etal\ 2002), 193 (Winston
\etal\ 2010)  & 27.94& 30.74 &29.12\\
Orion & \nodata& \nodata& using lit.; see text & 27.58 & 32.62 & 29.72 \\
Mon R2          & 492 & 167 & 154 (5$\sigma$; Kohno \etal\ 2002) & 29.58&31.16&30.16\\
GGD 12-15       & 229 & 172 & \ldots & 29.66 & 31.77 & 30.33\\
NGC 2264  & \nodata& \nodata& using lit.; see text & 28.45 & 31.29 & 29.95\\
L1688           & 315 &  69 & 11 (Imanishi \etal\ 2001) & 27.48&31.30&29.51\\   
Serpens Main    & 204 & 161 & 85 (Giardino \etal\ 2007) &29.13&31.71&29.97\\
Serpens South   & 294 & 82  & \ldots & 29.00&31.27&29.82 \\
IRAS 20050+2720 & 348 & 239 & 239 (G{\"u}nther \etal\ 2012)&29.28&30.88&30.12\\
IC 1396A        & 185 & 129 & 415\tablenotemark{d} (Getman \etal\ 2012) &29.14&32.26&30.34 \\
Ceph C          & 200 & 97  & \ldots & 29.44&30.95&30.17\\
\enddata
\tablenotetext{a}{Number X-ray sources detected via our reprocessing at $> 2\sigma$.}
\tablenotetext{b}{Number of our X-ray sources matched to sources in the IRAC
cryogenic-era catalog.}
\tablenotetext{c}{For the sample of X-ray detected sources with YSOVAR
light curves (the subset of the standard sample for statistics
detected in X-rays), the mininum, maximum, and median log $L_x$, where
$L_x$ is in ergs s$^{-1}$.}
\tablenotetext{d}{The detections used in Getman \etal\ (2012) do not
require IR matches, and they go well below our typical significance cut
off.}
\end{deluxetable}

\section{Sample Definitions}
\label{sec:sampledefinition}

For most of our clusters, there are no well-established membership
lists; the exceptions are Orion and NGC 2264.  Moreover, the
membership lists in the literature use a wide variety of wavelengths
and survey depths to identify members, and the spectroscopic follow-up
of candidate members is uneven.  If we decided to depend on the robust
member identifications only from the literature, our sample would be
greatly reduced for most clusters and highly biased. And, certainly,
our variability survey will identify new candidate members based on
the light curve properties. To attempt to make fair comparisons
between clusters, we need to define a set of (candidate) members in
the same or at least consistent ways between clusters.

As discussed above, even within our survey, different clusters may
have different amounts of monitoring data beyond the `fast cadence'
data. Thus, for making comparisons between clusters, we need to define
a standard set of data that are used for calculating statistics and
identifying variables. 

Therefore, we define a ``standard YSOVAR sample,'' which is the sample
that is (primarily) discussed in this paper and is what forms the
common core of the papers planned for each cluster. Each cluster may
have an additional ``augmented sample'' as well, to take into account
additional member identifications from the literature or our own data
where possible and necessary. We now discuss the definitions of these
samples.

\subsection{Standard Set of Members}
\label{sec:standardsetofmembers}

Each cluster has an IR-selected sample of member candidates defined by
the Gutermuth \etal\ (2008a, 2010) and G09 selection algorithm, run on
the cryo-era catalogs created anew as per the methodology above
(Sec.~\ref{sec:cryodatareduction}).  A detailed description of the YSO
candidate color selection algorithm can be found in Appendix A of
G09.  A sample of YSOs selected this way is thought to be a
statistically well-defined sample (see statistical discussions in
Gutermuth \etal\ 2008a and G09), composed nearly entirely of members,
though some contamination from background galaxies or asymptotic giant
branch stars is always possible. Very few of these IR-selected
members have spectroscopic follow-up (or spectra in the literature
pre-dating the Spitzer observations), and as such, many should
technically be thought of as YSO {\em candidates}, though we include
them all in the set of members.

All clusters except AFGL 490 have X-ray data, and thus also an
X-ray-selected sample of YSOs. (As with the IR-selected sample, few of
these have spectra, so technically they are YSO candidates.) Since we
should have identified most of the YSOs with disks (at least disks
detectable at 3.6 to 24 \mum) in the IR selection process, we use the
X-ray data to identify additional young stars without disks (e.g.,
Class IIIs). Thus, we add to the set of IR-selected members the
X-ray-selected sample defined by the algorithm described above in
Sec.~\ref{sec:chandrareduction}, which can be summarized as objects
having an X-ray detection above a 2-$\sigma$ significance threshold,
having a match to an IRAC object in the cryo-era Spitzer catalog, and
having an SED shape consistent with it being a disk-free YSO or a star
(e.g., Class III; see Sec.~\ref{sec:sedslopes}). This sample is
designed specifically to find and add to our set of members those
members without disks. However, it should be noted that: (a) the
Galactic contamination rate is likely to be higher in this X-ray
selected sample than in the IR-selected sample; and (b) specifically
because of the contamination rates, members with disks are identified
as members from the IR excess, not the X-ray flux, though of course
members with disks can also have measured X-ray fluxes in our
database. 

We have thus defined our ``standard set of members'' to be the union
of all IR selected members with disks and X-ray selected members
without disks. There are provisions for adding additional objects; see
\S\ref{sec:augmentedsample} below.  Note that this definition can be
applied independently of whether or not there is a light curve, but of
course in the context of this discussion of YSOVAR data, we require a
light curve. Note also that the IR selection requires four bands of
IRAC, and thus both very faint and very bright previously-identified
members may be omitted from the standard set; objects such as these
known to be members via some other approach in the literature may be
added in the augmented sample (Sec.~\ref{sec:augmentedsample}).
Finally, note that because the data that go into our selection of
cluster members are of various depths, and because the clusters are at
a variety of distances, the effective mass limit reached by each set
of standard members varies from cluster to cluster.

\subsection{Standard Set for Statistics}
\label{sec:standardsetforstatistics}

All of the original YSOVAR light curves were obtained with very
similar HDR mapping observations (see
Sec.~\ref{sec:newspitzeroverview}) in a fast cadence (see
Sec.~\ref{sec:cadence}), though the length varies, and some clusters
have additional slow cadence observations and/or staring
observations.  The time sampling and total length of light curves
obtained within a single cluster field can also vary as the field of
view changes with time (Sec.~\ref{sec:footprints}). We have attempted
to remove all instrumental effects from the input data, and only
retained photometry where there were valid measurements on at least
three BCD frames (Sec.~\ref{sec:spitzernewdata}). 

We now define the ``standard set of data for statistics'' as follows.
Since the fast cadence is the most common (and most similar) among the
YSOVAR clusters, we used only these mapping fast cadence data, for
those light curves that have at least 5 viable epochs (with each
epoch obtained from at least 3 BCDs per epoch that are not obviously
compromised by instrumental effects or cosmic rays), to calculate
statistical quantities such as mean, median, etc., as well as Stetson
index and $\chi^2$ (discussed below, Section~\ref{sec:findingvars}).
We have defined statistical values calculated on the fast cadence data
as the `standard set of statistical values' for each cluster and
employ them to identify variables and compare values across clusters.
Finally, for stars fainter than [3.6]$\sim$[4.5]$\sim$16, noise tends
to dominate the light curves (Sec.~\ref{sec:brightnessdistrib} \&
\ref{sec:findingvars}). We have retained these faint sources, but
objects this faint are considered individually where relevant.

Note that the standard set for statistics is thus defined as all light
curves with at least 5 points, just the fast cadence. This is
independent of whether the target is identified as a member or not.

Elsewhere in this paper, we refer to ``all objects with a light
curve'' -- this means anything with a light curve in the standard set
for statistics.  (Essentially no YSOVAR-classic sources have only
points outside of the fast cadence.)

When available, additional epochs of data can be included for
additional calculations on a per-cluster basis in the corresponding
papers, and will clearly be indicated as such where relevant.

\subsection{Augmented Sample of Members}
\label{sec:augmentedsample}

We identified members above using IR and X-ray data, which implicitly
relies on the shape of the SED between 2 and 24 \mum. That sample is
still the best set of members that we will use to compare across
clusters, because that set of members is defined as similarly as
possible across all clusters. However, additional young objects may be
identified in the literature, and additional members may be suggested
based on our own data. The ``augmented sample of members'' is where
these additional likely members can be included. 

Some clusters (e.g., NGC 1333, L1688) have considerable literature
discussion of members, and others (e.g., AFGL 490, Mon R2) have far
less. We therefore cannot rely exclusively on the literature to select
members, but neither should we ignore members identified in the
literature and not selected above. Thus, each cluster paper may
include in the augmented sample the literature-identified sources that
are not already found using our IR or X-ray methods above.  

We can use our own data to identify new cluster members. While only
1-2\% of the field population may be variable, YSO variability is the
rule, not the exception. It is therefore possible to identify new
cluster members from light curve properties alone; one could identify
all variables as new members, or one could take just those with
certain properties such as amplitude above a threshhold.  We could
identify cluster members from either the standard set for statistics
(just the fast cadence), or, for those clusters with longer cadences,
from those additional data.  

The set of statistically selected variables (see
Sec.~\ref{sec:findingvars}) are those with Stetson index greater than
0.9, and/or with $\chi^2$ greater than 5, and/or with a significant
period, calculated over just the standard set for statistics (the fast
cadence). Often, these variables should also be identified
as cluster members, but these individual objects will be discussed on
a per-cluster basis (because, for example, they can be background
eclipsing binaries; Morales-Calder\'on \etal\ 2012). Variables
identified using data beyond the standard set for statistics (data
beyond the fast cadence data) will also be included on a per-cluster
basis, and will be discussed in the cluster papers. Those
newly-identified cluster members may also be included in the augmented
sample of members.

While variability-identified objects will make an important
contribution to our understanding of the complete membership of each
cluster, they should not be used in calculations of, e.g., variability
fractions; that sample should be selected on the basis of a parameter
distinct from variability, such as disk excess or X-ray emission. This
is why the new candidate members we identify from our data are
included in the augmented sample of members, and not in the standard
set of members. 

This augmented set of members is only used (where it is clearly
identified) in the individual cluster papers, not in the remainder of
this paper.

\section{Ensemble Analysis}
\label{sec:ensembleanalysis}

In this section, we present analysis of the entire set of data we
used for our clusters, independent of variability, which is discussed
in Sec.~\ref{sec:findingvars}.

\subsection{Cluster Parameterization}
\label{sec:clusterparameterization}
\label{sec:parameterization}

\begin{figure}[ht]
\epsscale{0.5}
\plotone{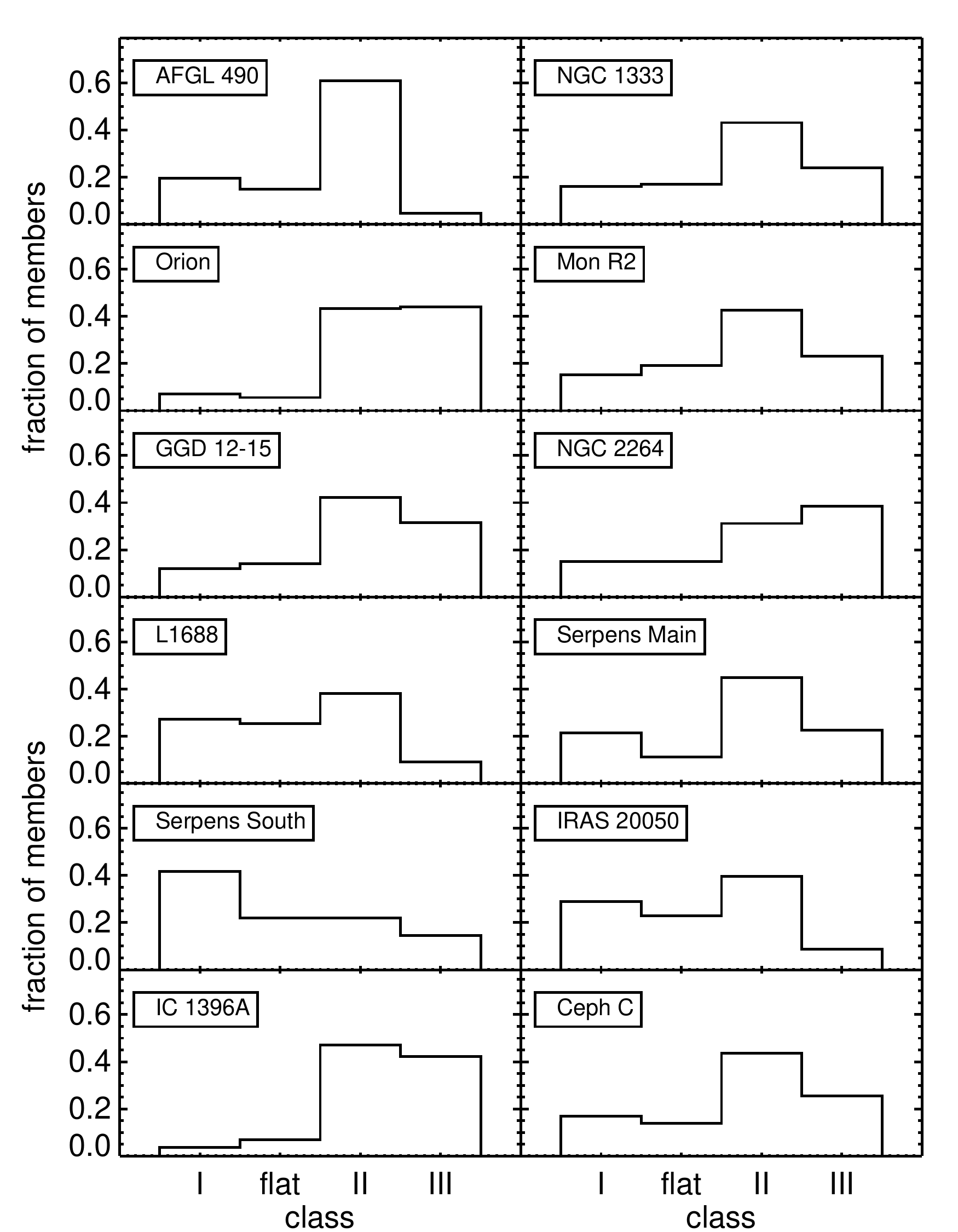}
\caption{Histograms of the relative fractions of fitted SED classes
(classes derived from SED fits as discussed in
App.~\ref{sec:sedslopes}) for the standard set of members with light
curves. Clusters appear in RA-order. By nearly any ratio of classes
used as a parameterization, IC 1396A and Orion have the highest
fraction of sources with more negative SED slopes (taken to be less
embedded sources), and Serpens South has the highest fraction of
sources with more positive SED slopes (taken to be most embedded
sources). AFGL 490 is noticably incomplete in the Class III bin; there
are no X-ray observations available for it, and the Class III objects
in that bin have small IR excesses and SED slopes consistent with
Class III (see App.~\ref{sec:sedsection} for description of classes
and class selection).  }
\label{fig:classratios}
\end{figure}

In order to compare results among clusters, it is useful to be able to
place clusters in some sort of relative order that could, in the most
useful (though hypothetical) case, be tied to age. There are various
ways of parameterizing the evolutionary state of these clusters, and
we considered several, all aimed at capturing the relative numbers of
sources in various SED class bins (see App.~\ref{sec:sedsection} for a
brief definition of SED classes and our placement of sources therein).
Such ratios in some sense capture the relative ``degree of
embeddedness'' of sources in these regions, perhaps with an ultimate
(though undefined here) link to cluster ages. Formally, we are binning
by SED slope, so the parameterizations are, strictly speaking,
relative fractions of sources with a given SED slope, indicative of
the amount of circumstellar material, e.g., how self-embedded a given
source may be. We interpret clusters with more sources with large SED
slopes to, on average, contain more sources that are more embedded.

G09 chose to use the ratio of Class II to Class I sources. These
ratios were all obtained internally consistently, e.g., sources
selected and categorized according to the same series of IR color cuts
and data reduction. Because G09 was working with Spitzer data over a
relatively large region for each cluster, this Class II to Class I
ratio could be calculated for the whole region and for subsections of
the region. The Class II to Class I ratios as calculated for the cores
of these clusters (e.g., Table 6 in G09) appear in our
Table~\ref{tab:clusterproperties} and in the discussion in
Sec.~\ref{sec:clusterproperties}. 

In YSOVAR, we have clusters not included in G09, and moreover, even
for the clusters included in G09, we generally have light curves for
only a small subset of the sources G09 considered, because we observed
a smaller region.  To calculate a Class II to Class I ratio for the
portion of each cluster sampled by the YSOVAR monitoring, we performed
the same calculation for objects in the standard set for statistics
(e.g., having YSOVAR light curves) by reducing the cryogenic data in
the same way and performing the same series of G09 color cuts and
classification; see Sec.~\ref{sec:cryodatareduction}. Then, we
recalculated the ratio of Class II to Class I sources specific to the
YSOVAR data using the classes assigned via the G09 algorithm, just for
those sources with light curves. Those Class II to Class I ratios also
appear in Table~\ref{tab:clusterproperties} and in the discussion in
Sec.~\ref{sec:clusterproperties}. The values are provided in the
present work in part as a link back to the G09 analysis; note that
they are calculated in the same way as G09, e.g., with the IR-selected
sources alone, not on the standard set of members per se.

The G09 parameterization is based only on IR-selected sources. For
most of our clusters, we have X-ray data as well (see
Sec.~\ref{sec:chandrareduction}), so we at least have some information
on the Class III population. It is, however, true that we do not
always have complete Chandra coverage of our fields (further
discussion of coverage appears in Sec.~\ref{sec:chandrareduction} and
Figs.~\ref{fig:afglfootprints},
\ref{fig:n1333footprints}--\ref{fig:cephcfootprints}), and even for
clusters where we have Chandra coverage, the Chandra sensitivity is a
strong function of location on the array.  Nonetheless, we would like
to include the information we have, and simply using the Class II to
Class I ratio does not incorporate information about the Class III
population. 

We explored several alternate parameterizations of the relative
fractions of embedded sources, all of which involved various ratios of
classes (or groups of classes) to the total or other classes (or
groups of classes). Histograms of the relative fractions of the SED
classes for the standard set of members
(Sec.~\ref{sec:standardsetofmembers}) for objects with light curves in
the standard set for statistics
(Sec.~\ref{sec:standardsetforstatistics}) in each cluster appear in
Figure~\ref{fig:classratios}. AFGL 490 can be seen to be deficient in
a complete sample of Class III objects because it has no X-ray data;
the Class III objects identified here via X-rays have small IR
excesses (or sufficient reddening at 2 \mum) and thus SED slopes
consistent with Class III. By many metrics, Serpens South has the
highest fraction of sources with more positive SED slopes, which we
take to be most embedded sources. Similarly, both Orion and IC 1396A
have the highest fraction of sources with more negative SED slopes,
which we take to be less embedded sources.

For further analysis here, we have settled on the fraction of Class I
sources to the total number of members, for objects with light curves
(objects in the standard set for statistics). These ratios are
also included in Table~\ref{tab:clusterproperties}. They include (as
part of the total number of members) objects selected via X-rays.
However, there are still the fundamental, systematic uncertainties
inherent in the classification approach, in the selection of members
without spectroscopic follow-up, in the completeness of the surveys
involved (in both area and depth), and the requirement that there be a
light curve (e.g., bright enough in the IRAC channels) as well as in
the relative paucity of Class I objects overall.  Additionally, for
AFGL 490, there are no X-rays to be used, so the ratio is calculated
with solely the IR-identified members. However, the Class III objects
appear only in the denominator, combined with all the other classes. 
We also note that our inventory of Class I sources must be incomplete,
since we lack the long wavelength coverage that would be needed to
find the most embedded sources (e.g., Stutz \etal\ 2013), but such
extremely embedded sources will also not have a YSOVAR light curve. 

\begin{figure*}
\epsscale{0.8}
\plotone{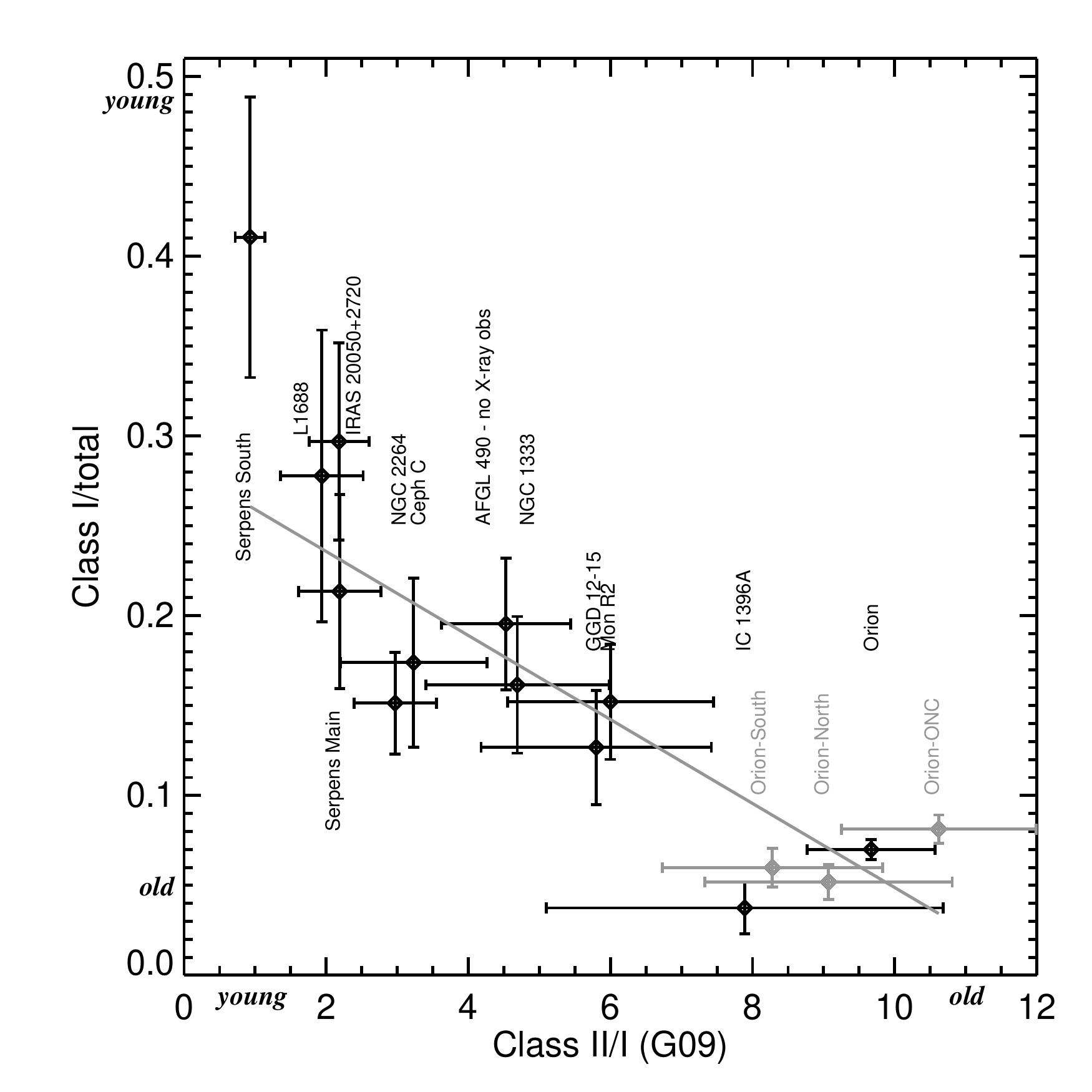}
\caption{Comparison of the Class II/I parameterization from G09 and
the Class I/total parameterization used here. All values are
calculated just for objects with light curves (the standard set for
statistics, Sec.~\ref{sec:standardsetforstatistics}); the G09 values
use just the IR-selected members from the G09 approach, and the Class
I/total values use the standard set of members
(Sec.~\ref{sec:standardsetofmembers}). The error bars are derived
assuming uncorrelated Poisson statistics. Cluster labels appear at the
Class II/Class I location corresponding to the cluster. The grey line
is the best-fit line, using errors in both directions on each point,
and has a slope of $-$0.024$\pm$0.005. Pearson's correlation
coefficient ($r$), calculated for these points is $-$0.86; the
calculated probability that the parameterizations are not correlated
is only 0.04\%. We take these values to be correlated. The additional
grey points are subregions of Orion, used here to show the scatter
inherent in the large Orion map (see the text). Notional `young' and
`old' annotations on the axes describe approximate relative ages that
may quantitatively correspond to large or small values of these
parameterizations. \label{fig:parameterizations}}
\end{figure*}

This parameterization using the Class I/total ratio should be related
to the G09 Class II/Class I ratio determined for the objects having
light curves. Figure~\ref{fig:parameterizations} plots these
parameterizations against each other.  The error bars  are derived
assuming uncorrelated Poisson statistics, because it is difficult to
quantify the additional systematic errors described above. The
best-fit slope of the line fit to this relation (taking into account
errors in both directions on each point) is $-$0.024$\pm$0.005. The
correlation coefficient, Pearson's $r$, calculated for these two
parameters is $-$0.87, and a probability that the parameterizations
are not correlated of only 0.04\%. We assume based on the statistics
and our underlying physical intuition that these values are indeed
correlated. 

Several individual points in Fig.~\ref{fig:parameterizations} merit
additional discussion. Despite having no X-ray data, AFGL 490 is
consistent with the trend shown in Fig.~\ref{fig:parameterizations}.
(If one assumes that there might be about as many Class IIIs as Class
IIs in this region, then the point could move down to about 0.1-0.15,
which would still broadly be consistent with the trend.) The NGC 2264
point is below the trend. The X-ray data obtained for NGC 2264 is
deeper than the X-ray data for the other clusters, except for the
central Orion region. This results in more of the fainter sources
being included in the total number of YSOs, and pushes the Class
I/total ratio towards lower numbers, as seen.  For both
parameterizations, Serpens South is selected as the cluster with the
most embedded sources, as expected from Fig.~\ref{fig:classratios}. It
is well above the fitted line in Fig.~\ref{fig:parameterizations};
perhaps an exponential decay rather than a simple line would be a
better fit to use, but in the absence of additional very embedded
clusters to constrain the most embedded end of the distribution (or,
indeed, spectroscopic vetting of the members and reduction of other
such uncertainties), a line is the simplest fit to use. IC 1396A,
based on Fig.~\ref{fig:classratios}, should be one of the clusters
with the fewest embedded sources. It is identified as the least
embedded using the Class I/total ratio; there is a large uncertainty
on the Class II/Class I ratio, and it is consistent with being the
least embedded within 1$\sigma$.  The Class II/Class I ratio formally
identifies Orion as the least embedded. However, aside from AFGL 490
where there are no X-ray data, Orion is the cluster with the poorest
match between the YSOVAR-monitored region and the existing X-ray data
coverage. Orion has a very deep X-ray pointing, but only in the
central Orion Nebula Cluster (ONC) region (Feigelson \etal\ 2005,
Getman \etal\ 2005). There are two shallower pointings that contribute
X-ray data (see Sec.~\ref{sec:chandrareduction} and
Fig.~\ref{fig:orionfootprints}), but the region of sky in Orion for
which we have IR light curves has the least fractional coverage in
X-rays of all our clusters (aside from AFGL 490). To investigate the
degree to which the uneven X-ray coverage affects the placement of
Orion in this diagram, Fig.~\ref{fig:parameterizations} also includes
points for Orion when broken into `North' (Declination $>-$05:05:25),
`South' (Declination $<-$05:33:15$\arcdeg$), and `ONC' (between those
two limits) fields. There is scatter, clearly, in these points, but
when those points are used instead of the single Orion point, the fit
is functionally indistinguishable from that using the single Orion
point. 

We use the Class I/total parameterization in subsequent discussions in
this paper, most notably Section~\ref{sec:discussion}.

\clearpage

\subsection{Brightness Distribution at J, [3.6], and [4.5]}
\label{sec:brightnessdistrib}
    
\begin{figure}[ht]
\epsscale{0.8}
\plotone{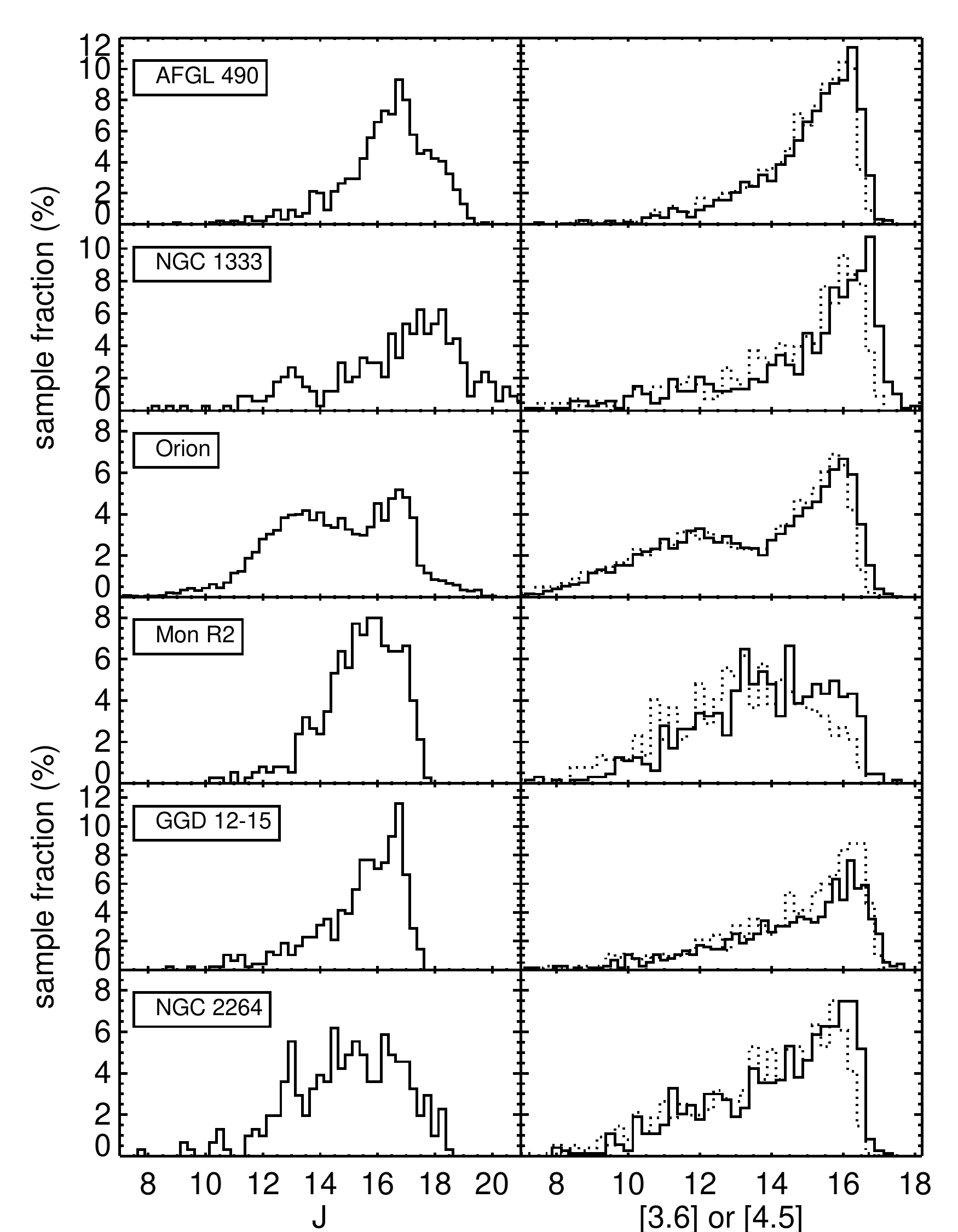}
\caption{Distribution of $J$ magnitudes (left) and IRAC magnitudes
(right; [3.6] in the solid line and [4.5] in the dotted line), for the
standard set for statistics (all objects with light curves), in units
of fraction of sample (in \%) for each cluster. This figure has AFGL
490, NGC 1333, Orion, Mon R2, GGD 12-15, and NGC 2264. See text for
discussion. As a result of these plots, we are cautious about objects
fainter than [3.6]$\sim$[4.5]$\sim$16, both because they are low
signal-to-noise in our monitoring data and because they are likely
dominated by non-members. }
\label{fig:jdistpart1}
\end{figure}
    
\begin{figure}[ht]
\epsscale{0.8}
\plotone{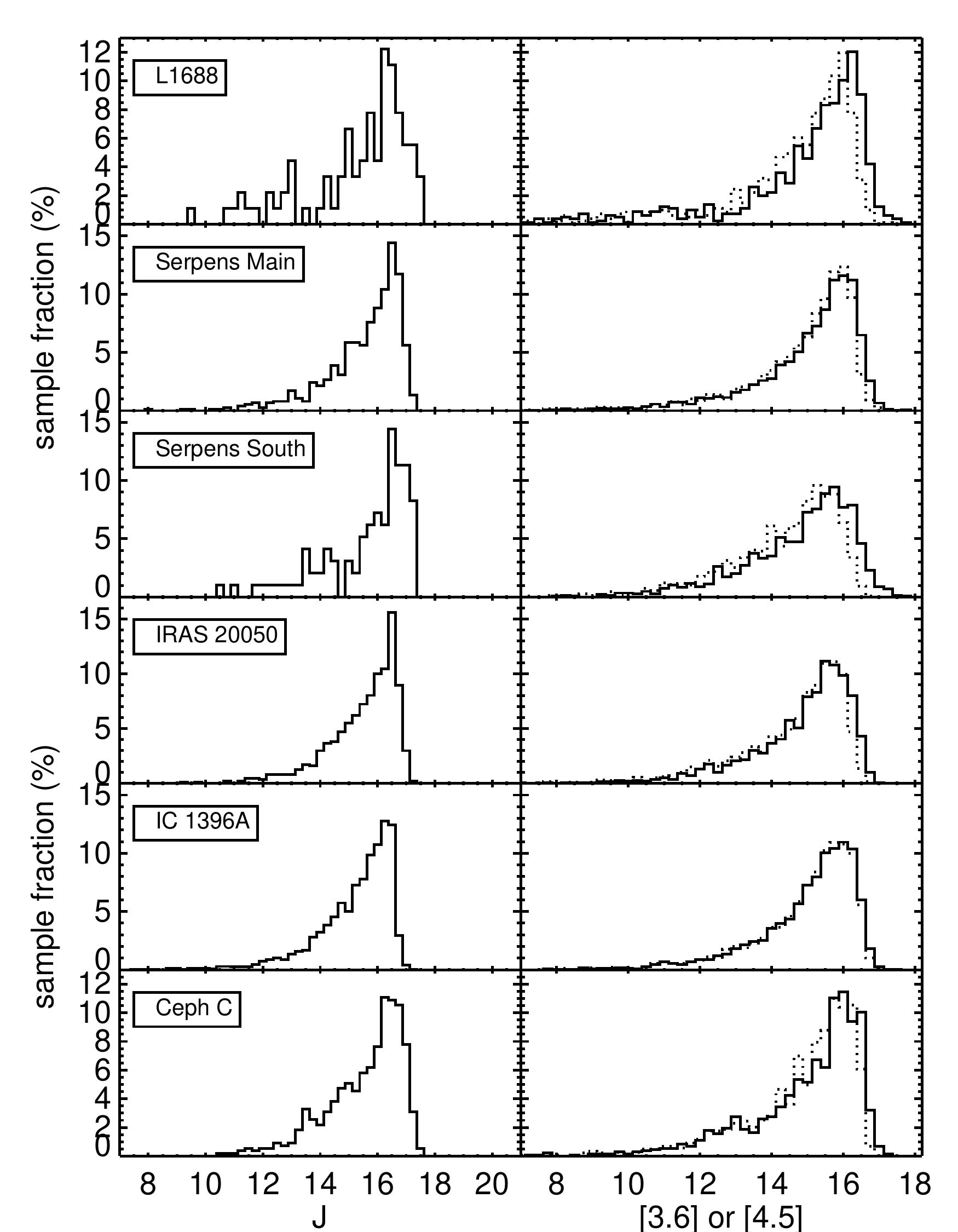}
\caption{As for Fig.~\ref{fig:jdistpart1}, but for L1688, Serpens
Main, Serpens South, IRAS 20050+2720, IC 1396A, and Ceph C.  }
\label{fig:jdistpart2}
\end{figure}

Figures~\ref{fig:jdistpart1} and \ref{fig:jdistpart2} show the
distribution of $J$, [3.6], and [4.5], in units of magnitudes and
percent of sample, for the standard set for statistics (all objects
with at least 5 points in the YSOVAR fast-cadence light curve,
including cluster members and field objects). The total number of
objects portrayed in the figures ranges from $\sim$100 ($J$: L1688 or
Serpens South) to $\sim$7200 ([3.6] and [4.5]: Orion).  The $J$ data
are generally from 2MASS; notably, NGC 1333 is deeper than the other
clusters in $J$ because additional data from the 6$\times$ 2MASS
survey have been included (Sec.~\ref{sec:otherir}).  Generally, more
objects are available at [3.6] or [4.5] than at $J$, largely due to
the greater effective depth of Spitzer. In several cases (most notably
L1688, Serpens South, and NGC 2264), a relatively small fraction of
the objects with Spitzer light curves have $J$ counterparts, since
these clusters are on average, generally more embedded than the
others. For most of the clusters, for most of the objects, $JHK_s$
data are not available for objects with [3.6] or [4.5] fainter than
about 15th mag.

The Orion maps extend out beyond the edges of the cluster, and include
a higher proportion of field stars and other contaminants than do the
other smaller-field clusters. This can be seen in the structure in the
Orion [3.6] and [4.5] histogram, which is double-peaked; the brighter
peak is likely dominated by the cluster members, and the fainter peak
is likely dominated by contaminants. For similar reasons, if the
UKIDSS data (\S\ref{sec:otherir}) are included, the Serpens Main $J$
histogram extends to $J\sim$20 and is also double-peaked, but the
L1688 $J$ histogram is not so obviously double-peaked, likely due to
the higher obscuration levels of the background population.

Mon R2 seems to be different from the other clusters in that the
fainter end of the [3.6] and [4.5] histograms are considerably
flatter. This is most likely a symptom of the difficulty of obtaining
Spitzer light curves for faint sources in the presence of high and
spatially variable background; there are some very IR-bright sources
in the IRAC field of view (see, Fig.~\ref{fig:monr2footprints}), and
the scattered light is substantial, coupled with intrinsically bright
outflows and PAH features. No UKIDSS $J$ mags in this region were in
the public archive at the time we checked (in 2013 Sep). 

As can be seen from the turnover at about 16th mag in the [3.6] and
[4.5] histograms in Figures~\ref{fig:jdistpart1} and
\ref{fig:jdistpart2}, our data do not extend much fainter than
[3.6]$\sim$[4.5]$\sim$16 mag. By inspection, all of these faint
objects appear to be legitimate point sources on the images. For stars
fainter than this, noise tends to dominate the light curves (see
Sec.~\ref{sec:findingvars}). It is hard to completely reject these
fainter sources -- we could drop those whose cryogenic-era [3.6]$>$16,
though the YSOVAR epochs could vary above and below that boundary; or
we could discard those where the mean during the YSOVAR campaign is
$>$16, but there are some objects for which the mean [3.6]$<$16 but
the mean [4.5]$>$16.   As noted in Sec.~\ref{sec:sampledefinition}, we
have retained these faint sources, but in general, the large
uncertainties associated with their Spitzer photometry preclude strong
statements about their variability; objects this faint are considered
individually where relevant. 

Similar histograms for just the standard set of members (identified
through IR excess and X-ray emission; see
Sec.~\ref{sec:sampledefinition}) are considerably less populated,
and brighter; the peaks are around [3.6]$\sim$[4.5]$\sim$12 mag. This
is consistent with the location of the brighter peak in Orion, and
several of the tails of the distributions seen in
Figs.~\ref{fig:jdistpart1} and \ref{fig:jdistpart2}.

\subsection{X-ray Brightness Distribution}
\label{sec:xraybrightnessdistrib}

\begin{figure}[ht]
\epsscale{0.8}
\plotone{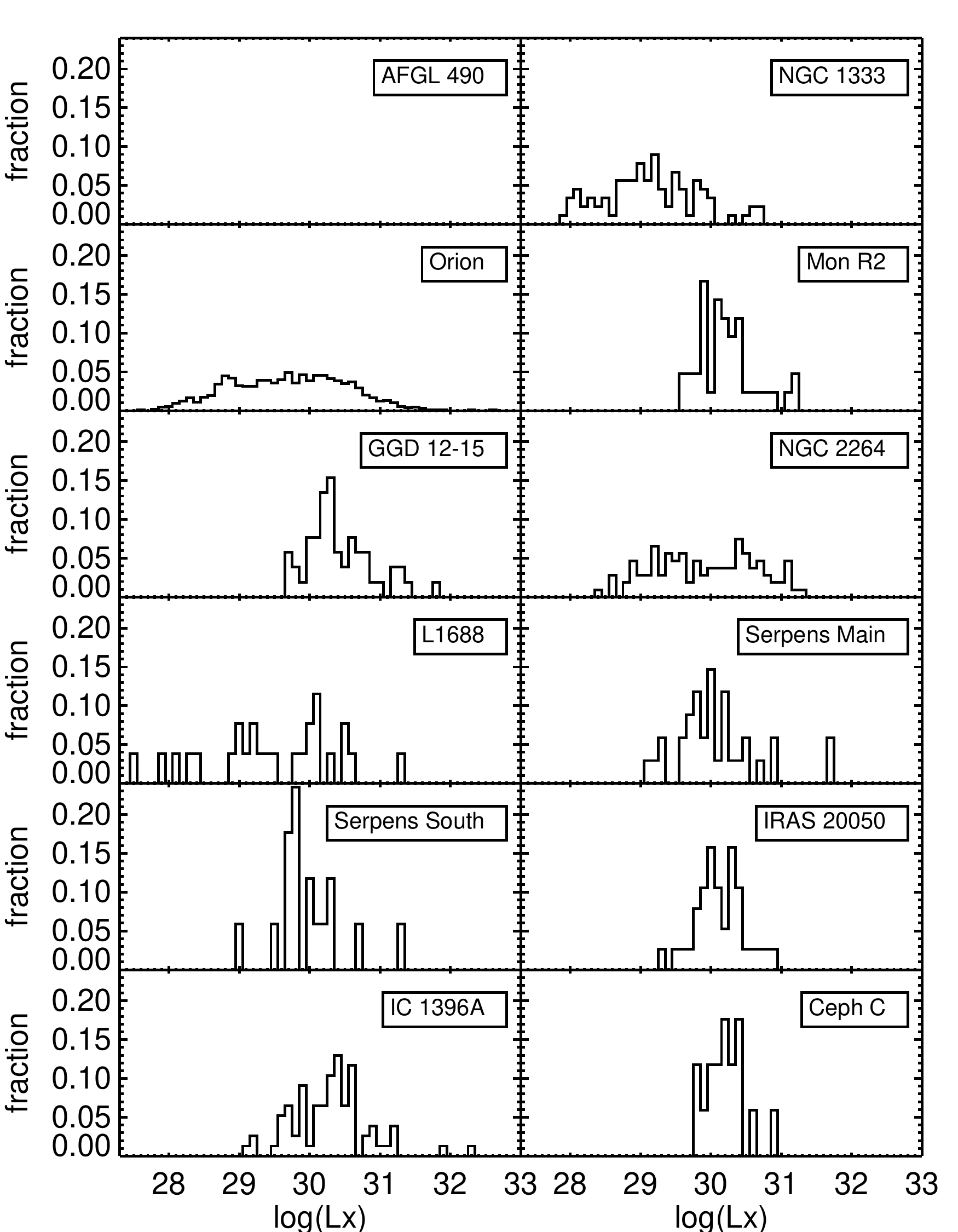}
\caption{Distribution of log($L_x$) values (in ergs sec$^{-1}$) for
the objects in the standard set for statistics (with $L_x$ detections)
in our clusters.  Note that there are no data available for AFGL 490,
and we use literature data for NGC 2264 and Orion; see text. L1688 and
NGC~1333 are the closest clusters, and so relatively faint $L_x$
measurements are obtained with even relatively shallow observations.
NGC 2264 and Orion have the deepest integrations and therefore also
include relatively faint $L_x$. \label{fig:loglxdist}}
\end{figure}

Figure~\ref{fig:loglxdist} contains histograms of the log of the X-ray
luminosities (log $L_x$, where $L_x$ is in ergs s$^{-1}$) for those
objects with light curves (standard set for statistics) and bright
enough in flux ($F_x$) to have a calculated $L_x$
(Sec.~\ref{sec:chandrareduction}). The Orion and NGC 2264 histograms
reach fainter values in $F_x$ than those of the other clusters, because
those integrations were considerably deeper. Moreover, essentially the
entire NGC 2264 field considered here has X-ray data, whereas the
fractional X-ray coverage of the Orion field is relatively low
compared to NGC 2264 or the other clusters here (aside from AFGL 490,
where there is no X-ray data).   Because this Figure has incorporated
distance (distances are listed in Table~\ref{tab:clusterproperties})
to the clusters in the calculation of $L_x$, the two closest clusters
(NGC 1333 and L1688) have histograms reaching the faintest $L_x$.

\clearpage

\begin{figure}[ht]
\epsscale{0.8}
\plotone{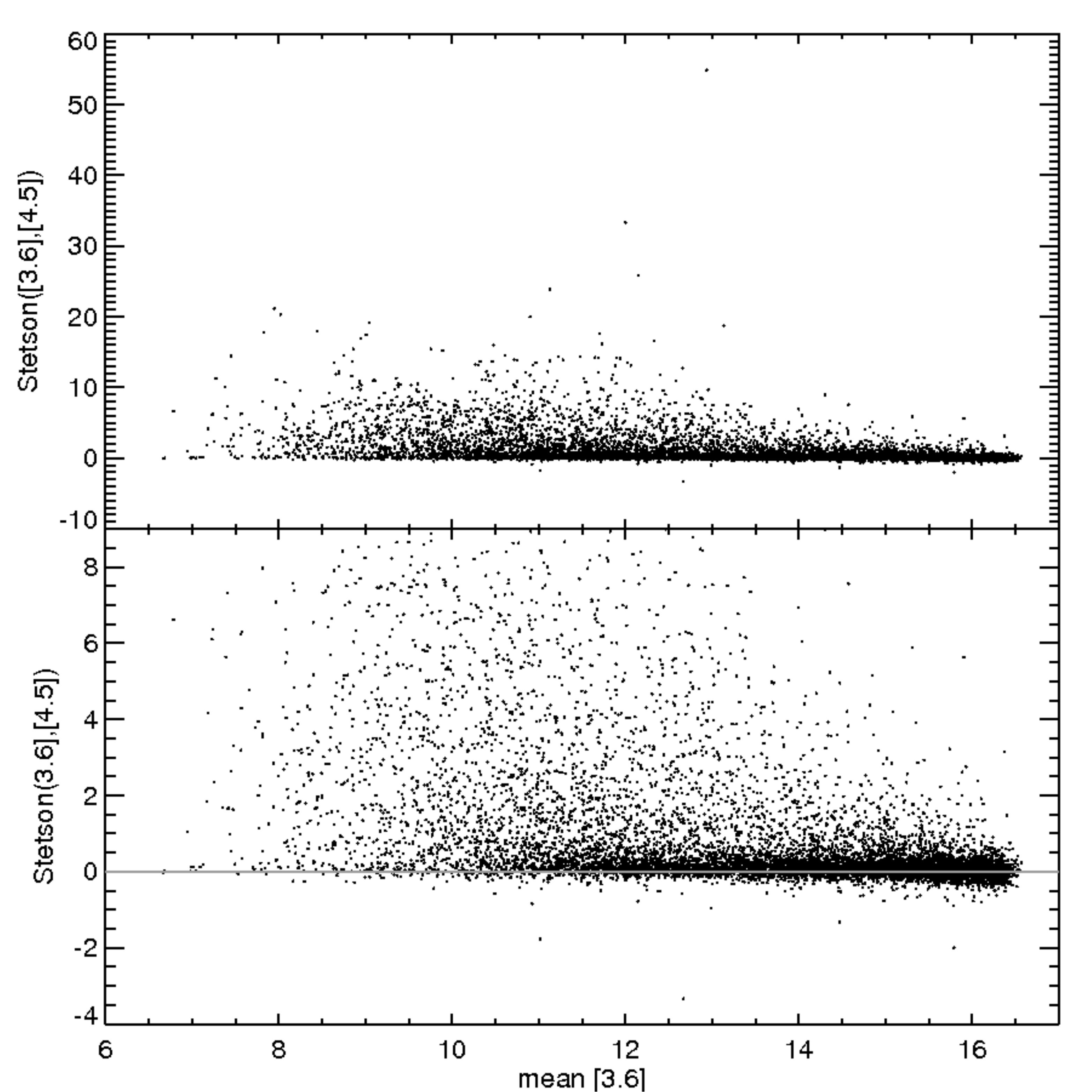}
\caption{Stetson index as a function of mean [3.6] in magnitudes for
all clusters, for the standard set for statistics (YSOVAR-classic fast
cadence data). The lower panel is an expanded view of the top panel,
with an additional grey line at Stetson index=0. There is no indication of
correlated noise here, as would be the interpretation of a change in
the distribution of Stetson index for non-variables as a function of
brightness.  There is an increase in frequency of large Stetson index
values for brighter [3.6], but that is likely because brighter objects
are more likely to be legitimately young cluster members, and as such
more likely to be variable (with variability correlated between the
two IRAC channels). Note that essentially no variables are identified
fainter than [3.6]$\sim$16, consistent with our observation that those
light curves are particularly noisy.}
\label{fig:stetsoni1mag}
\end{figure}
  
\begin{figure}[ht]
\epsscale{0.8}
\plotone{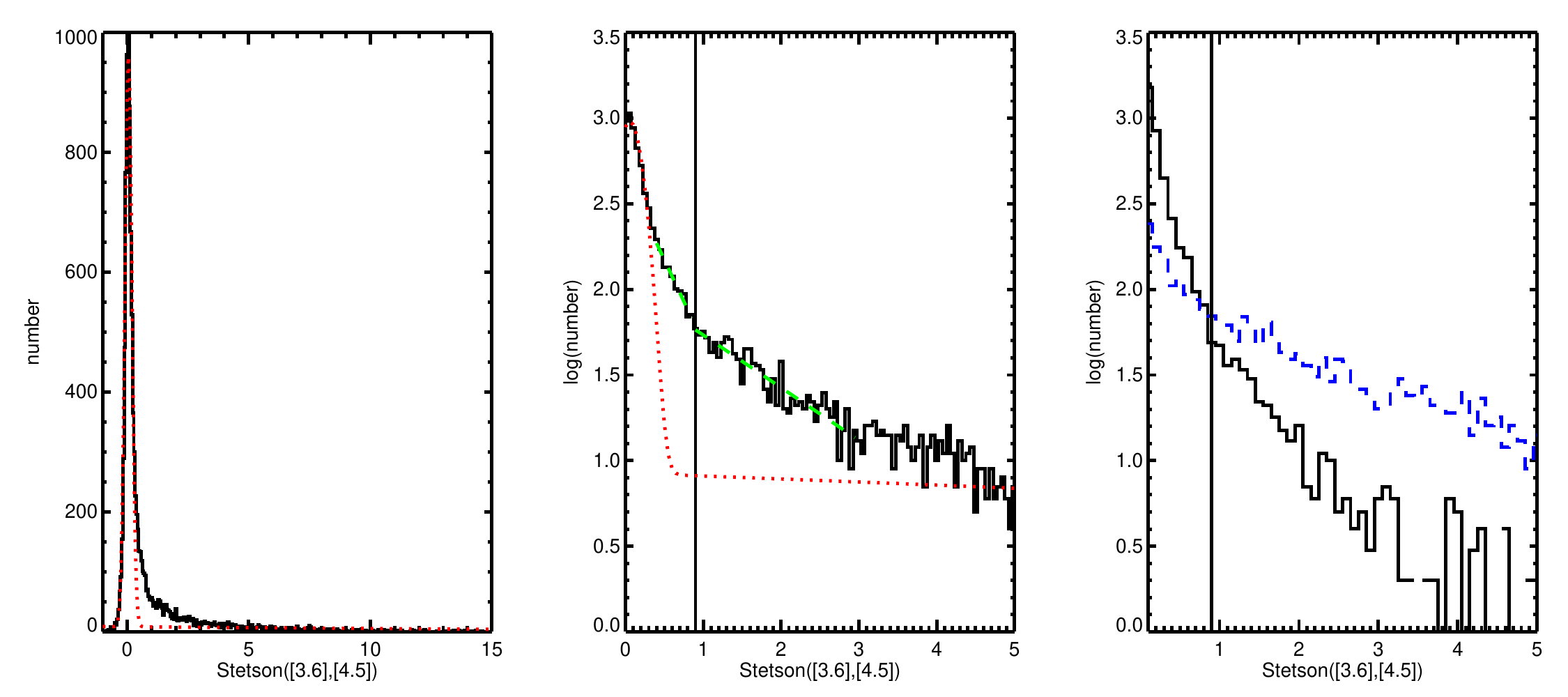}
\caption{Histograms of the Stetson indices calculated for the standard
set for statistics, the subset of which has sufficient points in both
IRAC bands. Left: linear histogram. The red dotted line indicates a
Gaussian fit to the histogram, showing deviations from Gaussianity
towards higher values of the Stetson index, as expected for a
population of identified variable stars.  Middle: A
zoom of the Stetson values between 0 and 5, with log ordinate. The red
dotted line is the Gaussian fit from the left panel. The green, dashed
lines with two different slopes show that there is a break in the
distribution defining our cutoff between variable ($\geq$0.9) and
non-variable ($<$0.9), with the black vertical line at 0.9. 
A value of 0.9 corresponds to
about 6$\sigma$ for the Gaussian fit to the distribution. Right:
Histograms of the Stetson indices for objects in our standard set of
members (blue dashed line) and likely non-members (black solid line).
While this division is imperfect, the distributions cross at
$\sim$0.9, suggesting that relative populations of members/non-members
is the dominant effect in the break in the slope of the entire
distribution at $\sim$0.9.  We conclude that the Stetson index cutoff
of 0.9 is indeed a sensible boundary for demarcating variables from
the general population.}
\label{fig:stetson}
\end{figure}

\section{Identifying Variables}
\label{sec:findingvars}

There are many ways discussed in the literature of identifying
variables in time series data. We tested several methods, and settled
on three primary ones, which we now discuss in separate subsections.
Recall that we are calculating statistics  for the standard
statistical sample, e.g., on just the fast cadence data, for those
objects with at least 5 viable data points in the light curve
(Sec.~\ref{sec:sampledefinition}).

\subsection{Stetson Index}
\label{sec:findingstetson}

The first way we identify variables is the Stetson index (Stetson
1996), which quantifies correlation of variability in two (or
more) bands. The Stetson variability index is
computed for each object as: \begin{equation}
S=\frac{\sum\limits_{i=1}^N g_i \times sgn(P_i) \times
\sqrt{|P_i|}}{\sum\limits_{i=1}^N g_i} \end{equation} where $N$ is the
number of pairs of observations for a star taken at the same
time\footnote{The I1 and I2 maps are taken of any given source
(providing it falls within the region with 2-band coverage) typically
within $\leq$12 minutes of each other. For these observations taken on
a few-epochs-per-day cadence, we are not sensitive to timescales of
minutes, and the data points are effectively simultaneous.},
$P_i=\delta_{j(i)}\delta_{k(i)}$ is the product of the normalized
residuals of two observations, and $g_i$ is the weight assigned to
each normalized residual. In our case the weights are all equal to
one. The normalized residual ($\delta$) for a given band is computed
as: \begin{equation}
\delta_i=\sqrt{\frac{N}{N-1}}\frac{mag_i-\overline{mag}}{\sigma_i}
\end{equation} where $N$ is the number of measurements used to
determine the mean magnitude and $\sigma_i$ is the photometric
uncertainty.  Objects with larger values of the Stetson index are
typically taken to be variable. Since errors are included in the
calculation, light curves that are just noisy are not identified as
variable. Objects with variability in different bands that is not
correlated will not be identified via this method; physically, we
expect most YSOs to have similar variations in the two IRAC channels
since it is hard to imagine processes that would make one IRAC channel
vary without the other. However, this method will not find variables
in cases where one IRAC channel is compromised, e.g., due to
instrumental effects, and the other is not.

If there is correlated noise between the two channels used for the
Stetson index, especially at the faint end, one would expect the
Stetson index to be correlated with source brightness, as seen in,
e.g., Plavchan \etal\ (2008b) or CHS01. Figure~\ref{fig:stetsoni1mag}
shows the distribution of all calculated Stetson indices against mean
[3.6] (for the standard statistical sample, e.g., all objects with
light curves, over all clusters, fast cadence only). Unlike the
analogous figures found in, e.g., Plavchan \etal\ (2008b) or CHS01,
here we have no substantial change in the bulk of the distribution of
Stetson index towards fainter [3.6] magnitudes, so we do not appear to
have correlated noise between the two channels. There is an increase
in frequency of large Stetson index values for brighter [3.6], but
that is likely because brighter objects are more likely to be
legitimately young cluster members, and as such more likely to be
variable (with variability correlated between the two IRAC channels). 
Essentially no large values of the Stetson index are
identified for objects fainter than [3.6]$\sim$16; the intrinsic
error on each point is sufficiently large that these objects do not
have a large Stetson index. (This is consistent with the discussion in
Sec.~\ref{sec:brightnessdistrib} regarding the number of objects
in our data set falling off rapidly fainter than
[3.6]$\sim$[4.5]$\sim$16.)

The specific location of the cutoff between variable and non-variable
can be unique to each data set, as it is affected by the sampling
length and rate of the light curves.  This is the primary  reason
behind our decision to calculate statistics over only the fast cadence
window.  We now discuss how we chose this cutoff value for the Stetson
index. The left panel of Figure~\ref{fig:stetson} shows a histogram of
the Stetson indices for all objects in the standard set for statistics
having at least five points in both I1 and I2. The bulk of the
distribution about 0 are the non-variables, and that part of the
distribution can be reasonably well-fit by a Gaussian.  There are
substantial deviations from Gaussianity towards higher values of the
Stetson index, as expected for a population of identified variable
stars.  There is a change in the distribution of the Stetson index
above and below 0.9; from where the distribution deviates from a
Gaussian to a Stetson index of 0.9, the slope in the middle panel of
Figure~\ref{fig:stetson} is $\sim-1.0$, and  from a Stetson index of
0.9 to $\sim$3, the slope is $\sim-0.3$. Based on this, we take 0.9 as
the cutoff for variability in our data set.  The value of 0.9
corresponds to about 6$\sigma$ for the Gaussian fit to the
distribution.

We note that, in the analogous histograms for each individual cluster,
typically each has a small gap in the Stetson index distribution
at $\sim$0.9. Orion, however, does not, and Orion contributes about
half of the $\sim$11,000 viable 2-band light curves for which the
Stetson index appears in Fig.~\ref{fig:stetson}.

To check that the Stetson index cutoff of 0.9 is sensible, we
conducted a series of Monte Carlo tests. For random light curves (with
a Gaussian distribution of points) using the same time sampling as the
real data, the distribution of Stetson indices is well-described by a
Gaussian with a width typically of 0.1-0.2, comparable to the
left-hand side of Fig.~\ref{fig:stetson}.  For Orion, where we have a
reasonably well-defined set of members and non-members (from MC11; not
just disked and non-disked, but confirmed membership lists), we can
compare the distributions of Stetson indices for the members and
non-members. In Orion, the distribution for non-members is generally
fairly well-described by a Gaussian centered on 0, but there is a
small `shoulder' asymmetry towards larger Stetson index (likely
legitimate field variables or as-yet unidentified members). The
distribution for members, in contrast, is not well-described by a
Gaussian. It is asymmetric with a substantial excess of objects with
high Stetson values. This is as expected, since members are more
likely to have large amplitude, correlated variability.  

Similarly, we can examine the distribution of the Stetson index for
our standard set of members that are also in the standard set for
statistics (and having sufficient points in both bands) and compare it
to the Stetson index distribution for the remaining objects not
selected as members (but still in the standard set for statistics, and
having sufficient points in both bands). We obtain a similar result in
the right panel of Figure~\ref{fig:stetson}; the Stetson distribution
for members crosses that for non-members at about 0.9, or perhaps a
little below that level, suggesting that this division is the dominant
cause of the break in the entire distribution at about that level. 
Even if our separation between members and non-members is imperfect
(which it certainly is), these distributions are consistent with our
selection of 0.9 as the cutoff. We conclude that a Stetson index
cutoff of 0.9 is a sensible boundary for our data set.  

While the objects with Stetson indices of $\lesssim$0.4 have a very
low chance of having legitimate correlated variability, and objects
with Stetson indices $>$0.9 have a high liklihood of correlated
variability, there is a continuum between these values. Objects with
Stetson indices $\gtrsim$0.4 and $\lesssim$0.9 have low-confidence for
correlated variability.  In MC11, we took a different Stetson index of
0.55 as the cutoff, based on the distribution of Stetson indices for
that particular data set. As such, some of the identified
low-confidence variables may have changed between this and the initial
analysis. We also note that the cutoff in Stetson index for CSI 2264 is
very different (Cody \etal\ 2014), but that program has a
substantially different observing cadence than the YSOVAR-classic data
discussed here. In general, the appropriate Stetson index cutoff must
be determined for the individual data set, and there is no universal
value.


\begin{figure}[ht]
\epsscale{0.8}
\plotone{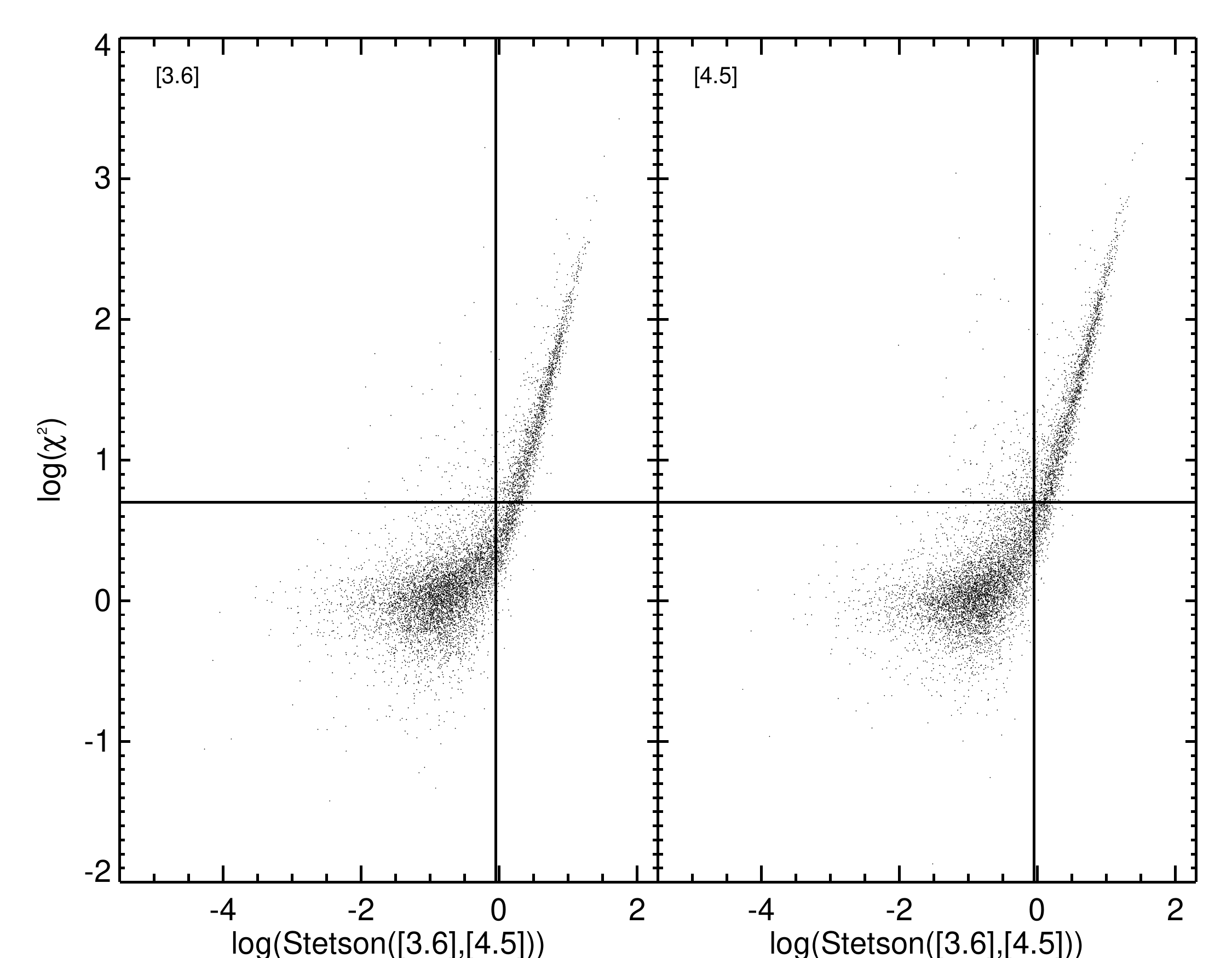}
\caption{Distributions of log $\chi^2_{[3.6]}$ (left) and log
$\chi^2_{[4.5]}$ (right) as a function of the log of the Stetson index.
For the objects within the standard sample for statistics where it is
possible to calculate both $\chi^2$ and the Stetson index, the values
are reasonably well correlated for the variables. The vertical line is
at a Stetson index of 0.9, our cutoff for selecting variable objects
based on the Stetson index. On the basis of this plot, we set a limit
of $\chi^2_{[3.6]}$ or $\chi^2_{[4.5]}\sim5$ (horizontal line) as the
cutoff for potential variability in those cases where only one $\chi^2$
can be calculated (e.g., where monitoring in only one band is
available).  }
\label{fig:chisqstetson}
\end{figure}

\begin{figure}[ht]
\epsscale{0.8}
\plotone{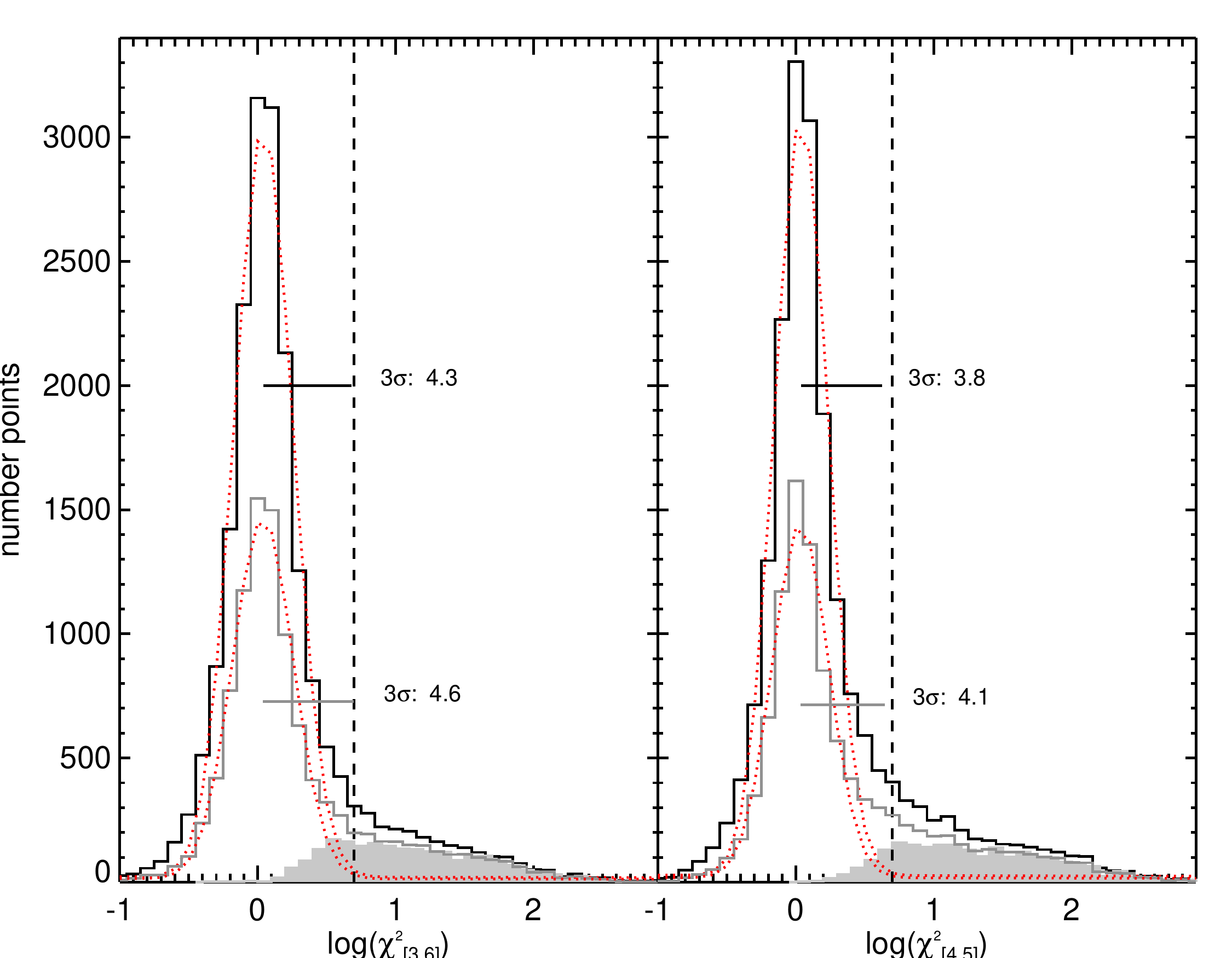}
\caption{Distributions of log $\chi^2_{[3.6]}$ (left) and log
$\chi^2_{[4.5]}$ (right), for all objects with $\chi^2$ values (black,
solid line histogram), for those objects with $\chi^2$ and Stetson
indices (grey, solid line histogram), and for those objects with
$\chi^2$ and a Stetson index $>$0.9 (light grey, filled histogram).
(All objects shown here are from the standard set for statistics.)
The red, dotted line is a Gaussian fit to the corresponding histogram.
The 3$\sigma$ values corresponding to that Gaussian, converted to
linear $\chi^2$, are indicated.  We identify a limit of $\chi^2_{[3.6]}$
or $\chi^2_{[4.5]}\sim5$ as a conservative cutoff for potential
variability in those cases where only one $\chi^2$ can be calculated
(e.g., where monitoring in only one band is available). That limit is
plotted as the vertical dashed line. We take objects with
$\chi^2_{[3.6]}$ or $\chi^2_{[4.5]} > 5$ as legitimately variable.}
\label{fig:chisq2}
\end{figure}

\subsection{Chi-squared test}
\label{sec:findingchi2}

A second method to identify variables is a chi-squared test
($\chi^2$), which, for a given band, is given by \begin{equation}
\chi^2=\frac{1}{N-1}\sum\limits_{i=1}^{N}
\frac{(mag_i-\overline{mag})^2}{\sigma_i^2} \end{equation} where
$\sigma_i$  is the estimated photometric uncertainty (corrected as per
our discussion of the YSOVAR noise floor in
Sec.~\ref{sec:spitzernewdata}).  

This test is used to identify objects with uncorrelated variability,
or variability in only one band (perhaps because data exist in only
one band). This makes the $\chi^2$ test more susceptible to
instrumental issues affecting only one band. However, to demonstrate
that it generally does a good job at recovering variables with large
Stetson indices, Figure~\ref{fig:chisqstetson} shows the distributions
of $\chi^2_{I1}$ and $\chi^2_{I2}$ as a function of Stetson index. For
those objects in the standard set for statistics where it is possible
to calculate both $\chi^2$ and the Stetson index, the values are
reasonably well correlated for the unambiguously variable objects.
Using this plot, we find that a limit of $\chi^2_{I1}$ or
$\chi^2_{I2}\sim5$ is an appropriate, conservative cutoff for
potential variability in those cases where only one $\chi^2$ can be
calculated (e.g., where monitoring in only one band is available).

For our largest data set (Orion), there are thousands of light curves
that meet the requirement imposed by the Stetson index of having data
in both IRAC channels. However, in the 11 smaller-field YSOVAR-classic
data sets, imposing such a 2-band restriction typically means that
more than half the viable light curves would be discarded. Thus, the
$\chi^2$ test is particularly useful in these cases where only one
band is available.

Figure~\ref{fig:chisq2} shows histograms of $\chi^2_{I1}$ and
$\chi^2_{I2}$, for the standard sample for statistics, as well as for
the subset of objects for which a Stetson index can be calculated
(e.g., the sample used in the prior figure), and the smaller subsample
of objects identified as variable using the Stetson index (Stetson
index $>$0.9). We fit a Gaussian to the sharply peaked distributions,
and found a 3$\sigma$ value of, in all cases, $\chi^2 \lesssim$4.5.
The bulk of the $\chi^2$ distribution for which the Stetson index is
$>$0.9 also has $\chi^2>$5. To be conservative, we thus set a limit of
$\chi^2_{I1}$ or $\chi^2_{I2}=5$ for identifying a candidate variable
object. In the ideal case where there are multiple bands of monitoring
data, one could assess each light curve with a large $\chi^2$ but
small Stetson index for physical plausibility.  To attempt to avoid
identifying false variability due to instrumental effects, the light
curves for each of these potentially variable objects will be examined
by hand in the context of the individual cluster analysis to come in
separate papers.

\clearpage

\subsection{Identifying Periodic Variables}
\label{sec:findingperiods}

Finally, as in MC11, there are still legitimately variable sources
within our standard statistical sample that fail both the Stetson
index test (perhaps because only one band is available) and the
$\chi^2$ test (perhaps because the amplitude of variability is small
and our limits for identifying variability were conservative by
design).  There are many mathematical tools available for identifying
periodic behavior in an unevenly sampled time series. The last test
for variability we run here is a periodogram analysis using the NASA
Exoplanet Archive Periodogram
Service\footnote{http://exoplanetarchive.ipac.caltech.edu/cgi-bin/Periodogram/nph-simpleupload}
(Akeson \etal\ 2013). This service provides period calculations using
Lomb-Scargle (LS; Scargle 1982), Box-fitting Least Squares (BLS;
Kov\'acs \etal\ 2002), and Plavchan (Plavchan \etal\ 2008b)
algorithms. These methods are varyingly more or less sensitive to
periodic behavior shaped like sinusoids or flat-bottomed transits,
and/or may be less sensitive to periodic behavior appearing in
addition to other behavior, such as a period superimposed on a
long-term trend. The expected periodic variability in our sample
includes anything repeated, from a sinusoidal-like signal originating
from hot or cool spots on a photosphere, to signals characteristic of
close binaries, to repeated dips in the signal (like `dippers' or AA
Tau; see, e.g., MC11), or even pulsations (e.g., Morales-Calder\'on
\etal\ 2009). 

Specifically because of the variety of expected light curve shapes,
and the weaknesses inherent in any of these methods for finding
periodicity, and noting the approach used by (and results from)
McQuillan \etal\ (2013ab), we also calculated the autocorrelation
function (ACF) for each light curve as a check on repeated patterns. 
We linearly interpolated the light curve onto evenly spaced times, and
then calculated the ACF using the following expression where $L$ is a
lag in days, and $x$ is the light curve (with elements $x_k$):
\begin{equation}
ACF_{x}(L)=ACF_{x}(-L)=\frac{\sum\limits_{k=0}^{N-L-1}
(x_k-\overline{x})(x_{k+L}-\overline{x})}{\sum\limits_{k=0}^{N-1}(x_k-\overline{x})^2}
\end{equation}
We experimented with several different timescales as obtained from the
ACF, and settled on the location of the first peak, providing that the
peak was above an ACF value of 0.2. For those objects with signficant
periods, this coherence time should be well-matched to the period.

We looked for periods in light curves where we had at least 20 points
(more restrictive than our standard statistical sample); we ran all
four methods (LS, BLS, Plavchan, and ACF) on not just the 3.6 \mum\
and 4.5 \mum\ light curves, but also, where possible, the
[3.6]$-$[4.5] light curves. In some cases, a long term trend
(astrophysical, not instrumental) is present in the individual I1 or
I2 light curve, masking a periodic signal, but the color exhibits the
periodic signal.  We looked for periods between only 0.05 and 15 d,
given the overall sampling of our data, and we require at least 2
complete periods over the typically $\sim$40 day window of our
observations. We investigated phased light curves for those periods
calculated using all of these methods. Based on these many thousands
of results, we concluded that LS is the best, for our data set, for
finding reliable, plausible periods. BLS and the Plavchan algorithm,
while they look for a wider variety of shapes of signals, struggle
with light curves that typically have less than 100 points, as ours
do; the ACF approach finds only the strongest signals. Typically, if
the LS algorithm found a reliable period, those other three approaches
found comparable periods.

Thus, we filtered first on the LS results. We dropped candidate
periodic objects if the calculated false alarm probability (FAP; see,
e.g., Scargle 1982) was $>$0.03, or if the recovered period was
$>$14.5 d and the FAP for that period was $>$0.01, or if the period
was $<$0.1 d (slightly larger than the lower limit over which we
searched), or if the calculated period was exactly 15.0 d (by
inspection of the light curves, input and phased, a returned period
exactly equal to the upper limit of our search window was usually
indicative not of a true periodicity, but instead of a long-term trend
in the data). For each of the remaining objects, we investigated the
phased light curve. Perhaps unsurprisingly, given our overall FAP
cutoff of 0.03, about 3\% of the surviving candidate periodic light
curves did not produce physically plausible phased signals. Those
objects were omitted from the final set of periods, and will be
identified as such in the corresponding cluster papers. The planned
individual cluster papers may include a few additional periodic
objects not identified automatically due to the presence of outlying
photometry points which mask periodicity unless removed by hand.

We proceed to include all of the periodic objects identified via the
LS algorithm (dropping the candidate periodic objects as described) in
the set of variable objects, even if they fail the other (Stetson,
$\chi^2$) variability tests. Individual objects will be discussed in
the papers dedicated to each individual cluster, but anywhere from 1
to 15 objects, typically $\leq5$, were added to the list of likely
variables for each of the smaller-field clusters.  For any given
object, we wish to assign a single period to that object.  We take any
period derived from the [3.6] data first ([3.6] is less noisy than
[4.5]), then, only if there is no [3.6] period of sufficient power, we
take the period derived from [4.5], and finally, if no other period of
sufficient quality is available, then we take that derived from
[3.6]$-$[4.5].

A preliminary list of periodic objects in Orion appeared in MC11. The
approach we are now using to search for periods is more stringent than
that in MC11.  Our current approach recovers the bulk of the objects
that MC11 reported as periodic, but does not, for example, recover the
objects reported as having $P>15$d. A complete list of the $\sim$800
periodic variables in Orion as derived from this YSOVAR data reduction
will appear in a later paper.

\subsection{Detection limits for Variability}
\label{sec:varlimits}

Above, we described how variable sources are identified using the
Stetson test, the $\chi^2$ test, and a search for periodic
variability. The cut-off values for those tests were chosen to yield a
conservative list of variable sources and to reject those sources
where the variability stems mostly from instrumental artifacts. We now
present Monte-Carlo simulations to quantify how much variablity
is required to meet those criteria.

Several different physical effects may contribute to the observed
variability and this can lead to very complex patterns in the
lightcurve. In the absence of a theoretical model to explain the
different contributions, we concentrate on simple analytical
prescriptions for lightcurves so as to gain a sense of the sensitivity
of our statistical tools. First, we consider a source that has two
states, a bright and a faint state. The lightcurve switches randomly
between those two states. We simulate lightcurves for a different
fraction of time spent in the upper state (0\%, 10\%, 20\%, 30\%,
40\%, and 50\%) and we expect that variability is more easily found
for sources with a larger amplitude between the two states and an
equal chance to find the source in each state. Second, we simulate
sinusoidal lightcurves with different periods -- 0.1 to 2 days (in
steps of 0.1 days), and 2 to 20 days (in steps of 2 days). We add
Gaussian noise to each lightcurve and vary the ratio of the signal
amplitude and the noise (from 0 to 10 in steps of 0.1, where a
relative amplitude of 0 means a constant lightcurve with noise only).
The sensitivity of the Stetson test, the $\chi^2$ test and the
Lomb-Scargle periodogram is independent of the magnitude of a source
(that is not noise-dominated); only the relative amplitude of the
signal and the noise plays a role. Thus, it is equally possible to
detect strong variability in a weak source (with large photometric
uncertainties) as weak variability in a bright source (with small
photometric uncertaintes).

\begin{figure}[ht]
\epsscale{0.7}
\plotone{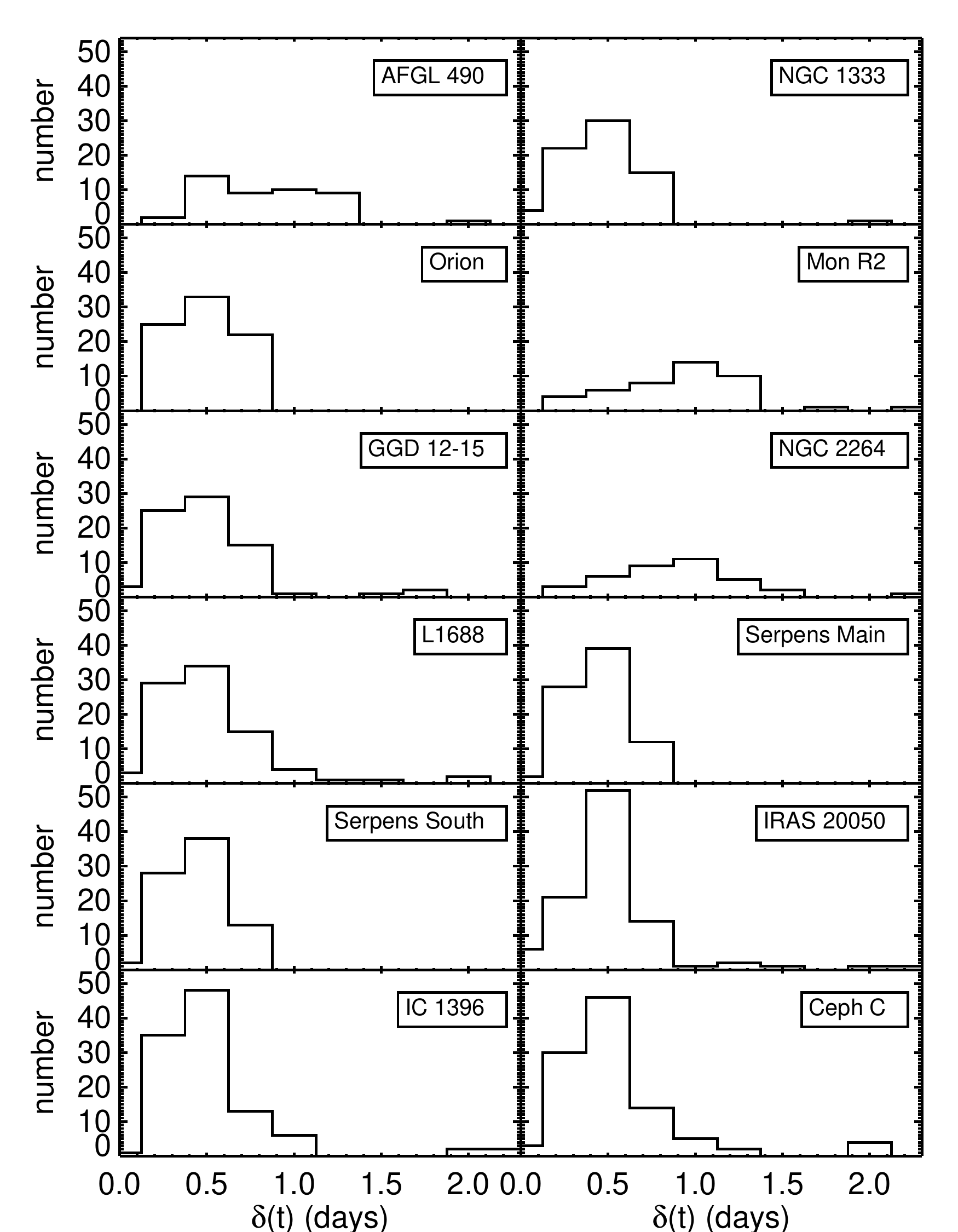}
\caption{Histograms of the time steps between observations for each
cluster. In all cases, the time between observations is non-uniform to
avoid aliasing for a specific period. AFGL~490, Mon~R2 and NGC~2264
have a lower overall sampling rate than all other
clusters. \label{fig:dtpercluster}}
\end{figure}

\begin{figure}[ht]
\epsscale{1}
\plotone{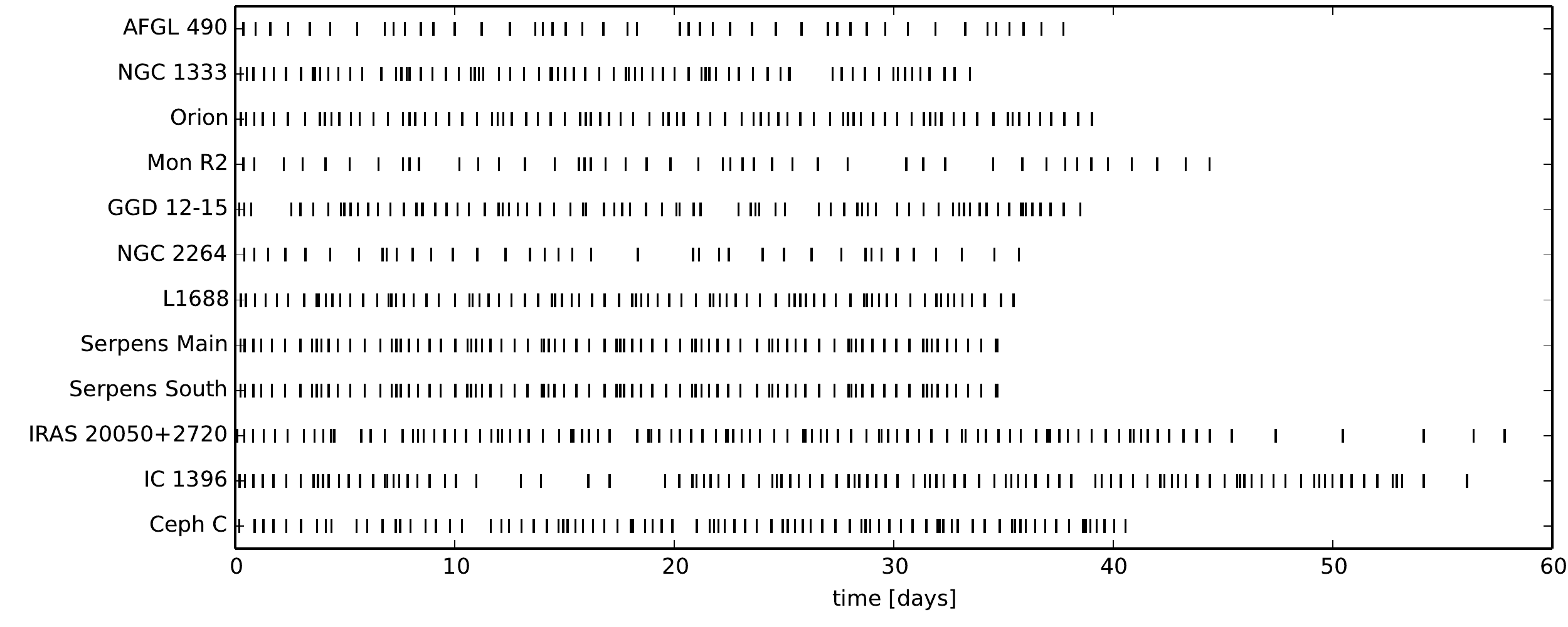}
\caption{Representation of the relative time steps for the fast
cadence observations, in a similar format to Fig.~\ref{fig:cadence},
with a ``$|$'' denoting a time step (from the red sections of
Fig.~\ref{fig:cadence}). AFGL~490, Mon~R2, and NGC~2264 can also be
seen here as having a lower overall sampling
rate.\label{fig:fastcadence}}
\end{figure}

For each grid point in relative amplitude and fraction of time in
upper level or period, for each band, we simulate 10,000 lightcurves
for each grid point for each band. The lightcurves are sampled at the
time intervals of the actual YSOVAR observations.
Figure~\ref{fig:dtpercluster} shows the histogram of the time steps
between observations for each cluster (the typical min and max $\Delta
t$ were given above in Table~\ref{tab:programs2}). For all clusters,
the sampling is non-uniform to avoid aliasing for a specific period.
However, the histograms fall in two groups, with AFGL~490, Mon~R2, and
NGC~2264 having a lower overall sampling rate than the other clusters.
Figure~\ref{fig:fastcadence} is another representation of the sampling
rate. In the style of Fig.~\ref{fig:cadence}, it visually represents
the sampling rates for the fast cadence monitoring. Here, too,
AFGL~490, Mon~R2, and NGC~2264 can be seen to have a lower overall
sampling rate than the other clusters. We ran all Monte Carlo
simulations using the actual time sampling from L1688, one of the
clusters with a high sampling rate and Mon~R2, a representative
cluster with a lower sampling rate.

\begin{figure*}[h]
\epsscale{1.0}
\plotone{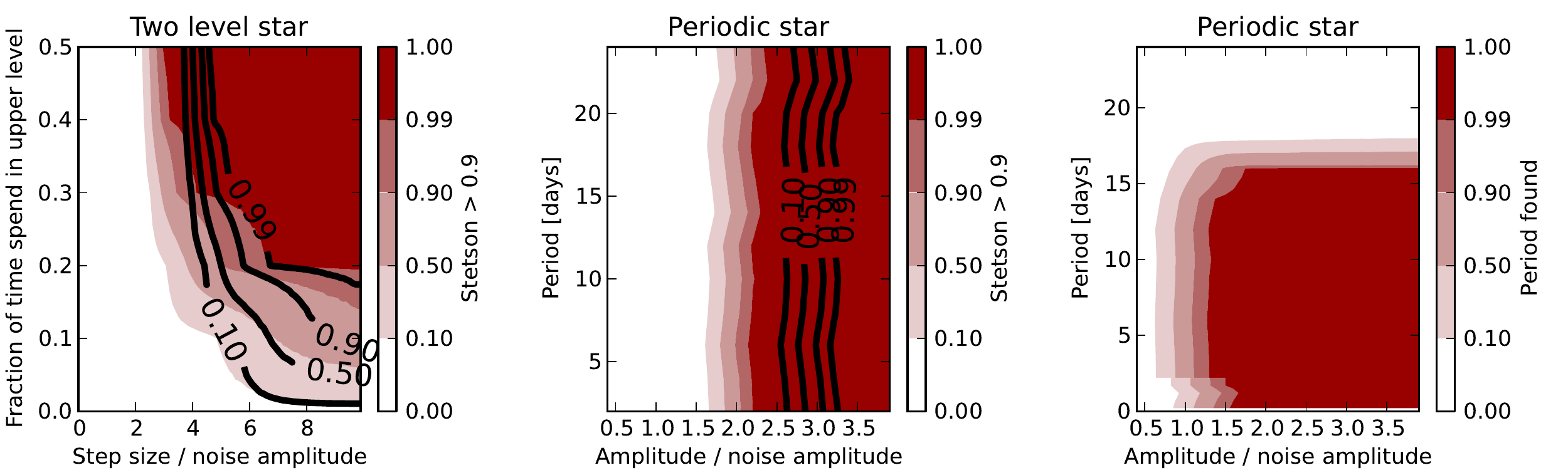}
\caption{Efficiency for variability detection. The left panel shows
simulations for lightcurves with two distinct states (bright and
faint); the other two panels show results for a sinusoidal lightcurve.
The contours indicate the fraction of sources in that part of the
diagram that return $\chi^2>5$. In the first two panels, colors
indicate probability that the Stetson index $>$0.9 (and note that the
color scale is non-linear). In the right panel, colors indicate the
simulated period that was recovered in a Lomb-Scargle periodogram. 
The time sampling is from the observations of L1688. 
\label{fig:detectionlimits_L1688} }
\end{figure*}

Figure~\ref{fig:detectionlimits_L1688} shows results from the Monte
Carlo simulations using the sampling of the L1688 cluster. The left
panel presents the detection efficiency for lightcurves from an object
with two distinct luminosities. If the star is found in each state
half the time, the Stetson test will identify it as variable in
almost all cases (99\%), if the step size is at least three times
larger than the noise level. Since the $\chi^2$ test uses data from
one band only, the step size must be larger (five times the noise
level) to reach the same detection efficiency. If the star spends less
than 20\% of the time in either state, there is a resonable
chance that the variability will not be found, even for larger step
sizes, since the sampling might catch only few datapoints in this
state. For a periodic lightcurve (middle panel), variability is again
found more easily in the Stetson test than in the $\chi^2$ test, and
the period of the variability does not influence the detection
efficency, since those two tests do not consider the time ordering of
the observed data. Using the LS approach, we are sensitive to periods
($P$) between about 1 and 15 days. Even periods where the amplitude
($a$) of the lightcurve $a \sin(\frac{2\pi}{P}t)$ is only twice as
large as the noise level are easily detected, almost independent of
the period in the range 1 to 15 days. Such weak signals would not
necessarily show up as variable in the Stetson or $\chi^2$ test (see
middle panel). In general, the region where the tests detect
variability in some lightcurves but not in others is fairly narrow.

\begin{figure*}
\epsscale{1.0}
\plotone{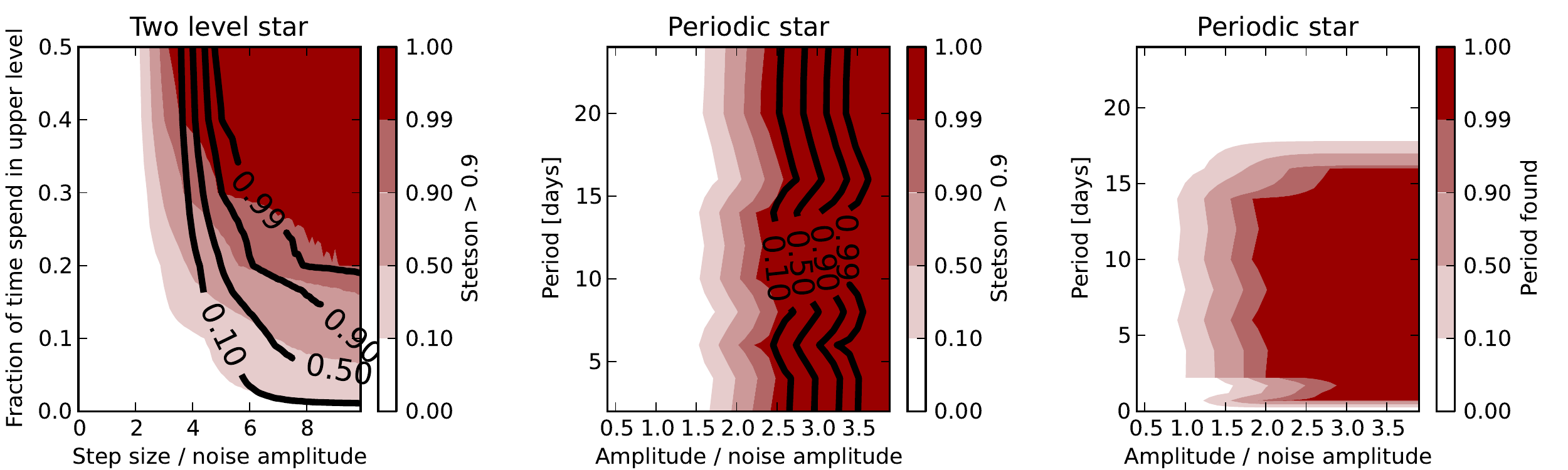}
\caption{Same as in figure~\ref{fig:detectionlimits_L1688}, but using
the time sampling of the Mon~R2 observations. The different time
sampling effectively means that slightly increased signal-to-noise
ratios are required. \label{fig:detectionlimits_MonR2}}
\end{figure*}

Figure~\ref{fig:detectionlimits_MonR2} shows the same plots as
Figure~\ref{fig:detectionlimits_L1688} but for the time sampling of
Mon~R2, one of the clusters with a lower cadence in the observations.
The general shape of the regions in the parameter space where
variability can be detected is the same, but due to the lower number
of observations and the larger time span between observations, a
larger amplitude is required, and we are less sensitive to periods
below about 2 days.

To explore more complicated lightcurves, we combined several effects,
e.g., a sinusiodal periodicity overlaid on a long-term trend. In
these cases, generally the strongest effect determines how the
variability will be seen. If the magnitude of the trend is large and
the amplitude of the sine wave is small, then the lightcurve will be
marked as variable, but the periodicity might not be detected. 

Given the possible complexity of real lightcurves, it is not possible
to cover the entire parameter space with Monte Carlo simulations.
However, the scenarios presented here show that in general we can
detect periodicity with an amplitude just twice the level of the
noise, variability at 3-5 times the level of the noise with the
Stetson test and about 6-10 times the level of the noise for
single-band lightcurves with the $\chi^2$ test. Similar results for
the relative sensitivity of the Stetson and the $\chi^2$ test were
also found by Flaherty \etal\ (2013), although they used different
cut-off levels than this work.

As seen above in Fig.~\ref{fig:rmsmag} and in MC11, there is a
reasonably strong correlation between the mean magnitude of a source
and its mean error. This overall relation affects our ability to find
variability or periodicity.  For sources brighter than 13th mag, the
total uncertainty is dominated by the error floor introduced in
Sec.~\ref{sec:spitzernewdata}. For those brighter sources, we can
detect periodicity if the amplitude is larger than about 0.02~mag and
variability if it is larger than about 0.03-0.1~mag (with the exact
number depending on the signal shape). 


\subsection{Identifying Variables: Cryo-to-Post-Cryo (6-7 years)}
\label{sec:ctopc}

\begin{deluxetable}{lclllclll}
\tabletypesize{\scriptsize}
\rotate
\tablecaption{Statistics on Variable Objects on the Longest
Timescales\label{tab:longtermstats}}
\tablewidth{0pt}
\tablehead{
\colhead{Cluster\tablenotemark{a}} & 
\colhead{I1 faint cutoff } &
\colhead{I1 center$\pm\sigma$\tablenotemark{b}}&
\colhead{I1 vars/LCs\tablenotemark{c}}&
\colhead{I1 memb.vars/LCs\tablenotemark{d}} &
\colhead{I2 faint cutoff} &
\colhead{I2 center$\pm\sigma$\tablenotemark{b}}&
\colhead{I2 vars/LCs\tablenotemark{c}}&
\colhead{I2 memb.vars/LCs\tablenotemark{d}} \\
 & \colhead{(mag)}
 & \colhead{(mag)} 
 & \colhead{(mag)} 
 &
 &\colhead{(mag)}
 & \colhead{(mag)} 
 & \colhead{(mag)} 
 }
\startdata
AFGL 490& 15.8& 0.001$\pm$ 0.04& 86/     907= 0.09& 43/     120= 0.36&
15.8& 0.017$\pm$ 0.05& 112/     904= 0.12& 58/     155= 0.37\\ NGC
1333& 16.0& 0.033$\pm$ 0.05& 43/     289= 0.15& 24/     115= 0.21&
16.0& 0.035$\pm$ 0.05& 82/     393= 0.21& 31/     112= 0.28\\ Orion&
16.0& 0.008$\pm$ 0.04& 738/    5313= 0.14& 439/    2326= 0.19& 16.0&
0.013$\pm$ 0.04& 920/    5921= 0.16& 476/    2426= 0.20\\ Mon R2&
16.0& 0.020$\pm$ 0.06& 57/     431= 0.13& 32/     170= 0.19& 16.0&
0.031$\pm$ 0.11& 24/     211= 0.11& 10/      90= 0.11\\ GGD 12-15&
16.0& 0.014$\pm$ 0.05& 56/     336= 0.17& 35/     127= 0.28& 16.0&
0.046$\pm$ 0.06& 44/     400= 0.11& 28/     138= 0.20\\ NGC 2264&
15.8& 0.011$\pm$ 0.06& 47/     365= 0.13& 29/     142= 0.20& 15.8&
0.017$\pm$ 0.09& 58/     492= 0.12& 27/     199= 0.14\\ L1688& 15.5&
0.017$\pm$ 0.05& 34/     264= 0.13& 18/      48= 0.38& 15.5&
0.039$\pm$ 0.06& 47/     312= 0.15& 13/      43= 0.30\\ Serpens Main&
15.5& 0.008$\pm$ 0.04& 130/    1112= 0.12& 25/      61= 0.41& 15.5&
0.038$\pm$ 0.05& 135/    1161= 0.12& 24/      83= 0.29\\ Serpens
South& 15.5& 0.013$\pm$ 0.04& 59/     587= 0.10& 28/      68= 0.41&
15.5& 0.031$\pm$ 0.05& 80/     706= 0.11& 24/      78= 0.31\\ IRAS
20050+2720& 15.8& 0.004$\pm$ 0.05& 123/    1372= 0.09& 31/     114=
0.27& 15.8& 0.021$\pm$ 0.06& 112/    1415= 0.08& 32/     106= 0.30\\
IC 1396A& 15.5& -0.007$\pm$ 0.03& 117/    1122= 0.10& 21/      73=
0.29& 15.5& 0.003$\pm$ 0.04& 185/    2116= 0.09& 44/     175= 0.25\\
Ceph C& 15.8& 0.004$\pm$ 0.04& 51/     531= 0.10& 27/      71= 0.38&
15.8& 0.030$\pm$ 0.05& 61/     553= 0.11& 33/      82= 0.40\\
\enddata
\tablenotetext{a}{The values we used for $\Delta t$, the time laps
between the cryo and post-cryo observations, are included in
Table~\ref{tab:cryoprograms}.}
\tablenotetext{b}{Mean I1 or I2 cryo-to-postcryo offset, e.g., center (peak
location) and $\sigma$ of Gaussian fit to distribution of differences
in magnitudes brighter than the faint cutoff.}
\tablenotetext{c}{Number of identified long term variables over all of the
objects with light curves / number of light curves (for all objects) =
long term variability fraction for all objects.}
\tablenotetext{d}{Number of identified long term variables over just the
standard set of members / number of light curves (for members) =
long term variability fraction for members.}
\end{deluxetable}

In addition to the variability probed on the YSOVAR monitoring
timescale of $\sim$40d, we also are interested in the evidence for
longer-timescale variations between observations of these same
clusters in the Spitzer cryogenic epoch and the post-cryo (YSOVAR)
epochs. To identify the variables in this case, for every object with
a light curve, we can compare the average measurement from the
earliest cryo era (Table~\ref{tab:cryoprograms}), and the mean for
that object over the YSOVAR standard statistical sample
(Sec.~\ref{sec:standardsetforstatistics}).  

\begin{figure}[ht]
\epsscale{0.8}
\plotone{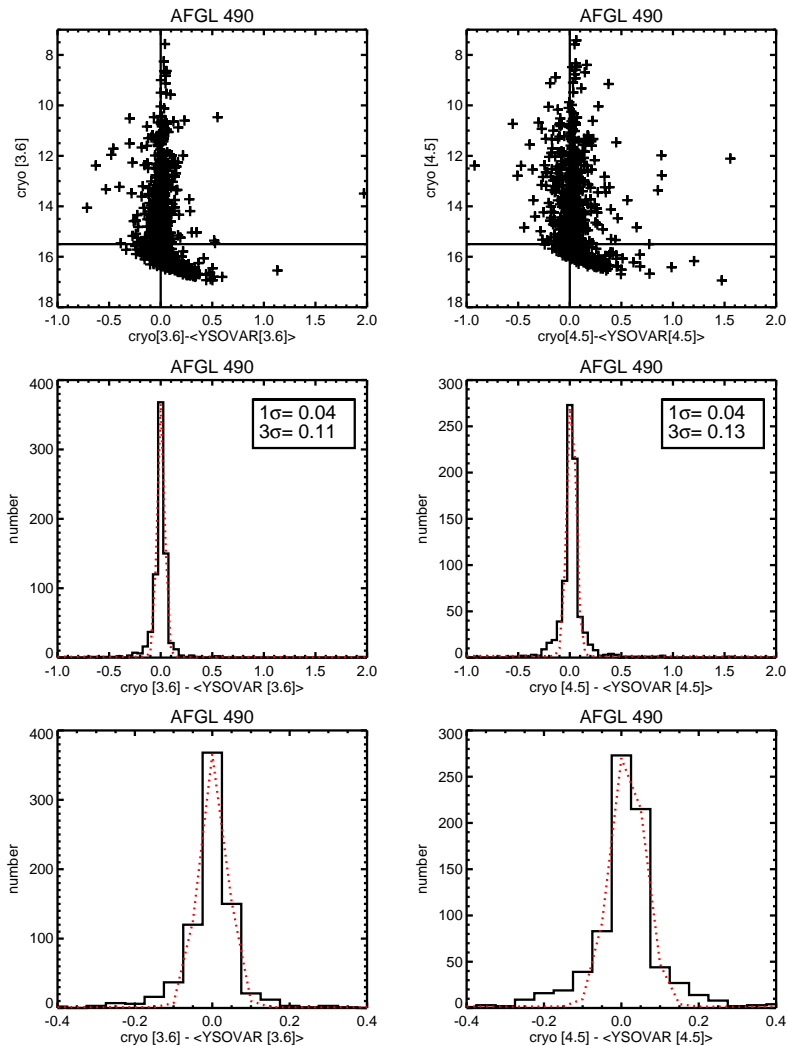}
\caption{An illustration of the process used to identify long term
variables in the clusters. Top: plot of the difference between the
cryogenic-era measurement and the mean measurement from the standard
set for statistics against the cryogenic-era measurement (left: [3.6],
right: [4.5]). Only objects brighter than 15.5 were used (in this
cluster) for the next step. Middle: Histograms of the difference in
magnitudes for objects brighter than 15.5. The solid line is the data;
the dotted line is a Gaussian fit to the histogram. Based on this fit,
we retained objects for which the difference in magnitudes is greater
than 3$\sigma$ away from the peak of the distribution as likely
variables. Bottom: Zoom in on the central portions of the prior two
histograms. A summary of this analysis searching for variables over
this long-term baseline is in Table~\ref{tab:longtermstats}.}
\label{fig:afgllongtermchanges}
\end{figure}

The process we used to identify the long-term variables in each
cluster is shown for AFGL 490 in Figure~\ref{fig:afgllongtermchanges}.
We plot the difference between the cryo and post-cryo measurements for
each object as a function of the cryo value and determine, for each
cluster, the brightness at which photometric noise clearly dominates.
This faintness limit is close to 16th mag, consistent with what we
noted in Figures~\ref{fig:jdistpart1}, \ref{fig:jdistpart2}, and
\ref{fig:stetsoni1mag}. We select this limit separately for each
cluster and consider only objects brighter than this limit.  We fit a
Gaussian to the histogram of the difference between the cryo and
post-cryo measurements, allowing the zero point as well as the height
and width of the Gaussian to be free parameters. We classify as
long-term variables all objects with cryo to post-cryo offset further
than 3$\sigma$ from the peak of the (fitted) distribution.   All
objects from the standard statistical sample, not just the members (or
just the variables), go into this process of defining the width of the
distribution. The fraction of objects in each field that are
classified as variable, and the subset of variables that are also
cluster members, are both identified after the long-term variables are
identified. A summary of the important parameters in these analysis
steps to search for variables over this long-term baseline is in
Table~\ref{tab:longtermstats}; the values we used for $\Delta t$, the
time lapse between the cryo and post-cryo observations, are included
in Table~\ref{tab:cryoprograms}. 

\clearpage
\section{Discussion}
\label{sec:discussion}

In this section, we present analysis of the distribution of rotation
rates as a function of IR excess, evidence (or lack thereof) for
transient IR excesses, evidence for skews over time towards more
brightening or fading sources, and how the long-term variability
fraction varies as a function of cluster parameterization (from
Sec.~\ref{sec:clusterparameterization}) or length of time baseline
sampled.

\subsection{Periodic variables}

Our YSOVAR map in Orion is far larger than the other cluster maps,
which focus on the most embedded (possibly youngest) objects in these
clusters. For these embedded objects, it is not generally possible to
obtain a rotation period from ground-based optical or NIR data due to
extinction. Moreover, for stars with more significant disks, it is
less likely that the IR light curve will be strictly periodic. Many
distinct processes can contribute to a YSO's mid-IR variability, and
it often results in a stochastic light curve (Cody \etal\ 2014). 

For the objects for which we can derive a period, we would like to
compare our values to those from the literature as a check on our
methodology. Since our clusters are, for the most part, very embedded,
there are not very many known periods in the literature. As discussed
in Sec.~\ref{sec:clusterproperties}, there are many periods for
Orion and NGC 2264, typically obtained in the optical, but with some
values from the NIR. Parks \etal\ (2014) reports on NIR periods from
objects in our region of L1688; there are other literature values for
rotation periods of objects elsewhere in L1688, beyond our monitored
region.  We can roughly compare to the MIR timescales reported in
Morales-Calder\'on \etal\ (2009) for IC 1396A. 

\begin{figure}[ht]
\epsscale{0.8}
\plotone{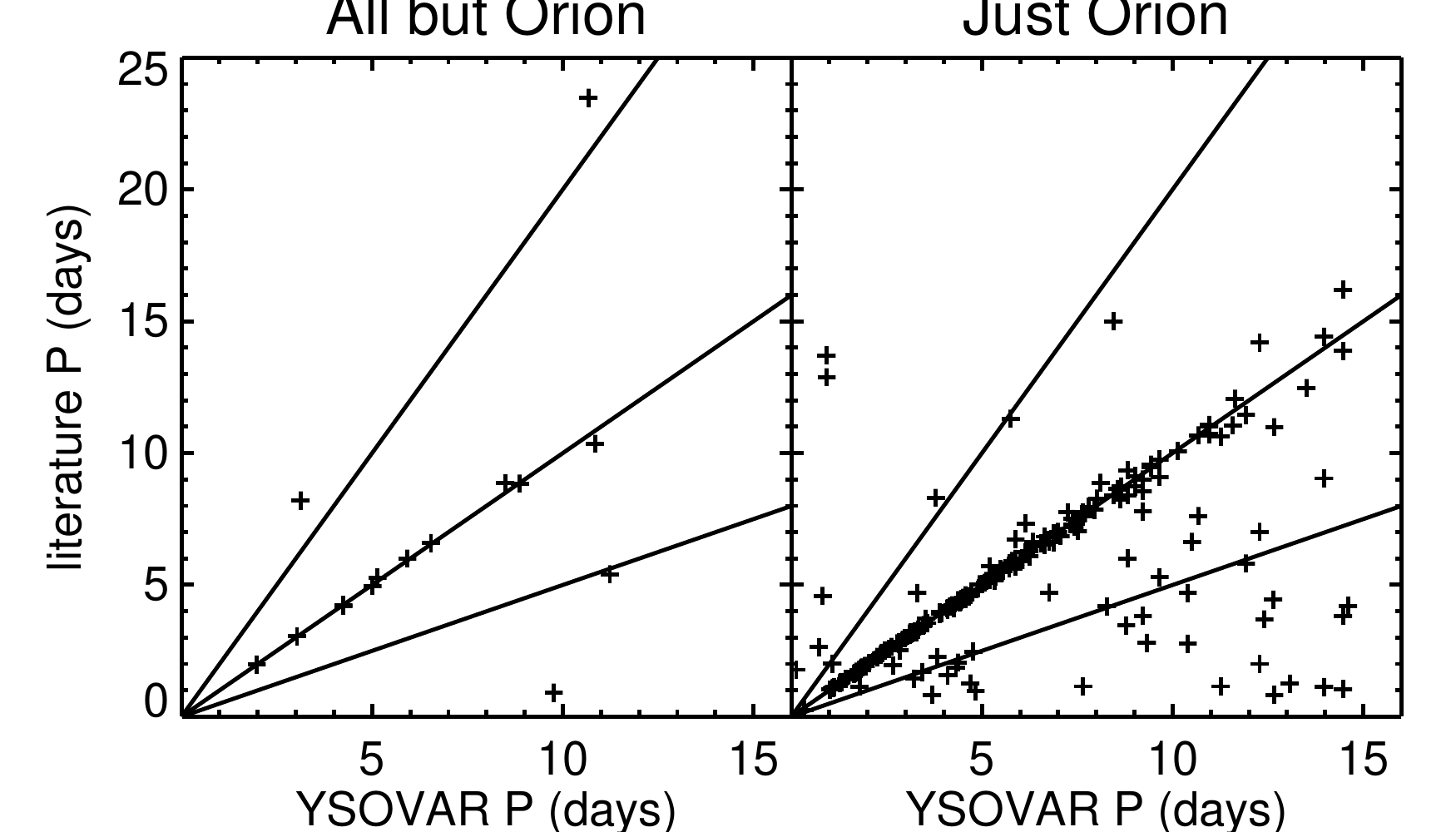}
\caption{The YSOVAR-determined period in days against the literature
period in days for (left) NGC 2264, L1688, and IC 1396A, and (right)
Orion alone. Solid lines indicate a 1-to-1 match, a 2:1 match, and a 1:2
match.}
\label{fig:periods2}
\end{figure}

Of the objects with periods in the literature, there are $\sim$200
that also have periods derived here in Orion (excluding those from
MC11, since those were derived from the same observations we use
here), and an additional $\sim$15 from NGC 2264, L1688, and IC 1396A.
About 75\% of those have period measurements that match to better than
10\%, so we have confidence that our period-finding approach is at
least well-matched to those in the literature.
Figure~\ref{fig:periods2} plots the YSOVAR-determined period against
the literature period for those objects where it is possible. The
clusters that are not Orion (NGC 2264, L1688, and IC 1396A) are
plotted separately simply because Orion dominates the statistics, and
it is useful to see if there are good matches outside Orion as well as
within Orion. Three of the four objects from the smaller-field three
clusters that are not well-matched to the YSOVAR-determined period are
close to likely harmonics, and the periods are of comparatively low
quality. The one that is most discrepant is from NGC 2264, SSTYSV
064101.40+093408.1, and is being compared to a period from Lamm \etal\
(2004). Our phased light curve looks correct (for our wavelength and
epoch of observation). Of the $\sim$50 Orion periods that do not agree
to 10\% (out of $\sim$200 Orion period comparisons total), it is
predominantly the case (by a ratio of 3 to 1) that the YSOVAR period
is longer than the literature period. About 60\% of these have
$[3.6]-[8]>0.8$.  All but 5 of those Orion periods were optically
determined. Because we are working in longer wavelengths, it is
possible that, particularly in those cases, we may be sampling a
different location in the star-disk system, e.g., futher away from the
photosphere, where Keplerian rotation periods are longer. 

\begin{figure}[ht]
\epsscale{0.8}
\plotone{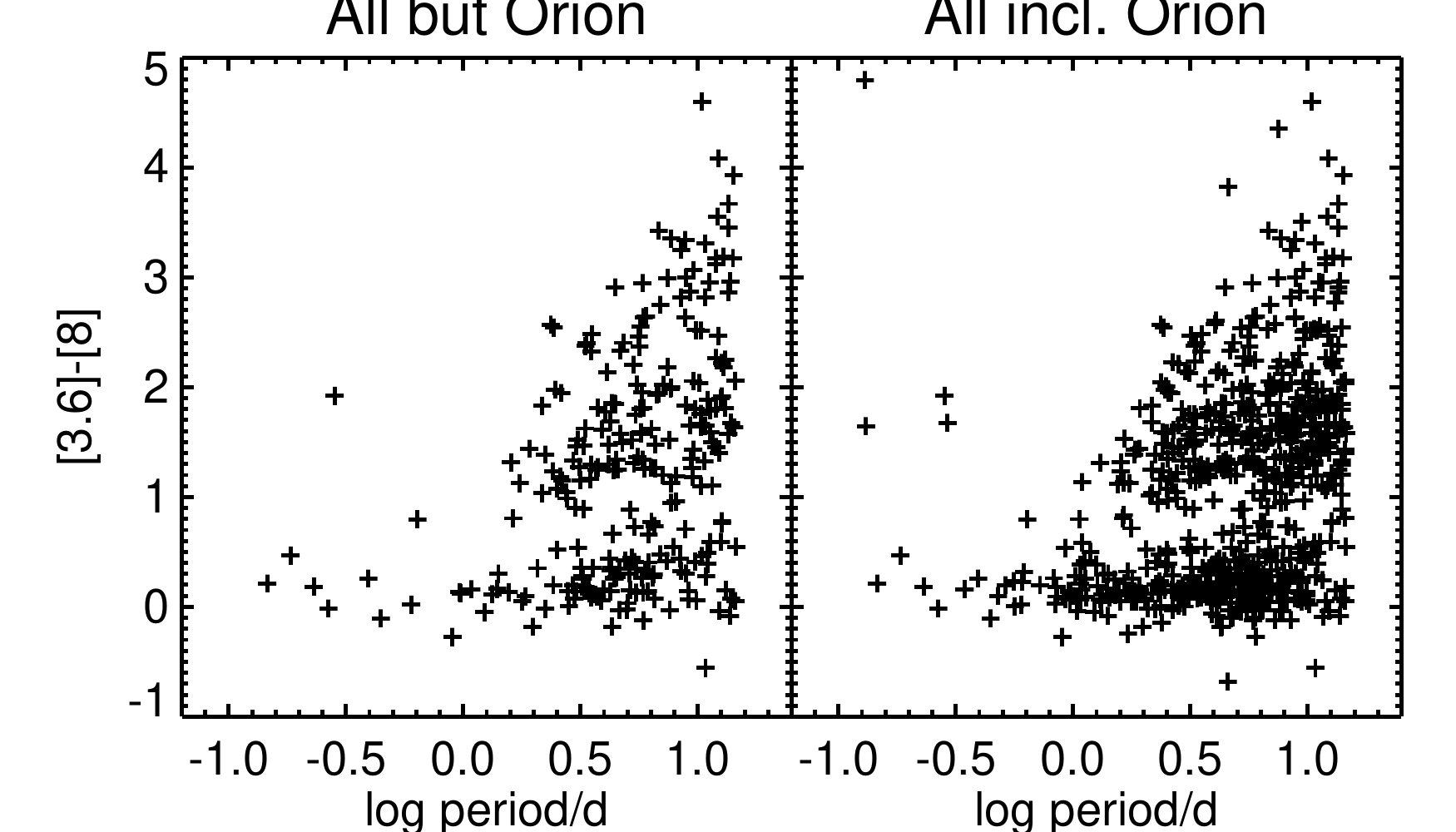}
\caption{IR excess ([3.6]$-$[8]) vs.\ log (period in days) for objects
in the YSOVAR clusters, using the periods derived from the YSOVAR data
as described in the text. Left: the 11 clusters, excluding Orion;
there are $\sim$250 objects. Right: all 12 YSOVAR clusters,
including Orion; there are $\sim$430 Orion objects plus the $\sim$250
objects from  left panel.  The plots are similar, both to each
other and to that obtained for Orion by Rebull \etal\ (2006), despite
the fact that most of these stars are on average thought to be younger
than those in Orion. }
\label{fig:periods}
\end{figure}

\begin{figure}[ht]
\epsscale{0.6}
\plotone{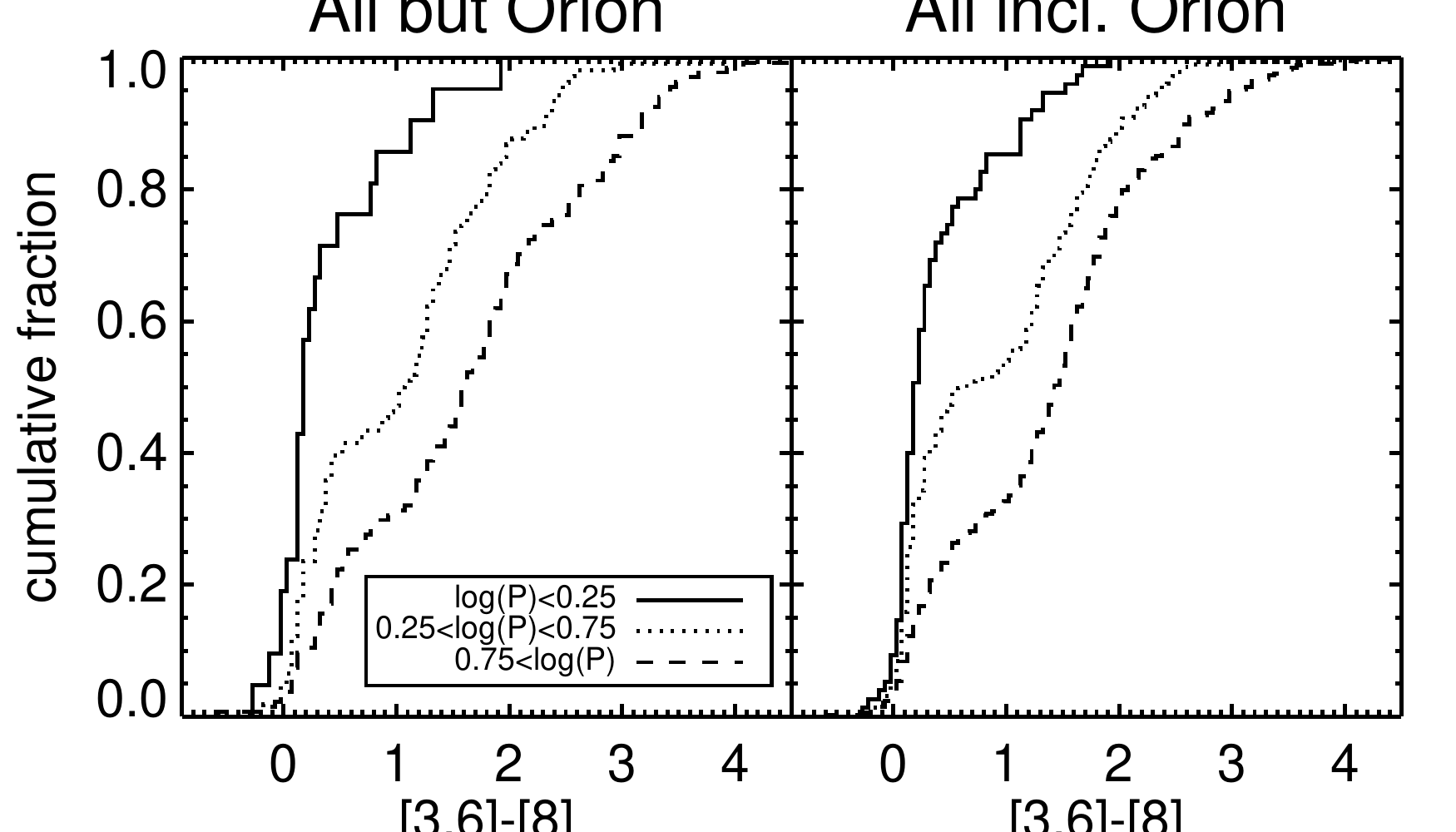}
\caption{Cumulative distributions of IR excess ([3.6]$-$[8]) for
objects with log ($P$)$\leq$0.25 (solid line),
0.25$<$log($P$)$\leq$0.75 (dotted line), and 0.75$<$log($P$) (dashed
line). Left: the 11 smaller-field clusters, excluding Orion. Right:
all 12 YSOVAR clusters, including Orion. The distributions are
significantly different according to a K-S test.}
\label{fig:periods4}
\end{figure}
\begin{figure}[ht]
\epsscale{0.6}
\plotone{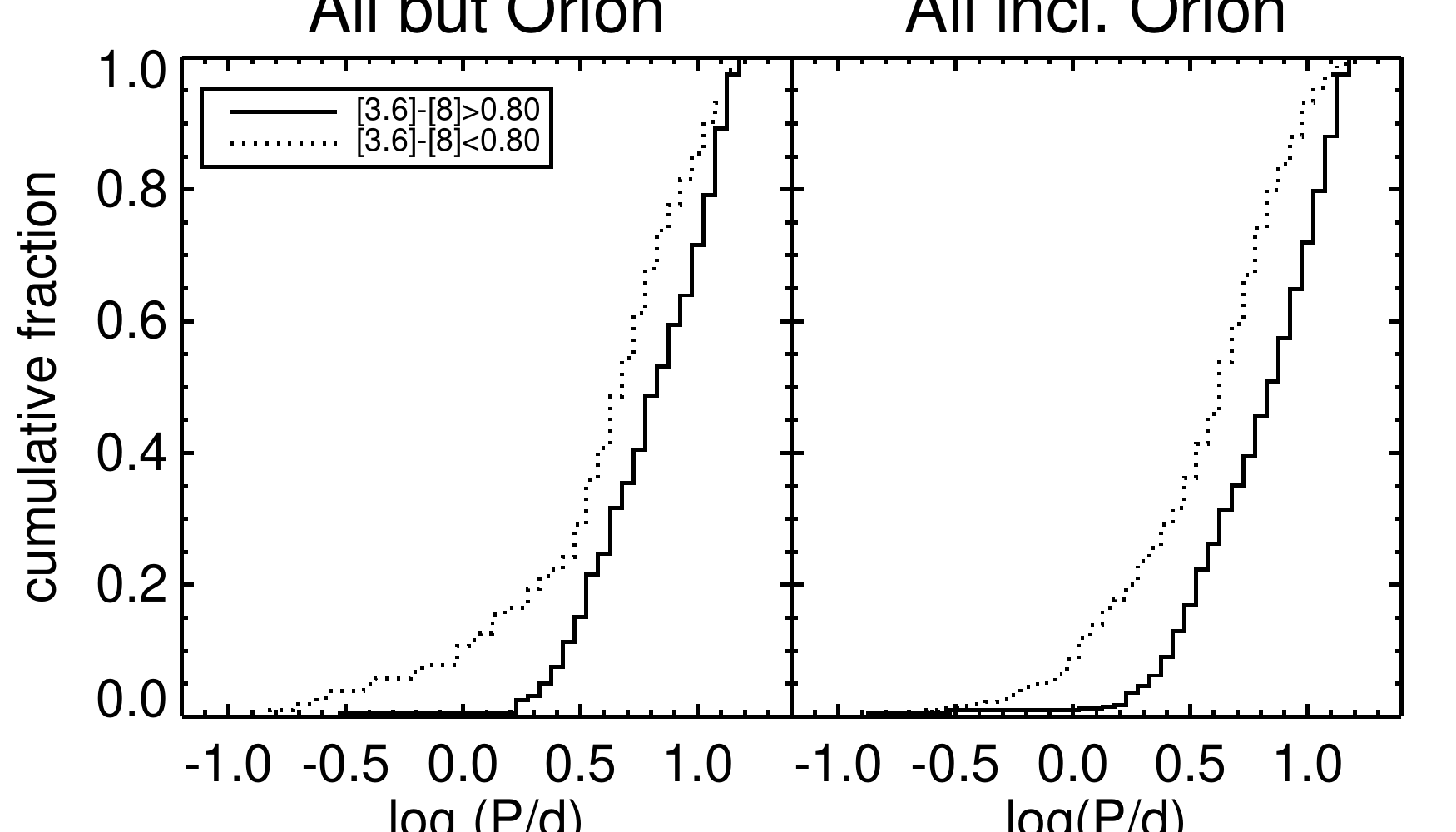}
\caption{Cumulative distributions of log (P)  for objects with
$[3.6]-[8]>$0.8 (solid line), and $[3.6]-[8]\leq$0.8 (dotted line).
Left: the 11 smaller-field clusters, excluding Orion. Right: all 12
YSOVAR clusters, including Orion. The distributions are significantly
different according to a K-S test. }
\label{fig:periods4b}
\end{figure}

A relation has already been found between IR excess and rotation rate
for young stars suggesting that IR excess and rotation rate are
related; out of our 12 clusters, this relation has been found in Orion
(Rebull \etal\ 2006) and NGC 2264 (Cieza and Baliber 2007). In the 11
smaller-field clusters (i.e., all but Orion), there are $\sim$350
stars with periods measured from YSOVAR light curves, but only
$\sim$250 of those also have cryogenic Spitzer measurements at 3.6 and
8 \mum\ from which we can get a clear indication of the IR excess in
these systems. There are $\sim$800 stars in Orion with measured YSOVAR
periods, but only $\sim$430 have [3.6] and [8] measurements. 

Figure~\ref{fig:periods} shows the relationship between IR excess
(specifically [3.6]$-$[8]) and YSOVAR-derived IR periodicity for these
sources.  In both cases, there is a gap near [3.6]$-$[8]$\sim$0.8,
which divides the disk candidates (above that cutoff) from the
non-disk candidates (below). There is also different behavior to the
left and right of $\log(P)\sim$0.25, or $P\sim$1.8d -- excesses do not
necessarily imply longer periods, but a star with a longer period is
more likely than those with shorter periods to have an IR excess.
Figure~\ref{fig:periods4} shows the cumulative distributions of
[3.6]$-$[8] for the same two panels as in Fig.~\ref{fig:periods}, for
three different bins of log($P$), divided at log($P$)=0.25 and 0.75
(1.78 days and 5.62 days, respectively). According to
Kolmogorov-Smirnov (K-S) tests, the distributions of [3.6]$-$[8] are
significantly different within each panel. The two distributions that
are the most similar are the full (all 12 clusters) distributions for
0.25$<$log($P$)$\leq$0.75, and 0.75$<$log($P$); the probability that
those populations were drawn from the same distribution is 4\%. The
probability that the populations were drawn from the same distribution
for log ($P$)$\leq$0.25 and 0.75$<$log($P$) (again, for all 12
clusters) is $\sim10^{-13}$; for log ($P$)$\leq$0.25  and
0.25$<$log($P$)$\leq$0.75, the probability that the populations were
drawn from the same distribution is $\sim10^{-10}$.  Similarly, 
Figure~\ref{fig:periods4b} shows the cumulative distributions of log
($P$) for the same two panels as in Fig.~\ref{fig:periods}, for two
different bins of [3.6]$-$[8], divided at [3.6]$-$[8]=0.8. Again,
according to K-S tests, the distributions of log($P$) are
significantly different within each panel; the probability that either
of the populations were drawn from the same distribution is
$<10^{-17}$. 

The plots in Fig.~\ref{fig:periods}  are very similar to that obtained
for Orion by Rebull \etal\ (2006), and that by other investigators in
other clusters, despite the fact that optically determined periods
were used there. The periods derived from our IR YSOVAR data may be
photospheric rotation rates, pulsation rates (see, e.g.,
Morales-Calder\'on \etal\ 2009), or inner disk rotation rates (see,
e.g., Artemenko \etal\ 2013); on the other hand, for those clusters
where there are periods available, we match the literature reasonably
well, and the literature for the most part is using optical data to
obtain periods. Therefore, it seems that we are, in most cases, not
sampling much different locations in the star-disk system. However, an
exception could be that the optical and IR observations are sampling
two separate places whose movements are locked together, such as
starspots on the photosphere and stellar-magnetosphere-driven disk
disturbances at the corotation radius.

It is also surprising that the results for the aggregate set of
clusters are so similar to that for Orion, because the clusters
should be for the most part substantially younger than Orion. This
could imply that disk locking may be in effect at even these young
ages, or that accretion-powered stellar winds are the dominant
mechanism to slow these objects.  However, it is likely that we can
obtain viable periods more easily for relatively unobscured stars,
e.g., with these periods, we are also sampling the older end of the
young star distributions in these clusters. Little is known about many
of the objects outside of Orion shown here; additional study of the
individual objects will help clarify matters. Individual objects will
be discussed in the corresponding YSOVAR cluster paper.

\subsection{Disks Don't Vanish or Appear}

There have been recent reports of debris disks undergoing significant
short-term changes; both Meng \etal\ (2012) and Melis \etal\ (2012)
report on systems that change significantly at wavelengths $>$10 \mum\
over timescales of years.  Our objects are much younger (a few Myr
rather than a few tens or hundreds of Myr) and our monitoring
wavelengths are considerably shorter. However, Rice, Wolk, \& Aspin
(2012) also note that 9 (36\%) of the stars in their sample of young
stars in Cygnus OB7 (comparable in age to our sample) have a transient
NIR ($JHK$) excess. We can use this first look at our data to
constrain the degree to which IR excesses in our sample vanish (or
appear) on the timescales of years, namely between the cryo-era
observation and that of our post-cryo observations.

Irrespective of whether or not objects have been identified as
variable above, we compared the cryo-era [3.6]$-$[4.5] color with the
maximum and minimum [3.6]$-$[4.5] color obtained during our YSOVAR
(fast-cadence) monitoring (standard set for statistics, the subset of
which have measurements in both channels). Out of $\sim$11,000 objects
(cluster members as well as background objects included) for which we
have [3.6]$-$[4.5] color light curves, there are at most 15 objects
that seem to have legitimate substantial changes to the [3.6]$-$[4.5]
color (changes of a size that might be consistent with big changes to
a disk), and these are all relatively faint ([3.6]$>$12 mag) objects.
At most, two of those cases have an IR excess that appears to be
possibly transient on these timescales, so at most, $<$0.02\%
frequency of occurrence. For the remaining 13 objects with plausibly
real changes in color, the disk is still clearly present, but the
brightness and color have changed substantially. Individual objects
will be discussed in the cluster papers. 

\clearpage

\subsection{Brightening as Likely as Fading}

\begin{figure}[ht]
\epsscale{0.5}
\plotone{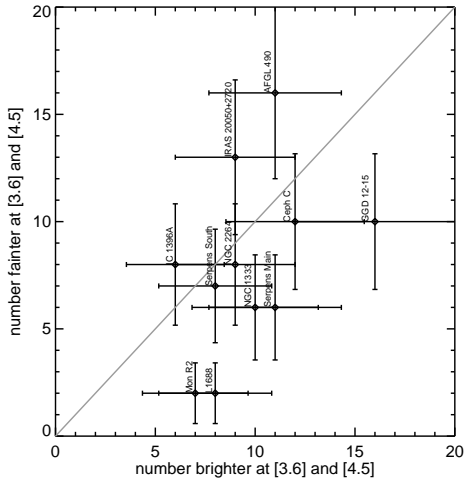}
\caption{Plot of the number of members that become fainter vs.\ the
number that become brighter in both [3.6] and [4.5] between the cryo
epoch and the YSOVAR epoch. Error bars are approximated by Poisson
statistics. The Orion point is to the far upper right, at (156, 153),
with errors of $\sim$12 in each direction. The grey line is the unity
relation. There are similar numbers of objects that become brighter as
become fainter; see text.}
\label{fig:brighterfainter}
\end{figure}

In the literature (e.g., Giannini \etal\ 2009,  Antoniucci \etal\
2014), constraints have been placed on the timescales for brightening
or fading by comparing how many sources are found to be getting
brighter or getting fainter (e.g., for each source, given the two
epochs, for how many cases is the second epoch brighter than the
first, and for how many cases is the first epoch brighter than the
second).  If there are random fluctuations in brightness, the same
number of sources should get brighter as get fainter. If, instead, 
there are more fading sources than brightening, then the type of
variability may be characterized by a short rise and a long fall.

We can make a similar comparison among our long-term variable sample.
To reduce scatter from noise, we consider just the standard set
of members (Sec.~\ref{sec:standardsetofmembers}), and consider just
those objects tagged as long-term variables independently in both
[3.6] and [4.5] (Sec.~\ref{sec:ctopc}).
Figure~\ref{fig:brighterfainter} compares the numbers of these
remaining sources that become brighter at both [3.6] and [4.5] with those
that become fainter at both [3.6] and [4.5] for 11 of the clusters;
errors are approximated by simple Poisson counting statistics.  A
significantly larger number of variables are found in Orion, with
essentially equal numbers of sources getting brighter as getting
fainter.  For the smaller-field clusters, the numbers of sources that
brighten is less consistently equal to the number of sources that
fade, but there are also far fewer sources to count. Summing up the 11
smaller-field clusters, there are 107 ($\pm$10) sources that brighten
and 88 ($\pm$9) sources that fade; these numbers are within 2$\sigma$
of each other, but both numbers have to be extended $\sim$1$\sigma$
towards each other.  Fitting a line to the points in
Fig.~\ref{fig:brighterfainter}, including Orion, results in a line of
slope 1.05$\pm$0.12, and an intercept of $-$2.8$\pm$3.3. This
is consistent with no difference between the brightening and fading
sources, but with only a $\sim$1$\sigma$ possibility that there are
slightly more brightening sources. With or without Orion, there are
very similar numbers brightening and fading.

In the case of several other papers in the literature, they were
looking for much more significant variability than we are finding
(e.g., FUors and EXors). They found internally consistent patterns of
sharp rises and long falls in brightness. For our sample, we are
apparently finding more varieties of variability, even on the
long-term, such that the timescales average out over 6-7 years to have
no significant bias towards brightenings or fadings.

\subsection{Long-Term Variability Fractions}

\begin{figure}[ht]
\epsscale{0.7}
\plotone{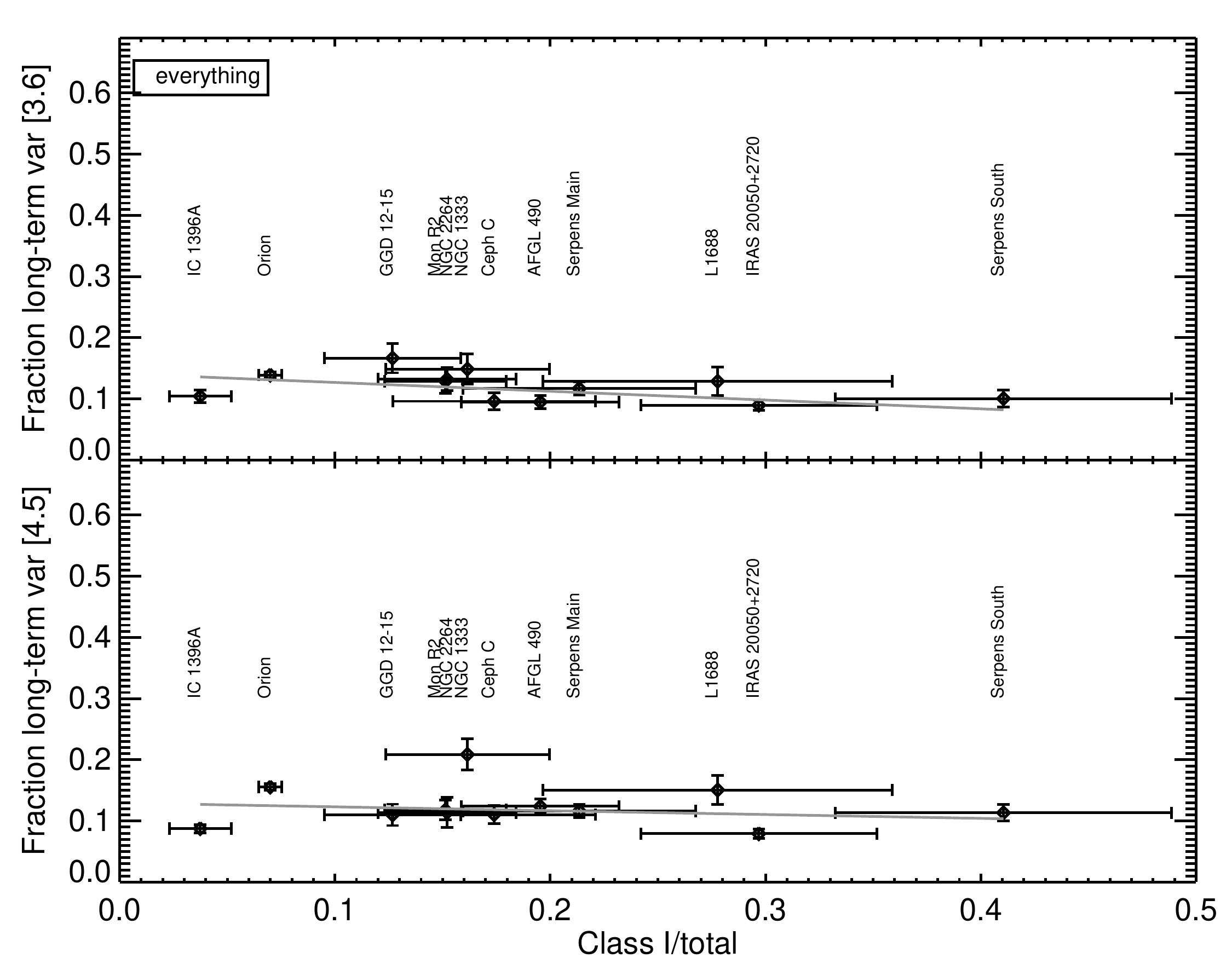}
\caption{Fraction of long-term variables from the standard set for
statistics, as a function of the ratio of Class I/total objects, for
{\em all objects} in the field, for [3.6] (top) and [4.5] (bottom),
for each cluster, as indicated. The range in $y$-axis is set to match
the range needed for the next Figure. Error bars are calculated
assuming Poisson counting statistics; Orion has by far the most
sources, and so has by far the smallest error bars. The grey lines are
the best-fit line, using errors in both directions for each point. The
long-term variability fraction for everything in the field is
relatively constant in this plot -- the slopes of these lines are, for
[3.6], $-$0.14$\pm$0.04, and for [4.5], $-$0.06$\pm$0.06. Correlation
coefficients suggest that there is no significant correlation here
([3.6]: $r=-0.41$; [4.5]: $r=-0.10$).
\label{fig:varfracLT-all}}
\end{figure}

\begin{figure}[ht]
\epsscale{0.7}
\plotone{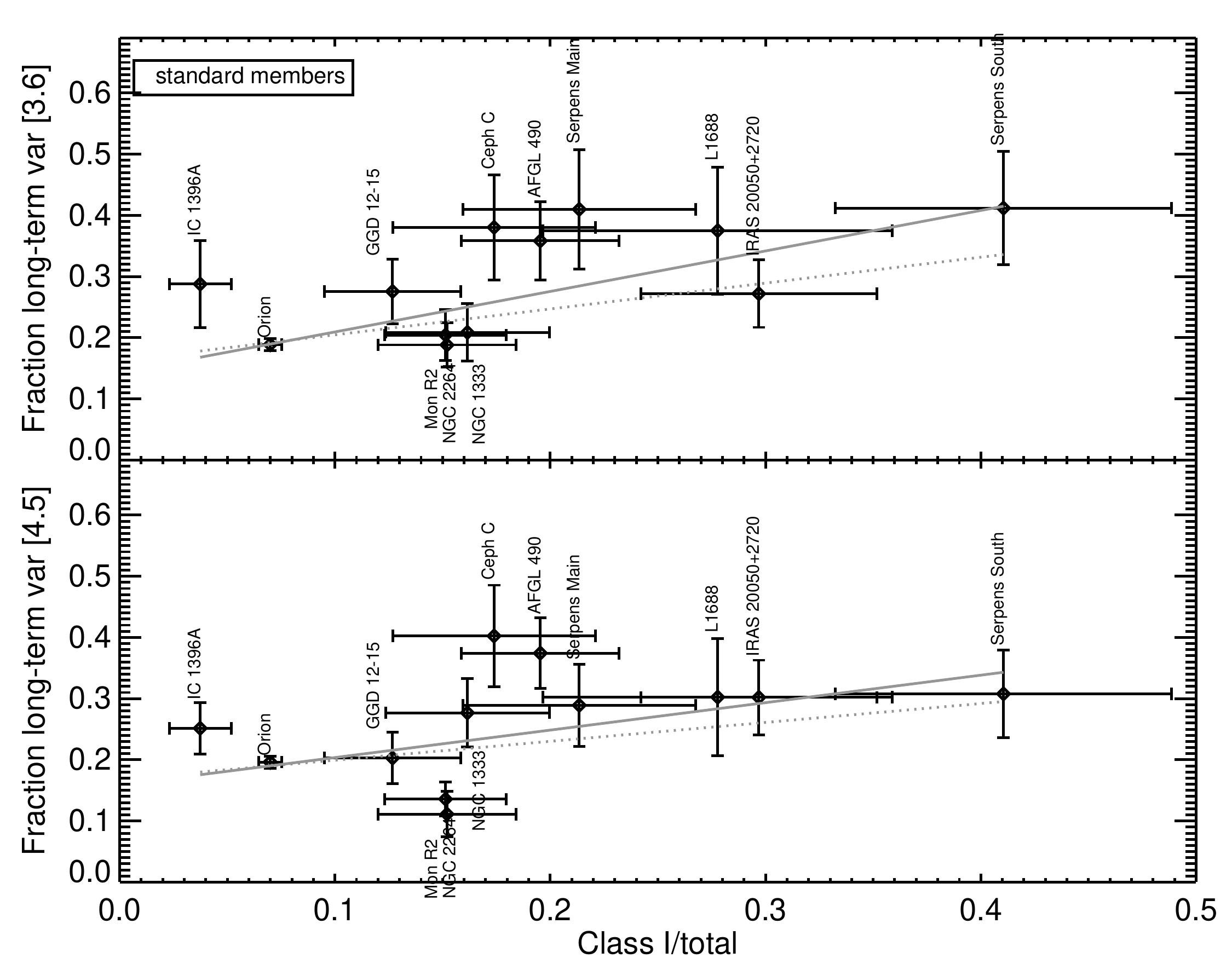}
\caption{Similar to Fig.~\ref{fig:varfracLT-all}, except the $y$-axis
is the fraction of long-term variable {\em members}. The slopes of the
grey best-fit solid lines are, for [3.6], 0.66$\pm$0.19, and for
[4.5], 0.45$\pm$0.18. Correlation coefficients are consistent with a
correlation ([3.6]:$r$=0.58; [4.5]: $r$=0.41).  Since Serpens South
has the largest fraction of embedded objects, we also tested fitting
this relation omitting this point, and this does weaken the
correlation. The dashed lines are these best fit values: for [3.6],
0.42$\pm$0.15, and for [4.5], 0.31$\pm$0.14. (Correlation coefficients
are [3.6]:$r$=0.45; [4.5]: $r$=0.42) The long-term variability
fraction increases significantly as the degree of embeddedness
increases (to the right). 
\label{fig:varfracLT-mem}}
\end{figure}

Having identified the long-term variables above (Sec.~\ref{sec:ctopc})
for each cluster, we can look at the fraction of objects that are
variable on the longest timescales we sample.
Figure~\ref{fig:varfracLT-all} shows the fraction of variables for all
objects in each field (the standard set for statistics), including
both members and likely field objects. The  $x$-axis is our
parameterization of the relative fractions of embedded sources (see
Sec.~\ref{sec:clusterparameterization}), the ratio of Class I/total
number of members. Fig.~\ref{fig:varfracLT-all} thus is somewhat
incongruous in that the standard set of members is used in the
$x$-axis, but the $y$-axis includes everything in the field of view.
About 10-20\% of all objects with light curves are tagged as variable
in the long-term.  However, as can be seen in the Figure,  the
variability fraction of everything in the field is not a strong
function of the Class I/total ratio -- the slopes of the best-fit
lines in Fig.~\ref{fig:varfracLT-all} are both small. (The slopes of
these lines are given in the figure caption.) Calculating Pearson's
correlation coefficient also suggests that there are no significant
correlations shown in Fig.~\ref{fig:varfracLT-all}. We expect that the
fraction of stars that are variable should be a function of Galactic
latitude (see approximate Galactic latitudes listed in
Table~\ref{tab:clusterproperties}), because the fraction of sources
that are background/foreground stars will be higher in the Galactic
plane, so the fraction of cluster members will be higher out of the
Galactic plane, and since any young star (cluster member) is more
likely to be variable, there should be a higher fraction of variable
objects at higher Galactic latitude. Indeed, the Pearson's correlation
coefficient suggests that the fraction of long-term variables for all
objects in each field is strongly correlated with the absolute value
of the Galactic latitude.

Figure~\ref{fig:varfracLT-mem} recasts Fig.~\ref{fig:varfracLT-all} by
using the long-term variability fraction of just the standard set of
members, rather than everything in the field. This time, there is a
sigificant correlation; the higher the fraction of embedded members,
the higher the fraction of long-term variables. If a cluster has 
more sources that are embedded and likely to be actively accreting and
interacting with their circumstellar material, it also has more
sources that are variable on timescales of years.  The slopes of the
best-fit lines shown in Fig.~\ref{fig:varfracLT-mem} are given in the
figure caption. Pearson's correlation coefficient suggests that there
is a correlation here, and it is stronger in I1 than I2 (consistent
with the calculated slopes).

Serpens South has the highest fraction of the most embedded sources;
to test if it is providing a significant ``lever arm'' on this fit, we
fit the remaining points omitting Serpens South, which does indeed
weaken the correlation, as can be seen in the figure.

We tested the correlations seen in Figs.~\ref{fig:varfracLT-all} and
\ref{fig:varfracLT-mem} using a variety of other SED class-based
parameterizations, and we found similar results -- there is no
significant correlation of the long-term variable fraction of all
sources in the field with any SED class-based parameterization, and
there is a correlation of the long-term variable fraction of the
members with any parameterization chosen that uses fractions of
various SED classes (or groups of classes). This correlation between
the fraction of members that are variable on these longest timescales
and the parameterization of `embeddedness' seems robust. We expected
that young stars were more likely to be variable than field stars. We
have found moreover that within the category of young stars, for a
higher fraction of embedded sources (a higher fraction of presumably
younger sources), we find a higher fraction of long term variables.
This is consistent with what has been found in individual clusters
(e.g., NGC 2264 -- Cody \etal\ 2014 -- and L1688 -- G\"unther \etal\
2014).

\begin{figure}[ht]
\epsscale{0.7}
\plotone{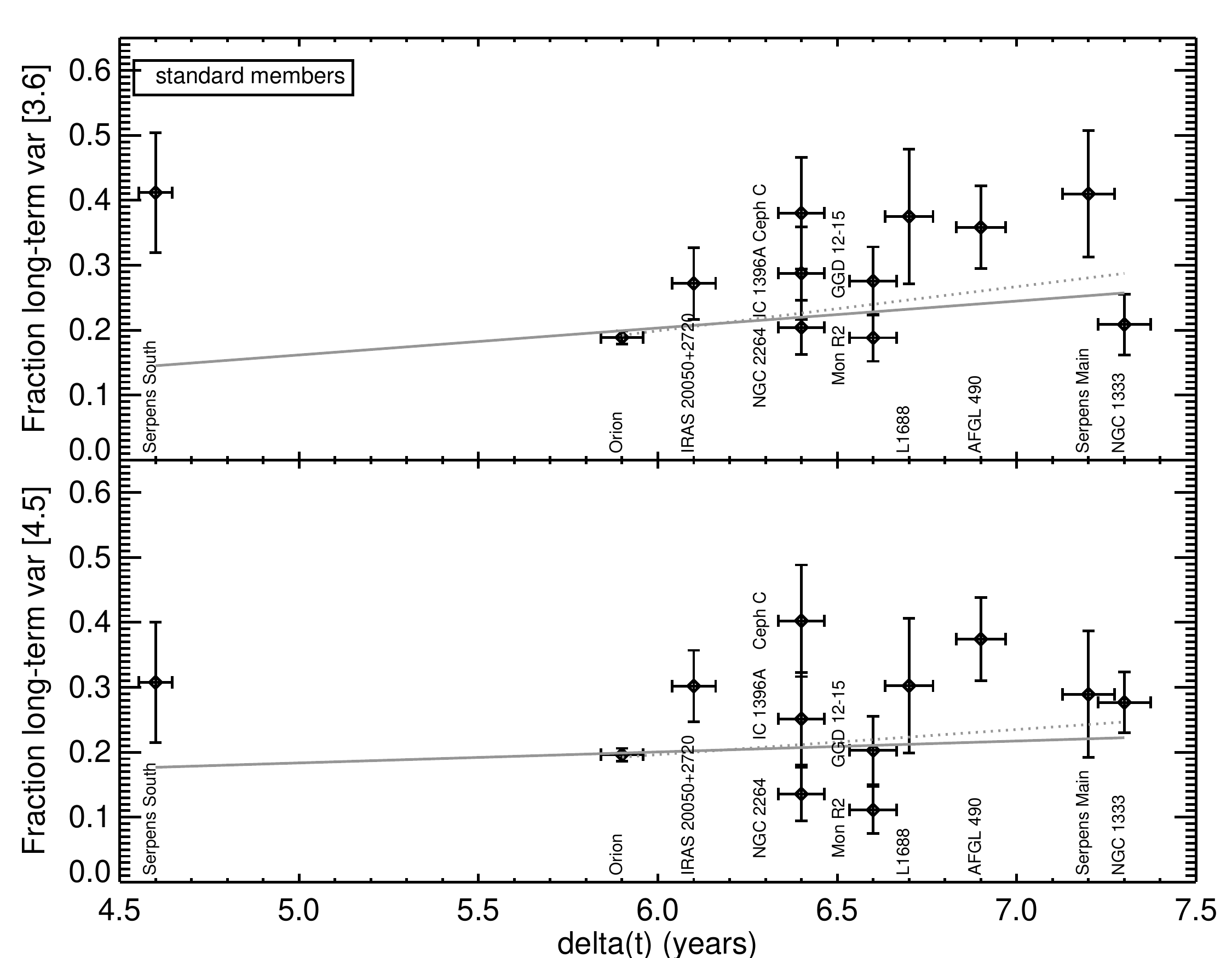}
\caption{The long-term variable fraction for the standard set of
members as a function of the time difference in years between the
cryo-epoch and YSOVAR fast-cadence epoch. The slopes of the grey
best-fit lines are, for [3.6], 0.04$\pm$0.02 ($r=-0.16$), and for
[4.5], 0.02$\pm$0.02 ($r=-0.0004$).  Dropping Serpens South as an
outlier, the slopes of the grey dashed best-fit lines are
$0.07\pm0.02$ and $0.04\pm0.02$, respectively. Correlation
coefficients for these options are consistent with there being no
significant correlation in either case ([3.6]: $r=0.35$;
[4.5]:$r=0.23$).  All of this evidence is consistent with no
significant effect of the timescale on the fraction of long-term
variables.  }
\label{fig:varfracLT-deltat}
\end{figure}

The difference between fractions of embedded objects among these
clusters is not the only potentially significant factor for this set
of observations. Here, we sample timescales of $\sim$4.5 to $\sim$7.5
years. If the amplitude of variability of the members increases as the
time baseline increases, then we expect more members to be selected as
variable in the long-term, and thus a higher long-term variability
fraction as the time baseline increases. However,
Figure~\ref{fig:varfracLT-deltat} shows this relationship between the
variability fraction and the time between epochs of observation (from
Table~\ref{tab:cryoprograms}). The best-fit lines and correlation
coefficients are consistent with no significant effect on the
variability fraction as a function of timescale sampled. Serpens
South, because it was observed (indeed, discovered; Gutermuth \etal\
2008b) comparatively late in the Spitzer mission, samples the shortest
timescales. If the Serpens South point is omitted from the fit, the
slopes become slightly steeper, but there is still no significant
correlation from the correlation coefficients. 

Scholz (2012), working in $K$-band, finds that the longer one monitors
a cluster, the larger the amplitude of variability, on timescales of
years. That work specifically investigated the amplitude of the change
in magnitude, for just one cluster at a time, and looked at the change
of the range of that distribution of amplitudes as a function of time
step. For the times we sampled on these longest timescales, analogous
plots do not show a signficant change in the median or the top
quartile or the 90th percentile. We tried using the most stringent set
of objects (just those in the standard set of members, standard set
for statistics, that had light curves in both IRAC channels); we still
did not find this effect. For this analysis, Scholz (2012) was working
in just $\rho$ Oph, over a larger area than we were, using slightly
different wavelengths, and a larger range of timescales, the maximum
of which ($\sim$2000 d) is comparable to the minimum $\Delta t$
($\sim$4.5 y$\sim$1600 d) we consider here. These things could account
for the observed differences.

\section{Conclusions}
\label{sec:conclusions}

We present in this paper the data collection and reduction for the
YSOVAR (Young Stellar Object VARiability) programs, representing
nearly 800 hours of Spitzer time studying the variability of young
stars in 12 different clusters (AFGL 490, NGC 1333, Orion, Mon R2, GGD
12-15, NGC 2264, L1688, Serpens Main, Serpens South, IRAS 20050+2720,
IC 1396A, and Ceph C).  We also describe the assembly of broad
collections of ancillary data for these clusters.  There are
$\sim$29,000 unique objects of any sort matched to 39,000 [3.6] or
[4.5] light curves in the YSOVAR data set. 

The goals of the broader YSOVAR program include the following: to
obtain the first extensive mid-infrared time-series photometry of
young stars to help reveal the structure of the inner disk region of
YSOs, provide new constraints on accretion and extinction variability,
assess timescales of mid-IR variability from seconds to years,
identify new young eclipsing binaries, help identify new very low mass
substellar members of the surveyed clusters, constrain the short and
long-term stability of hot spots on the surfaces of YSOs, and
determine rotational periods for objects too embedded for such
monitoring in the optical.

In this paper, we set the stage for several planned papers. We
establish here not only the data reduction approach, but also define
the standard sample on which we calculate statistics (the fast cadence
data, where there are at least 5 points per light curve), and a
standard sample of members (the union of all IR-selected members and
X-ray selected members), with a provision for adding members
identified in other ways (literature, or variability itself).

We use three mechanisms to identify variables in the standard set for
statistics (fast cadence data) -- the Stetson index (calculated using
both IRAC channels), the $\chi^2$ test (calculated for each channel
individually), and searching for significant periodicity (working on
light curves with at least 20 points, using primarily the LS approach,
independently on [3.6], [4.5], and [3.6]$-$[4.5]).  Based on
simulations, for these YSOVAR data, we find that we are sensitive to a
broad range of timescales and amplitudes.  If the star is found in one
of two states half the time, the Stetson test will identify it as
variable in almost all cases (99\%), if the step size between states
is at least three times larger than the noise level. If the star
spends less than 20\% of the time in either state, then there is a
resonable chance that the variability will not be found, even for
larger step sizes between states, since the sampling might catch only
few datapoints in this state. In general, we can detect periodicity
with an amplitude just twice the level of the noise, variability at
3-5 times the level of the noise with the Stetson test and 6-10 times
the noise for single-band lightcurves with the $\chi^2$ test.

We also identified variables on the longest timescales possible using
our dataset, timescales of 6-7 years, using a fourth method of
identifying variability. By comparing measurements taken early in the
Spitzer mission with the mean from our YSOVAR campaign (the standard
set for statistics), we can identify those objects that have changed
significantly between then and now.  We show that the overall fraction
of everything in each field that varies on these longest timescales is
independent of the ratio of Class I/total members in each cluster.
However, the fraction of members in each cluster that are variable on
these longest timescales is a function of the ratio of Class I/total
members in each cluster, such that clusters that have a higher
fraction of Class I objects also have a higher fraction of long-term
variables. We find no dependence of the fraction of members in each
cluster that are variable on these longest timescales with the time
step between the observations (between the cryogenic and
post-cryogenic observations). Among the most reliable of the long-term
variables, we find no strong preference for brightening or fading over
these timescales.

We find periods from our data in $\sim$1100 objects, $\sim$800 of
which are in Orion alone. About 650 of those have data in both 3.6 and
8 \mum\ ($\sim$430 of which are in Orion), enabling us to compare
[3.6]$-$[8] vs.\ log($P$) in a fashion similar to that found
previously in Orion (Rebull \etal\ 2006) and NGC 2264 (Cieza and
Baliber 2007).  Very similar results are obtained -- excesses do not
necessarily imply longer periods, but a star with a longer period is
more likely than those with shorter periods to have an IR excess. 
This is somewhat surprising in that (a) the periods are determined
from the IR, not the optical, as was done previously; and (b) the
clusters besides Orion are thought to be substantially younger than
Orion, suggesting that disk locking may be in effect at even these
young ages.  However, it is likely that we can obtain viable periods
more easily for relatively unobscured stars, e.g., the older end of
the young star distributions in these clusters. Little is known about
many of the objects outside of Orion shown here; additional study of
the individual objects will help clarify matters.

There have been recent reports of debris disks changing on timescales
of years; both Meng \etal\ (2012) and Melis \etal\ (2012) report on
systems that change significantly at wavelengths $>$10 \mum\ over
timescales of years.  Our objects here are younger and our monitoring
wavelengths are considerably shorter. Out of $\sim$11,000 objects
(cluster members as well as foreground/background objects included)
for which there is an essentially simultaneous YSOVAR measurement in
both [3.6] and [4.5], there are at most 15 objects that seem to have
legitimate substantial changes to the [3.6]$-$[4.5] color (changes of
a size that might be consistent with a disk appearing/disappearing),
and these are for the most part relatively faint ([3.6]$>$12 mag)
objects. At most, two of those cases have an IR excess that appears to
be possibly transient on these timescales, so at most, $<$0.02\%
frequency of occurrence. For the remaining 13 objects with plausibly
real changes in color, the disk is still clearly present, but the
brightness and color have changed substantially.

Details of individual objects of interest in each of the clusters will
appear in the forthcoming YSOVAR papers.

\acknowledgments

We are grateful to the SSC scheduling staff for accommodating this
complex cadence over the two years of the survey program. We thank
L.~Hartmann and M.~Skrutskie for early help with the program and for
helpful suggestions on the manuscript; we also thank N.~Calvet,
D.~Ciardi, J.~Holtzmann, J.~Muzerolle,  G.~Rieke, K.~Stapelfeldt,
H.~Sung, M.~Werner, B.~Whitney, E.~Winston, \& K.~Wood, for early help
with the program. We thank T.~Mazeh, L.~Tal-Or, and the Wise
Observatory for their help in preliminary observations of Orion YSOs. 

RG gratefully acknowledges funding support from NASA ADAP grants
NNX11AD14G and NNX13AF08G and Caltech/JPL awards 1373081, 1424329, and
1440160 in support of Spitzer Space Telescope observing programs.
H. Bouy is funded by the Ram\'on y Cajal fellowship program number
RYC-2009-04497.  This research has been funded by Spanish grants
AYA2012-38897-C02-01, AYA2010-21161-C02-02, CDS2006-00070 and
PRICIT-S2009/ESP-1496. 

This work is based in part on observations made with the Spitzer Space
Telescope, which is operated by the Jet Propulsion Laboratory,
California Institute of Technology under a contract with NASA. Support
for this work was provided by NASA through an award issued by
JPL/Caltech. The research described in this paper was partially
carried out at the Jet Propulsion Laboratory, California Institute of
Technology, under contract with the National Aeronautics and Space
Administration.  The scientific results reported in this article are
based in part on data obtained from the Chandra Data Archive
including observations made by the Chandra X-ray Observatory and
published previously in cited articles. 

This research has made use of NASA's Astrophysics Data System (ADS)
Abstract Service, and of the SIMBAD database, operated at CDS,
Strasbourg, France.  This research has made use of data products from
the Two Micron All-Sky Survey (2MASS), which is a joint project of the
University of Massachusetts and the Infrared Processing and Analysis
Center, funded by the National Aeronautics and Space Administration
and the National Science Foundation. The 2MASS data are served by the
NASA/IPAC Infrared Science Archive, which is operated by the Jet
Propulsion Laboratory, California Institute of Technology, under
contract with the National Aeronautics and Space Administration. 
This research has made use of the NASA Exoplanet Archive, which is
operated by the California Institute of Technology, under contract
with the National Aeronautics and Space Administration under the
Exoplanet Exploration Program

\appendix

\section{Naming Convention}

In our initial data release (MC11), we used a naming convention
following the IAU naming standards using an acronym (Initial Spitzer
Orion YSOVAR: ISOY) followed by the J2000 coordinates. 

For the final version of our YSOVAR catalog, discussed here, the
IAU-registered acroynm is SSTYSV, for Spitzer Space Telescope, Young
Stellar object Variability. We again follow it with the J2000
coordinates. Individual objects in this catalog need not be confirmed
young stars, but simply have a light curve in this data set. Detailed
data tables of cluster members for each cluster will be presented in
the individual cluster papers, and it is our intention to deliver a
final catalog of every object with a light curve to IRSA for general
distribution.

\section{SED classes}
\label{sec:sedsection}

\subsection{Background}
\label{sec:sedbackground}

In the context of understanding the evolution of young stars from a
very embedded state to a less embedded state, the community has chosen
to parameterize objects based on the slope of the SED (see, e.g.,
Wilking \etal\ 2001).  This classification is tied to the empirical
shape of the SED. Very embedded (presumably very young) objects will
have SEDs that peak at long wavelengths; as the object sheds its natal
cocoon, the peak of the SED moves to shorter wavelengths. Objects with
substantial circumstellar disks will emit more energy in the
IR (and longer wavelengths) than in the optical. Objects with less
substantial disks will have most of their energy emitted in
wavelengths shorter than the NIR, but there will be additional energy
contributions in the IR (and longer wavelengths) from the
circumstellar dust and/or debris, i.e., the SED has an IR excess. The
most embedded phase is referred to as Class 0, then proceeding (based
on SED shape) through Class I (rising SED), Flat (flat SED), Class II
(falling SED with an IR excess), and finally Class III SEDs, which are
objects with photospheric or near-photospheric SEDs, with little or no
IR excess, typically identified as young via other means such as
X-rays or H$\alpha$ emission. Sometimes additional classes such as
transition disks are added near the end of this sequence. Classical T
Tauri stars (CTTS) are often identified with Class IIs, and weak-lined
T Tauri stars (WTTS) are often identified with Class IIIs.
Nomenclature is difficult and inconsistent across the literature; see
Evans \etal\ (2009b) for a discussion of terminology.

We need to establish at least an internally consistent definition of
the SED classes such that we can investigate trends as a function of
SED class as a proxy for age. The reliability of the translation
between SED class and age has been discussed at length in the
literature (and will continue to be discussed in the future); other
factors such as inclination and multiplicity may play a large role. In
the context of our work, we wish to establish an internally consistent
approach that can be calculated for all sources in the YSOVAR fields.

\subsection{Definition}
\label{sec:sedslopes}

In order to assemble our SEDs for each object, we include all the data
described in Sec.~\ref{sec:obsanddatared} between $U$ and 25 \mum. For
the SED classes, only the IR bands are relevant. 

In order to define an internally consistent placement of the YSOVAR
objects into SED classes, in the spirit of Wilking \etal\ (2001), we
define the near- to mid-IR (2 to 24 \mum) slope of the SED, $\alpha =
d \log \lambda F_{\lambda}/d \log  \lambda$,  where  $\alpha > 0.3$
for a Class I, 0.3 to $-$0.3 for a flat-spectrum  source, $-$0.3 to
$-$1.6 for a Class II, and $<-$1.6 for a Class III.  For each of the
objects in our sample, we performed a simple  least squares linear fit
to {\bf all available photometry} (just detections, not including
upper or lower limits, but including archival and literature data)
{\bf as observed between 2 and 24 $\mu$m, inclusive}.  We included the
mean obtained from the standard set for statistics of the light curve
in the SED, in addition to the cryo-era Spitzer points. (This means
that there could be more than one point contributing to the fit at 3.6
\mum, one from the cryo era and one from the mean of the YSOVAR light
curves.) Formal errors (either from individual single-epoch
measurements or the mean and standard deviation from the IRAC light
curves) on the infrared points are so small as to not affect the
fitted SED slope. However, if the mean is calculated over more of the
light curve than our standard set for statistics (fast cadence)
sample, the mean may be different enough, in the cases of sparse SEDs,
that the class may change to an adjacent class; these cases will be
noted in the cluster papers where relevant. The linear fit is
performed on the {\bf observed} SED, e.g., no reddening corrections
are applied to the observed photometry before fitting.  

In the literature, the precise definition of $\alpha$ can vary, or the
distribution of slopes and classes can vary (e.g., Sung \etal\ 2009
lists Class I, II, II/III, pre-transition disk, transition disk
categories for NGC 2264), which may result in different
classifications for certain objects. Classification via our method is
provided specifically to enable comparison within this paper (and to
other YSOVAR papers) via internally consistent means. Our
classification, since it is based on observed SED, is possible for all
objects (not just those identified as YSO candidates). The formal
classification definition puts no lower limit on the colors of Class
III objects (thereby including those with SEDs resembling bare stellar
photospheres, and allowing for other criteria such as X-ray brightness
to define youth).  The SED slopes and classes for all of the sources
of interest will appear in the individual cluster papers. Histograms
of the relative fractions of each class for the standard set of
members for each cluster appear in Figure~\ref{fig:classratios}.

\subsection{Including or ignoring the 24 \mum\ point}

In terms of aggregate statistics over all data from all 12 clusters
(members and non-members together), we can constrain the fraction of
objects that change class depending on whether or not the 24 \mum\
point is included in the fit. There are about 21,000 objects with
light curves over all 12 clusters for which an SED slope can be fit
between 2 and 8 \mum. Out of those $\sim$21,000, there are only about
1760 with MIPS-24 detections (not limits), so only about 8\% of the
sources are affected. Admittedly, those sources that are detectable at
24 \mum\ are the ones with rising SEDs and thus are statistically more
likely to be true cluster members than a source selected at random
from the map. Out of the $\sim$1760 with MIPS-24 detections,
$\sim$72\% of them do not change class when the 24 \mum\ point is
included in the SED fit. The class bins of slopes are relatively large
and thus relatively insensitive even to a point at the far red end of
the SED. Of the $\sim$28\% that do change class, $\sim$85\% move only
one step, to an adjacent class. As expected, there is a bias, when
including the 24 \mum\ point, to move the objects to a more positive
slope, e.g., towards the more embedded end of the sequence; of the
ones that change class at all, $\sim$61\% move one step earlier
(towards more embedded, not necessarily in age) in the sequence,
$\sim$24\% of those move one step later (towards less embedded, not
necessarily in age) in the sequence.

We conclude that it does not make a significant difference for the
overwhelming majority of the sources if one uses the slope between 2
and 8 \mum\ or the slope between 2 and 24 \mum. (Note that this is
fitting all available points between thse values, not a simple
comparison of the two end points.) For any sources of interest in
which the class might change depending on the inclusion of the 24
\mum\ point, they will be noted in the individual cluster papers.

\subsection{Comparison to G09}

G09 provides placement into SED classes as part of the data presented
there, and the same algorithm has been applied to our entire
cryogenic-era catalog (Sec.~\ref{sec:cryodatareduction}). These
classes are identified based on {\em dereddened} colors, and are {\em
only} provided for objects thought to be young stars. Objects that are
not thought to be young stars are identified as other things such as
``PAH emission source''; objects missing bands such that the
classification scheme cannot be run are also not classified.  The SED
classes we are using here are based on fits to the observed SED, not
the dereddened SED, and are obtained for any source, regardless of its
true underlying nature.  To constrain the degree to which we might be
introducing a bias by fitting the observed SED, we can compare, for
some objects, the class obtained by G09 and by our mechanism above.

It is important to note that the classes provided by G09 are different
than the classes we use here. G09 has Class 0, I, II, and II/III, but
no Flat class. Here, we do not have Class 0; we have Class I, Flat,
Class II, and Class III. Over all 12 clusters, there are $\sim$3500
objects for which both classifications are available; 75\% of those do
not change class, even with the different bins that are defined. Out
of the $\sim$3500, about 14\% are identified as being from a later
(less embedded) class than G09, and about 11\% are identified as being
from an earlier (more embedded) class than G09. 

G09 also dereddens the SEDs before placing them in classes; we are
fitting observed SEDs. G09 does this to avoid a reddening bias towards
youth -- as discussed in Muench \etal\ (2007), the $K_s-$[3.6] color
can be affected by \av$\sim$40, though it takes \av$\gtrsim$200 to
affect the [5.8]$-$[24] color. However, in our case, because we are
fitting all available points between $K_s$ and [24], not just
subtracting two points in the SED, the influence of reddening on this
overall slope in the best case (where there are $K_s$, four bands of
IRAC, and one MIPS band), simulations suggest that we would need
$A_J\sim11$ or $A_V\sim40$ before a Class III object would be
misclassified as a Class II. Admittedly, this is a best case scenario,
where the SED is well-populated. For a source without MIPS 24 but just
$K_s$ through [8], we find that we need $A_J\sim6$ or $A_V\sim21$
before a Class III object would be misclassified as a Class II.

Our objects discussed in YSOVAR have to be bright enough to get
good-quality light curves at 3.6 and 4.5 \mum, which is effectively
brighter than 16th mag (see Figs.~\ref{fig:jdistpart1} and
\ref{fig:jdistpart2}), which means that our sources cannot generally
have extremely high extinction. Out of the objects for which we have a
Gutermuth-derived value for $A_K$ and a lightcurve, $\sim$6\% have
$A_K\gtrsim2$ (which corresponds to $A_J\gtrsim6$), and $\sim$1\% have
$A_K\gtrsim3.5$ (which corresponds to $A_J\gtrsim11$). We conclude
that any bias towards young objects is likely not substantial in our
data set.

We have opted to use most often our SED class definition as described
above for internal comparison and consistency. However, when
discussing the Class II/Class I ratio reported in G09, we are using
those classes from G09. In Sections~\ref{sec:clusterproperties} and
\ref{sec:parameterization}, we discuss the Class II/Class I ratio
derived in a very similar fashion as G09, using the G09 classes, but
only for those objects with light curves in the standard set for
statistics. This enables at least some comparison back to G09 and
related works.

\section{Observation Footprints}
\label{sec:footprintfigs}

In support of the discussion in Section~\ref{sec:footprints}, this
Appendix includes the approximate sky coverage for a summed-up image
consisting of all epochs of the YSOVAR observations for each of the
clusters (except for AFGL 490, included in the main body of the text).
In each case, footprint outlines are superimposed on a reverse
greyscale image of the cluster at 4.5 \mum\ obtained during the
cryogenic mission. The thicker blue solid line is 3.6 \mum\ and
the thicker red dashed line is 4.5 \mum. If there is substantial
field rotation during the YSOVAR campaigns, a single epoch of
observation is also indicated by thinner blue solid and red dashed
lines, with the difference between the single epoch and the larger
polygon due to (ecliptic latitude dependent) field rotation effects;
see Sec.~\ref{sec:footprints}. North is up and East is to the left in
each case. The distance between the farthest north and farthest south
coverage here is noted in each caption. The relevant Chandra coverage
in each cluster is shown as a yellow polygon.

\begin{figure}[ht]
\epsscale{0.5}
\plotone{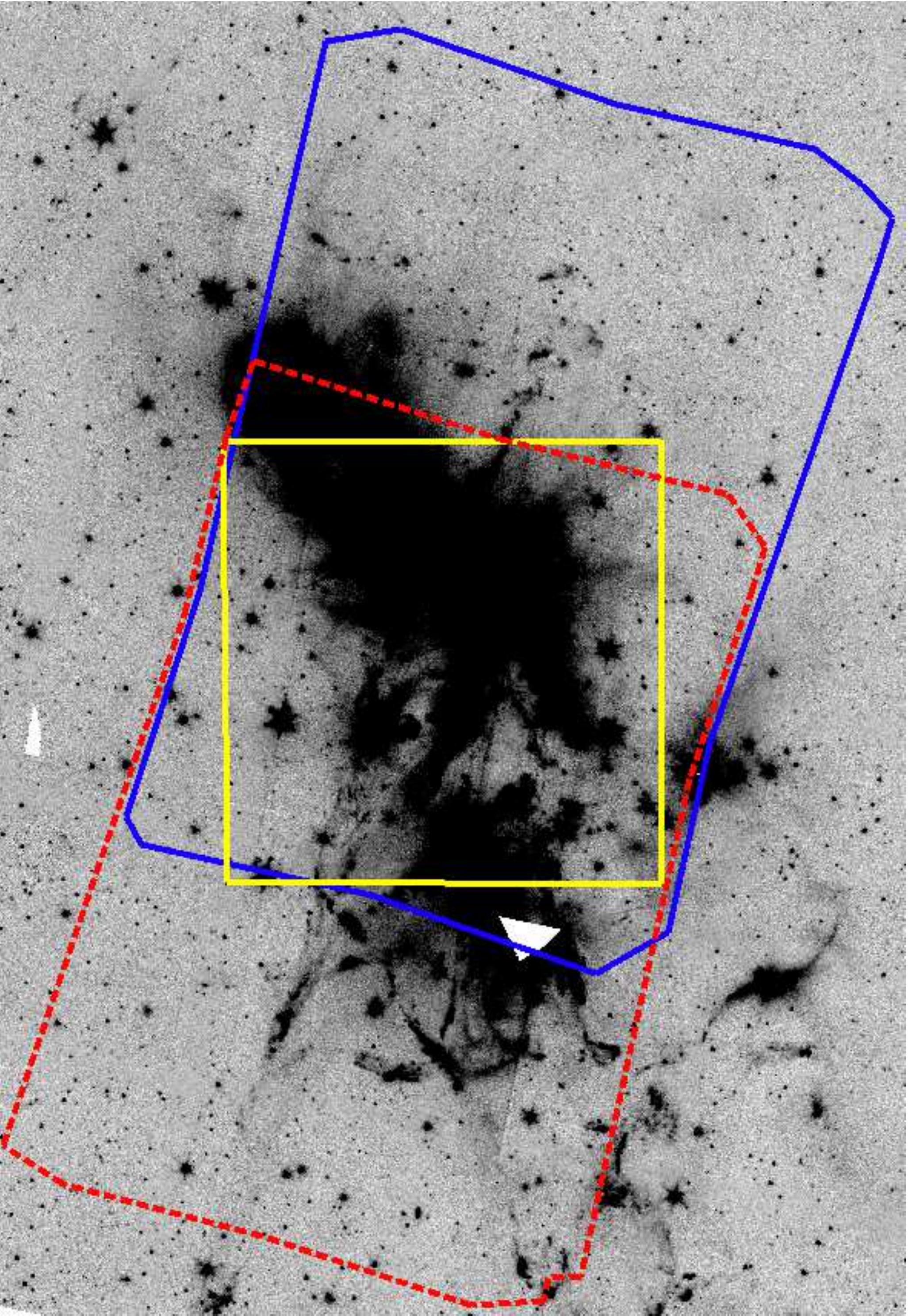}
\caption{As for Fig.~\ref{fig:afglfootprints}, but for our NGC 1333
observations, with the north-south boundary extremes separated by
$\sim$24$\arcmin$. The field rotation here is essentially zero. The
yellow polygon indicates the approximate region covered by Chandra
observations. }
\label{fig:n1333footprints}
\end{figure}

\begin{figure}[ht]
\epsscale{0.4}
\plotone{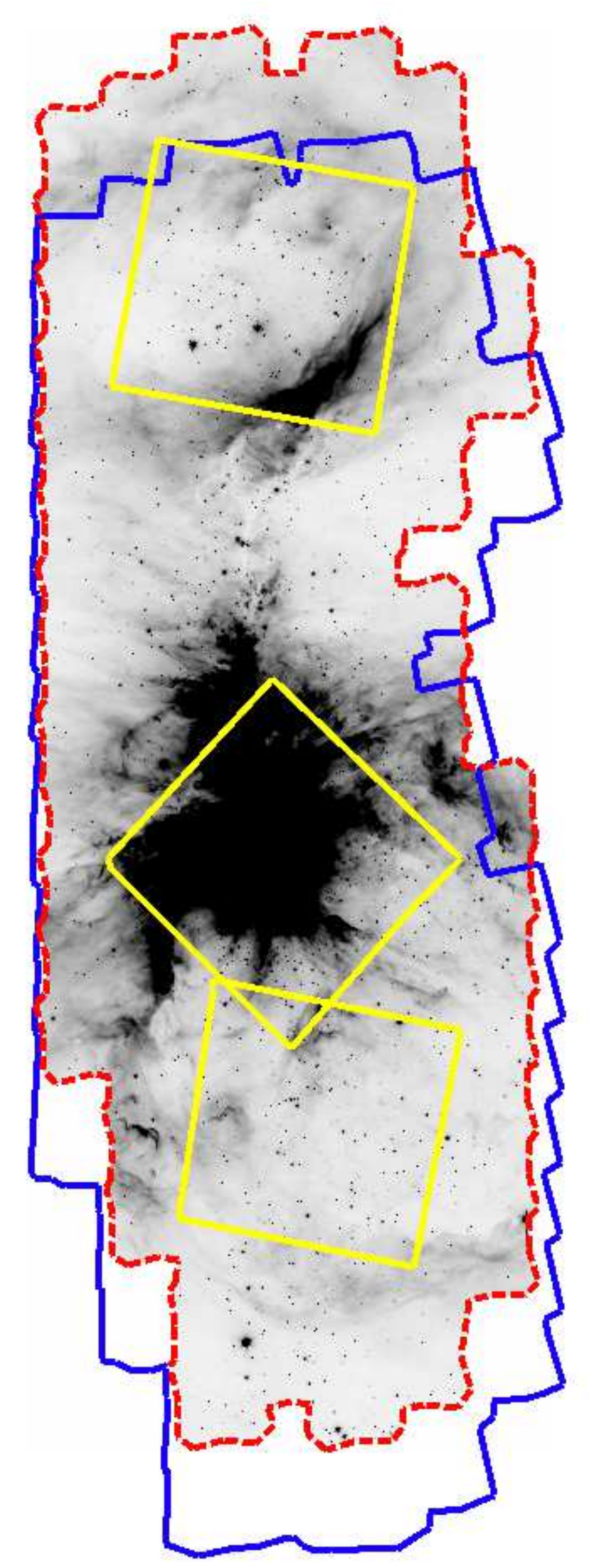}
\caption{As for Fig.~\ref{fig:n1333footprints}, but for our Orion
observations, with the north-south boundary extremes separated by
$\sim$1.6$\arcdeg$. The central Chandra (yellow) polygon is the
approximate footprint from the deep Orion Nebula Cluster observation
(Getman \etal\ 2005); the northern and southern Chandra pointings are
much shallower, and are taken from Ramirez \etal\ (2004b). The
background image here is an IRAC-2 image from the YSOVAR campaigns (as
opposed to a cryo-era image, as it is for most of the other figures
like this).}
\label{fig:orionfootprints}
\end{figure}

\begin{figure}[ht]
\epsscale{0.5}
\plotone{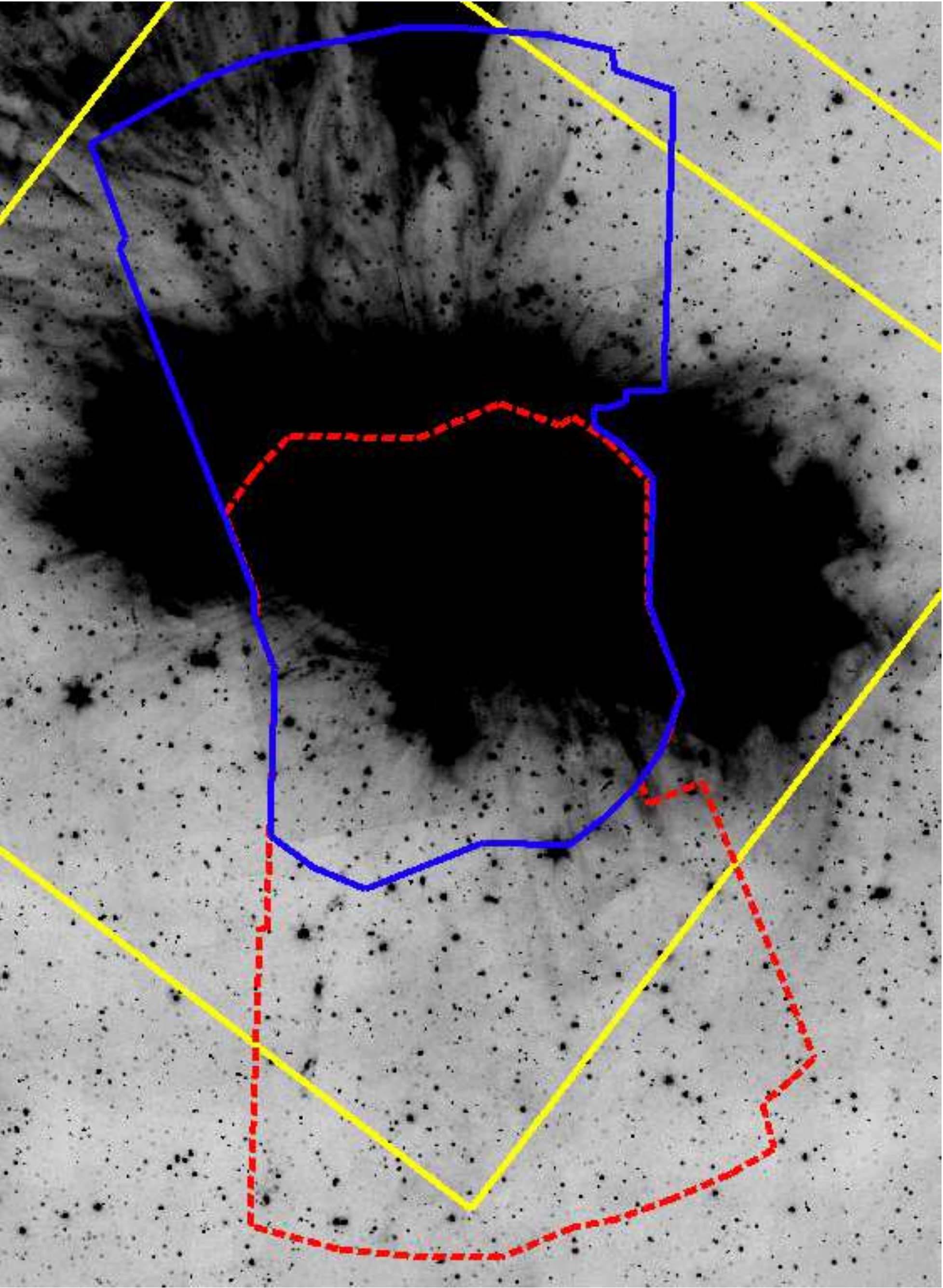}
\caption{As for Fig.~\ref{fig:n1333footprints}, but for our Mon R2
observations, with the north-south boundary extremes separated
by $\sim$21$\arcmin$.  Field rotation, while present, is not as
substantial for this field as it is for others of our clusters. There
is a small amount of additional X-ray data (not relevant for this
project) to the upper right, and a portion of that footprint can be
seen.}
\label{fig:monr2footprints}
\end{figure}

\begin{figure}[ht]
\epsscale{0.5}
\plotone{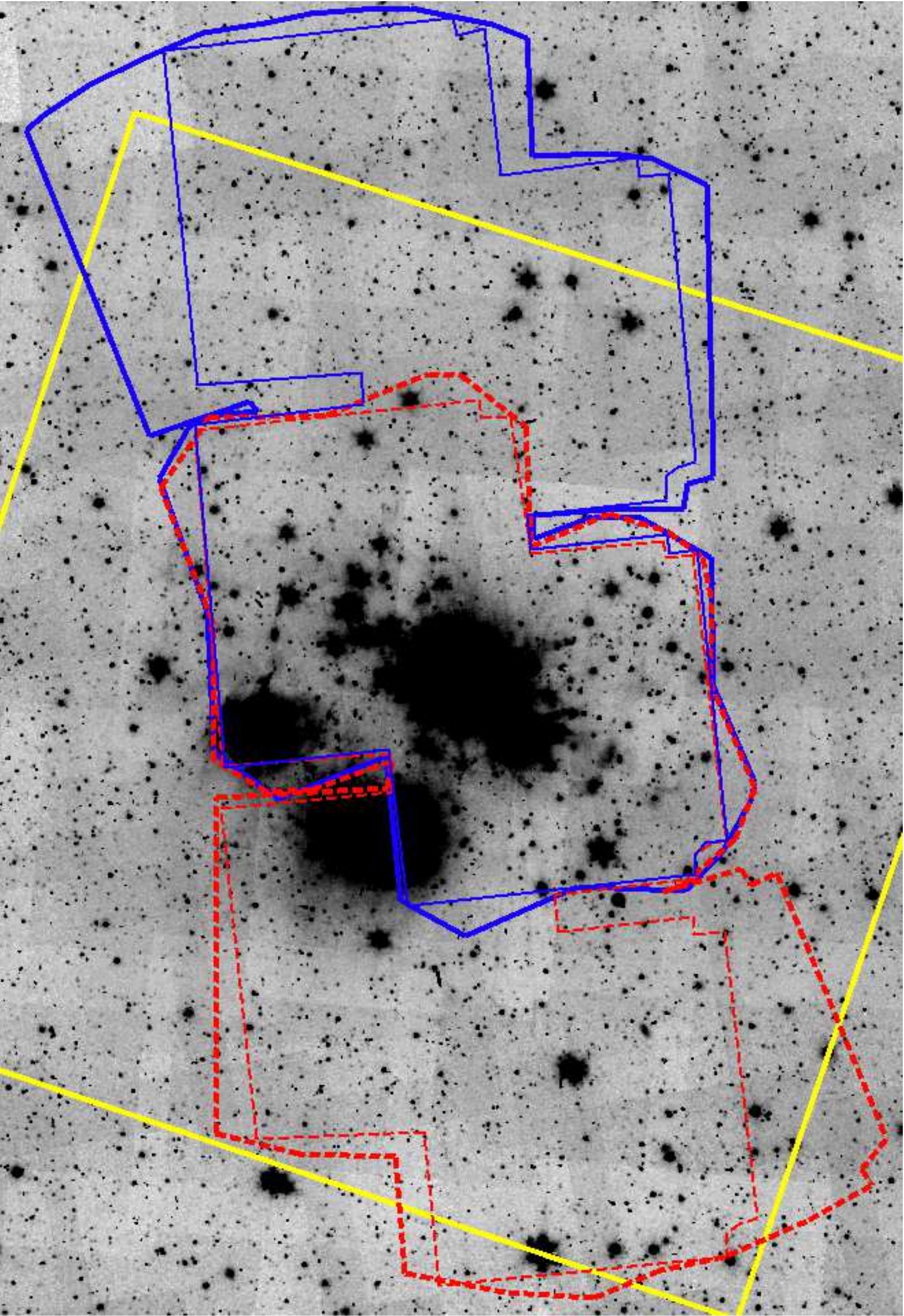}
\caption{As for Fig.~\ref{fig:n1333footprints}, but for our GGD 12-15
observations, with the north-south boundary extremes separated
by $\sim$23$\arcmin$.  A single epoch of observation is also indicated
by thinner blue solid (3.6 \mum) and red dashed (4.5 \mum) lines to
give an indication of the magnitude of the field rotation for this
field.}
\label{fig:ggdfootprints}
\end{figure}

\begin{figure}[ht]
\epsscale{0.5}
\plotone{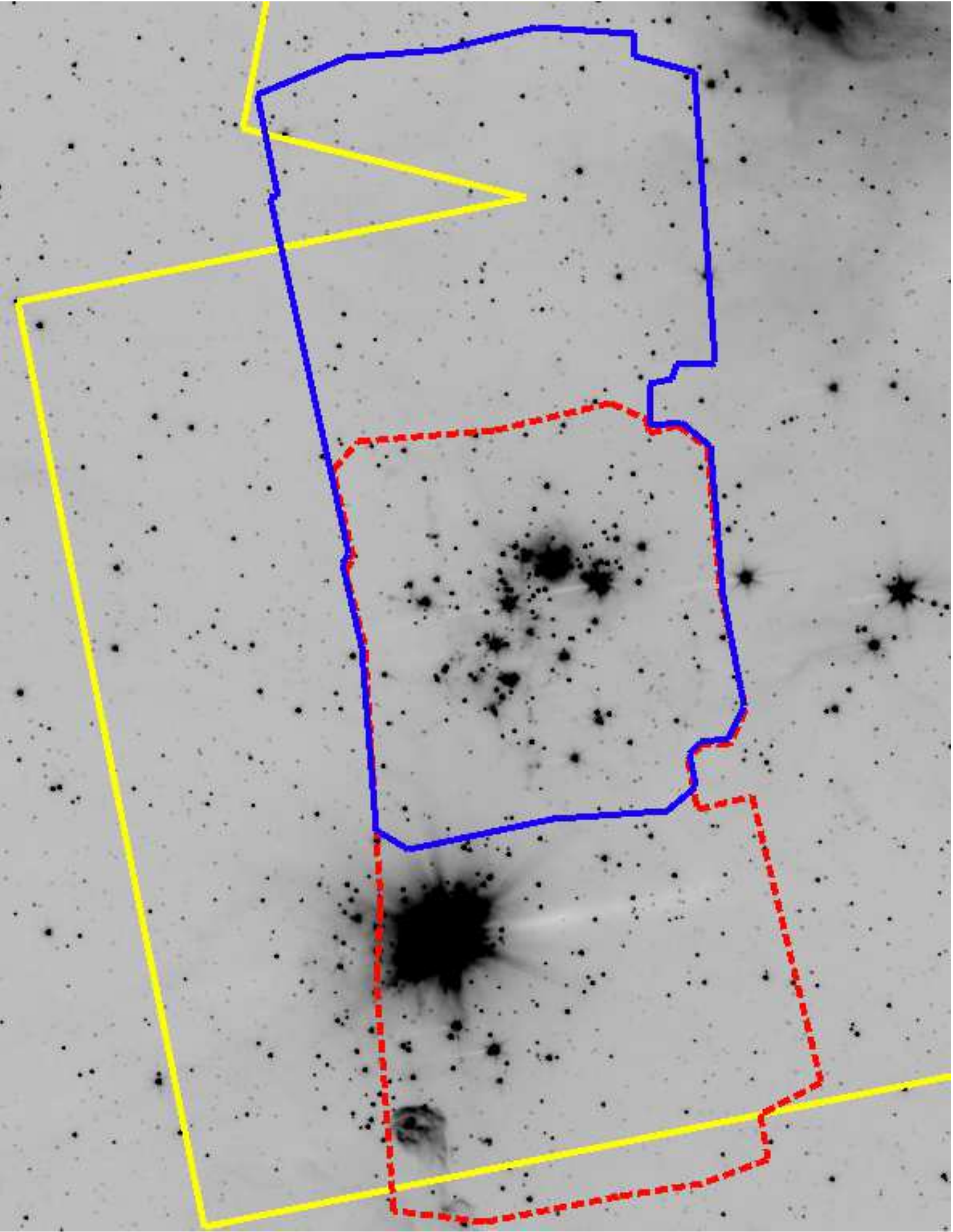}
\caption{As for Fig.~\ref{fig:n1333footprints}, but for our NGC 2264
observations, with the north-south boundary extremes separated
by $\sim$20$\arcmin$; the underlying image comes not from the cryogenic
era, but from the CSI 2264 observations.   Note that this is only the
field covered as part of the original YSOVAR observations, e.g.,
program 61027. For CSI 2264, see next figure. Field rotation, while
present, is not substantial. The background image here is an IRAC-2
image from the CSI 2264 campaign (as opposed to a cryo-era image, as it
is for most of the other figures like this). }
\label{fig:ngc2264footprints}
\end{figure}

\begin{figure}[ht]
\epsscale{0.6}
\plotone{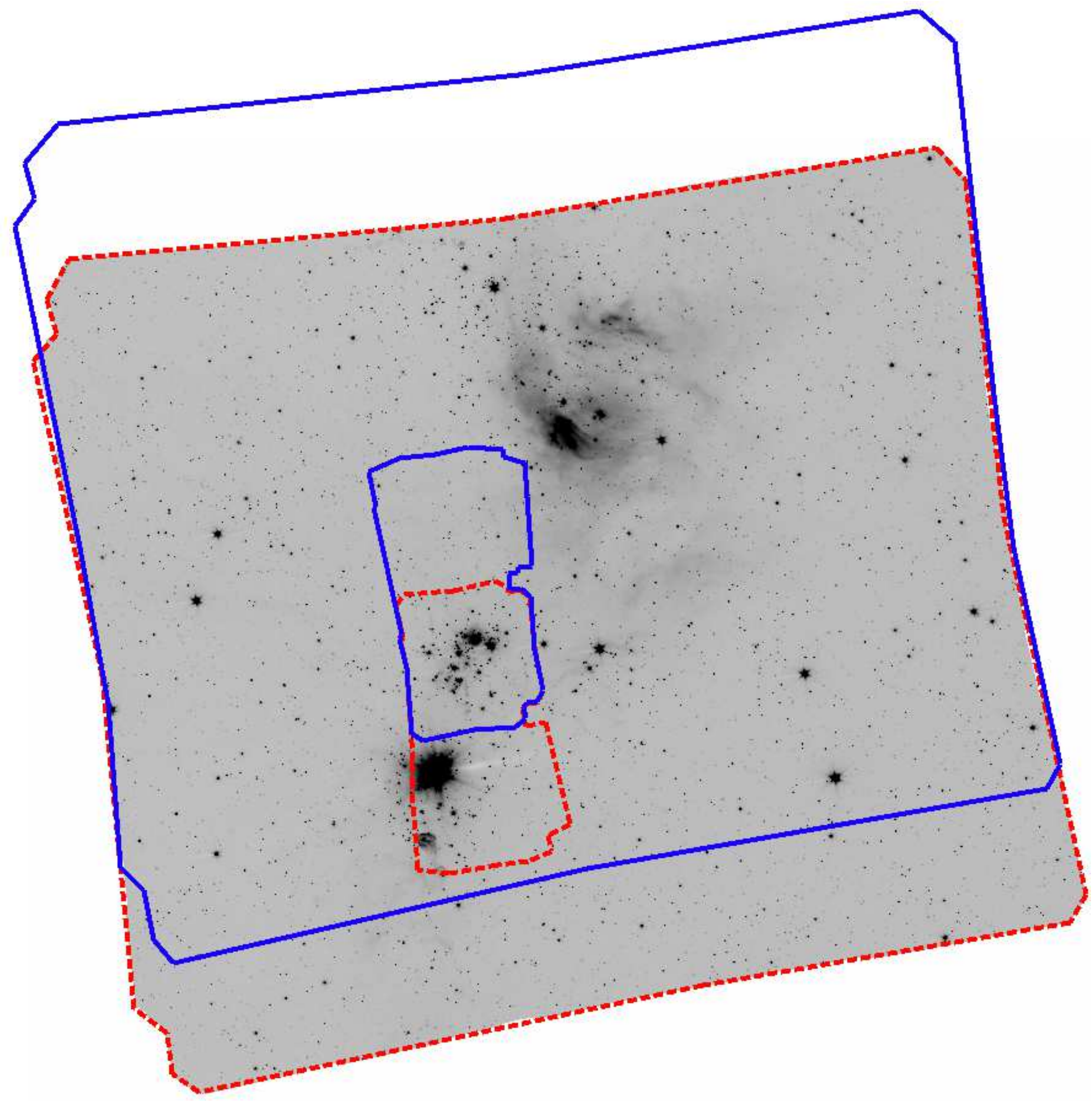}
\caption{As for Fig.~\ref{fig:afglfootprints}, but for our CSI 2264
observations, with the north-south boundary extremes separated by
$\sim$0.9$\arcdeg$. The underlying image comes from the CSI 2264
observations. The smaller footprints are the YSOVAR-classic monitored
region, from the previous figure.  }
\label{fig:csi2264footprints}
\end{figure}

\begin{figure}[ht]
\epsscale{0.6}
\plotone{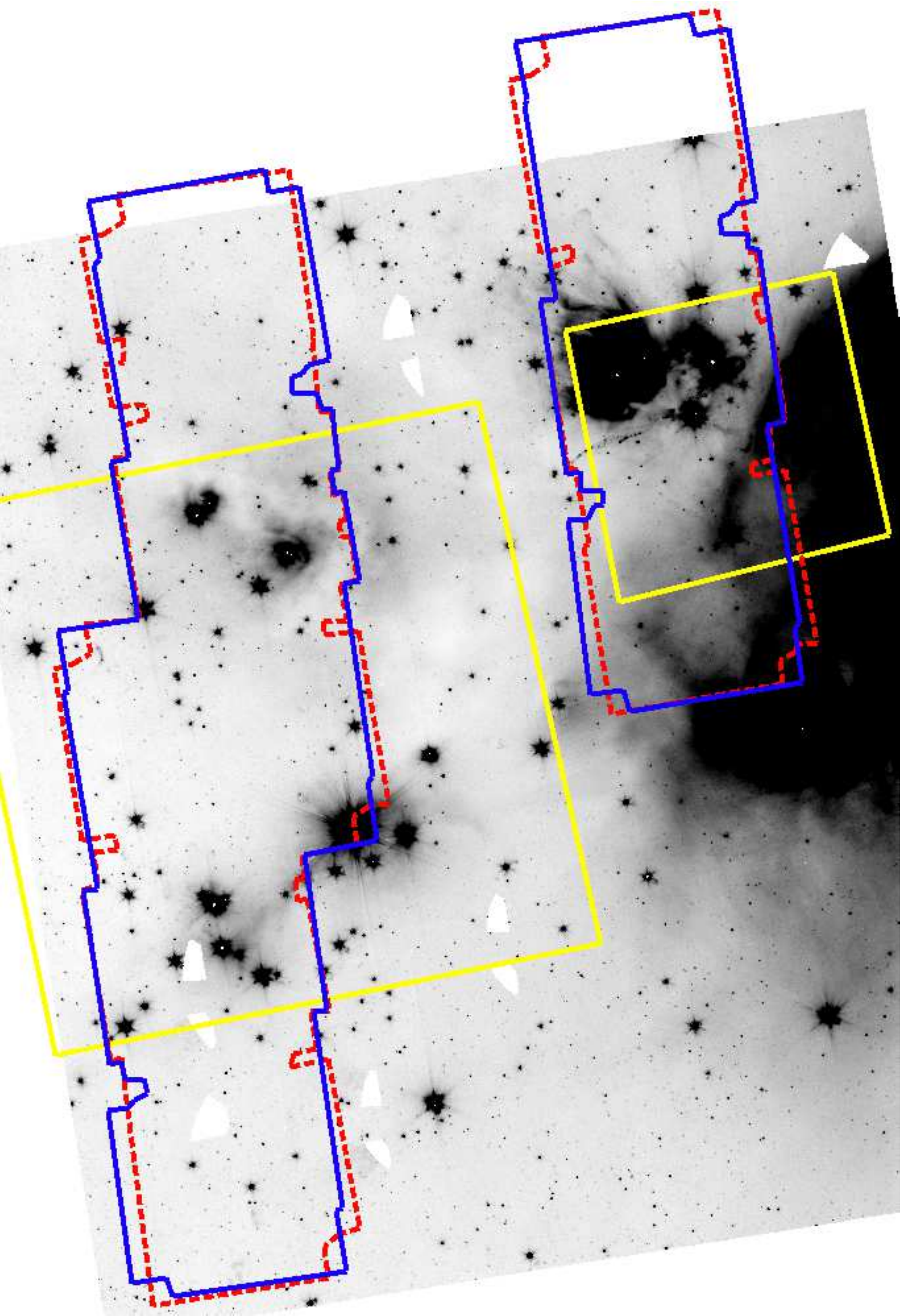}
\caption{As for Fig.~\ref{fig:n1333footprints}, but for our L1688
observations, with the north-south boundary extremes separated
by $\sim$38$\arcmin$.    The field rotation is essentially non-existent
during one $\sim$40 day window, and then flips by 180$\arcdeg$ for the
next $\sim$40 day window. The targets of observation are the three
regions with brighest stars and nebulosity in the centers of the three
`stripes' of coverage. }
\label{fig:l1688footprints}
\end{figure}

\begin{figure}[ht]
\epsscale{0.6}
\plotone{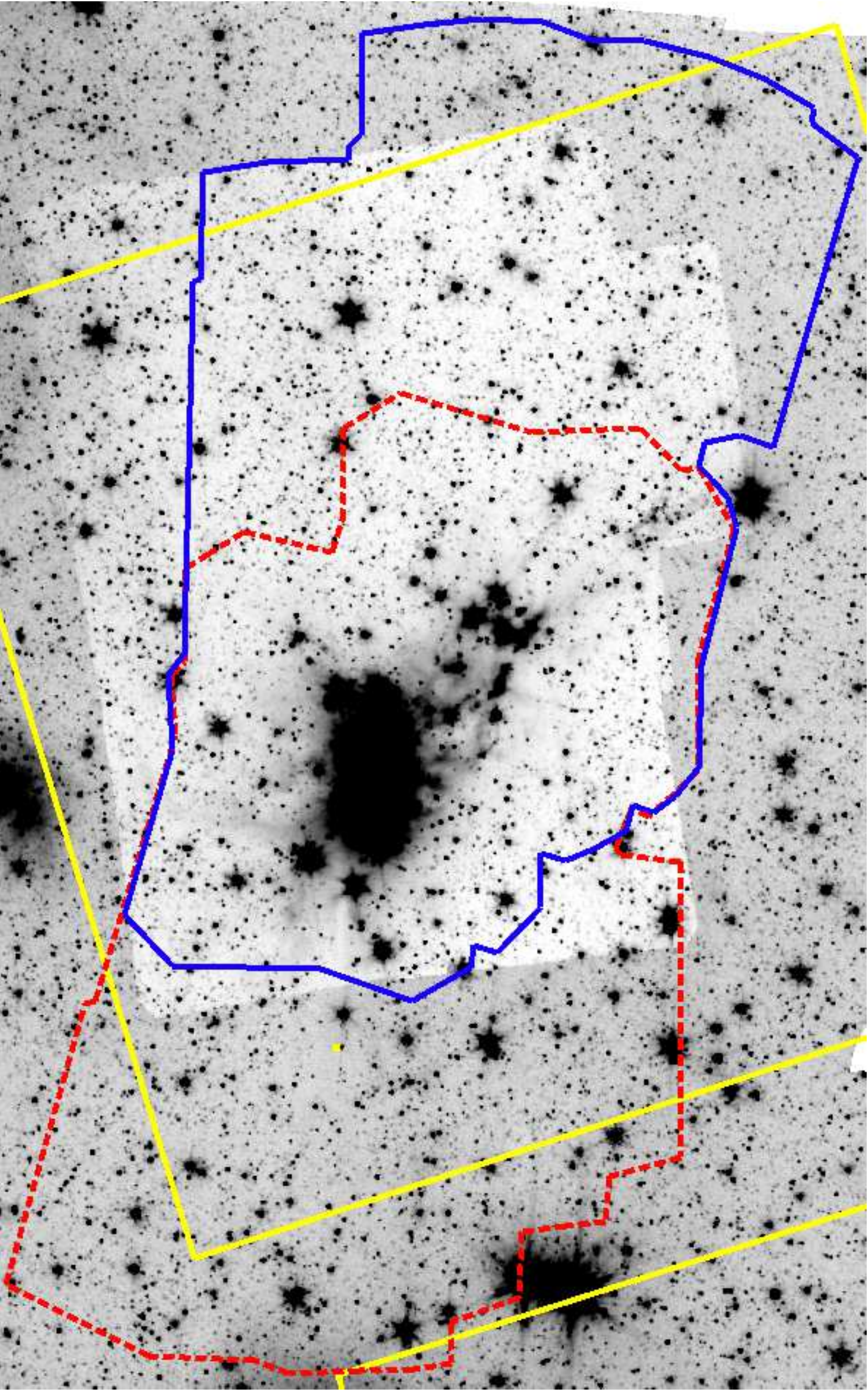}
\caption{As for Fig.~\ref{fig:n1333footprints}, but for our
Serpens-Main observations, with the north-south boundary extremes
separated by 20019990167$\sim$23$\arcmin$.   Field rotation, while present, is not
substantial. There is a small amount of additional
X-ray data (not relevant for this project) to the lower right, and a
portion of that footprint can be seen.}
\label{fig:serpensmainfootprints}
\end{figure}

\begin{figure}[ht]
\epsscale{0.6}
\plotone{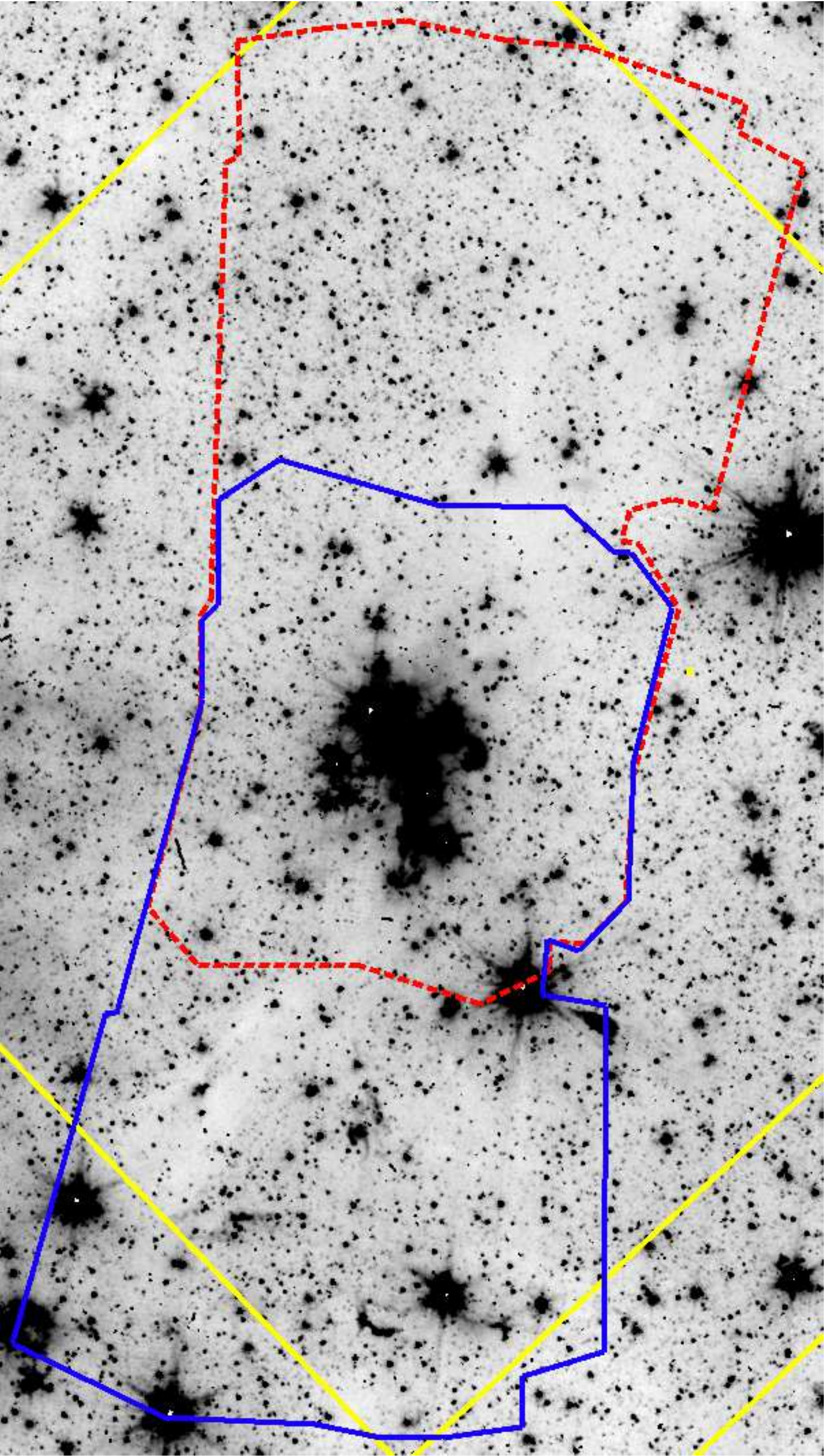}
\caption{As for Fig.~\ref{fig:n1333footprints}, but for our
Serpens-South observations, with the north-south boundary extremes
separated by $\sim$21$\arcmin$. Field rotation, while present, is not
substantial.  There is a small amount of additional
X-ray data (not relevant for this project) to the lower right, and a
portion of that footprint can be seen.}
\label{fig:serpenssouthfootprints}
\end{figure}

\begin{figure}[ht]
\epsscale{0.8}
\plotone{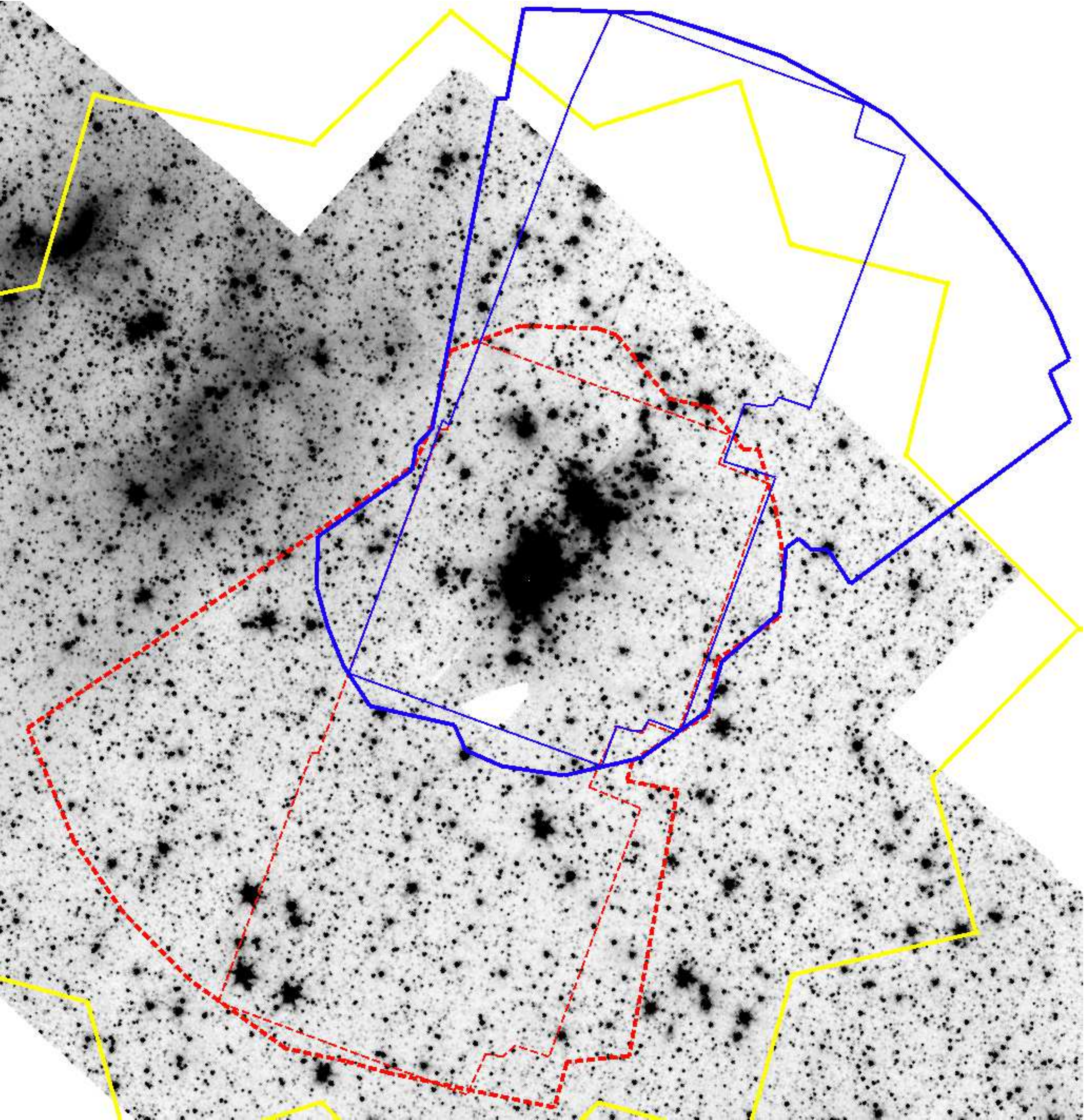}
\caption{As for Fig.~\ref{fig:n1333footprints}, but for our IRAS 20050
observations, with the north-south boundary extremes separated
by $\sim$21$\arcmin$. As for Fig.~\ref{fig:ggdfootprints}, a single
epoch of observation is also indicated by thinner blue solid (3.6
\mum) and red dashed (4.5 \mum) lines. Multiple visits of Chandra data
at multiple roll angles were obtained here, resulting in the
star-shaped Chandra coverage. }
\label{fig:iras20050footprints}
\end{figure}
                    
\begin{figure}[ht]
\epsscale{0.8}
\plotone{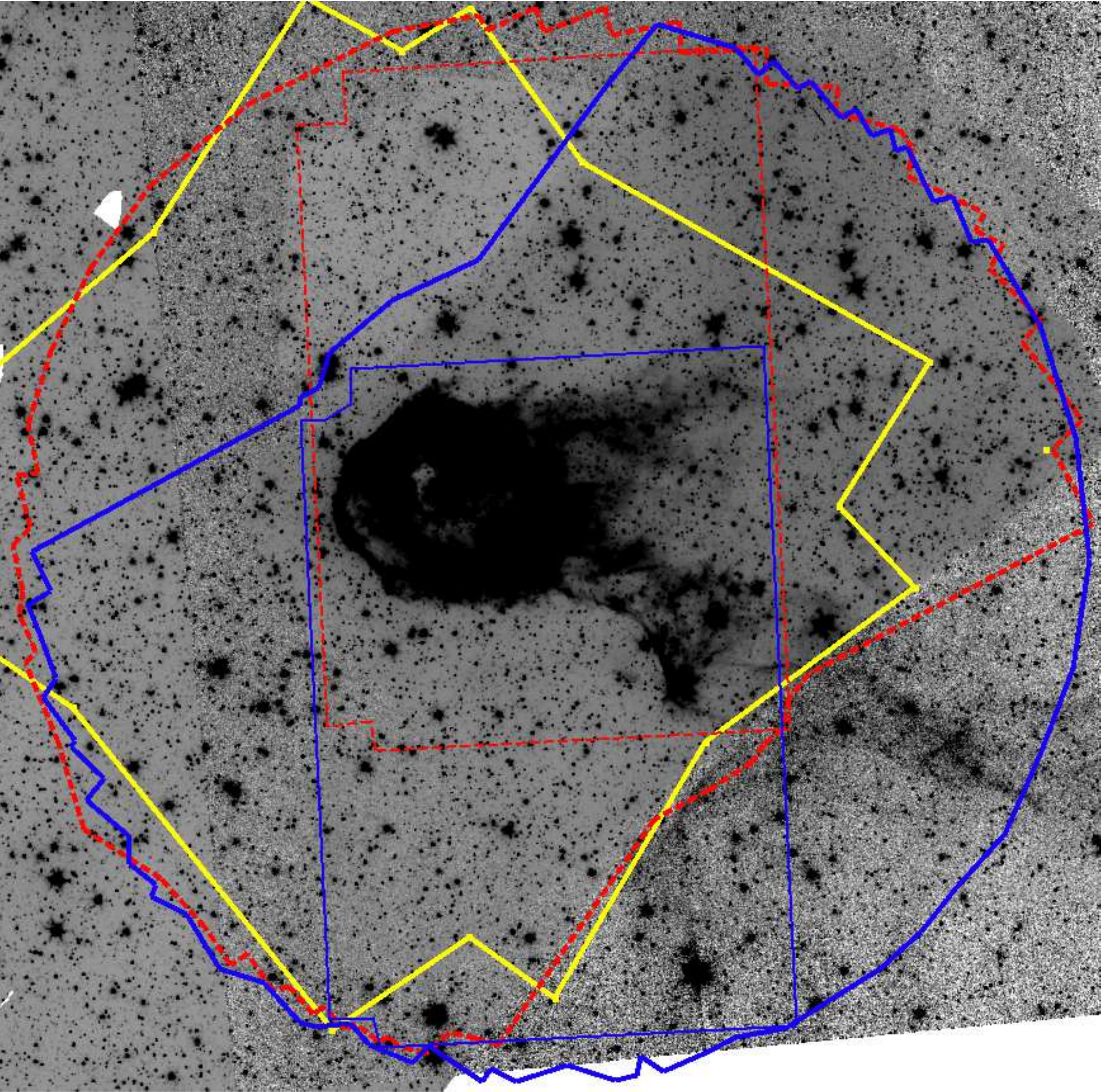}
\caption{As for Fig.~\ref{fig:n1333footprints}, but for our IC 1396A
observations, with the north-south boundary extremes separated
by $\sim$24$\arcmin$. As for Fig.~\ref{fig:ggdfootprints}, a single
epoch of observation is also indicated. Two pointings of Chandra data
at two different roll angles were obtained here, resulting in the
polygon of Chandra coverage.}
\label{fig:ic1396footprints}
\end{figure}

\begin{figure}[ht]
\epsscale{0.8}
\plotone{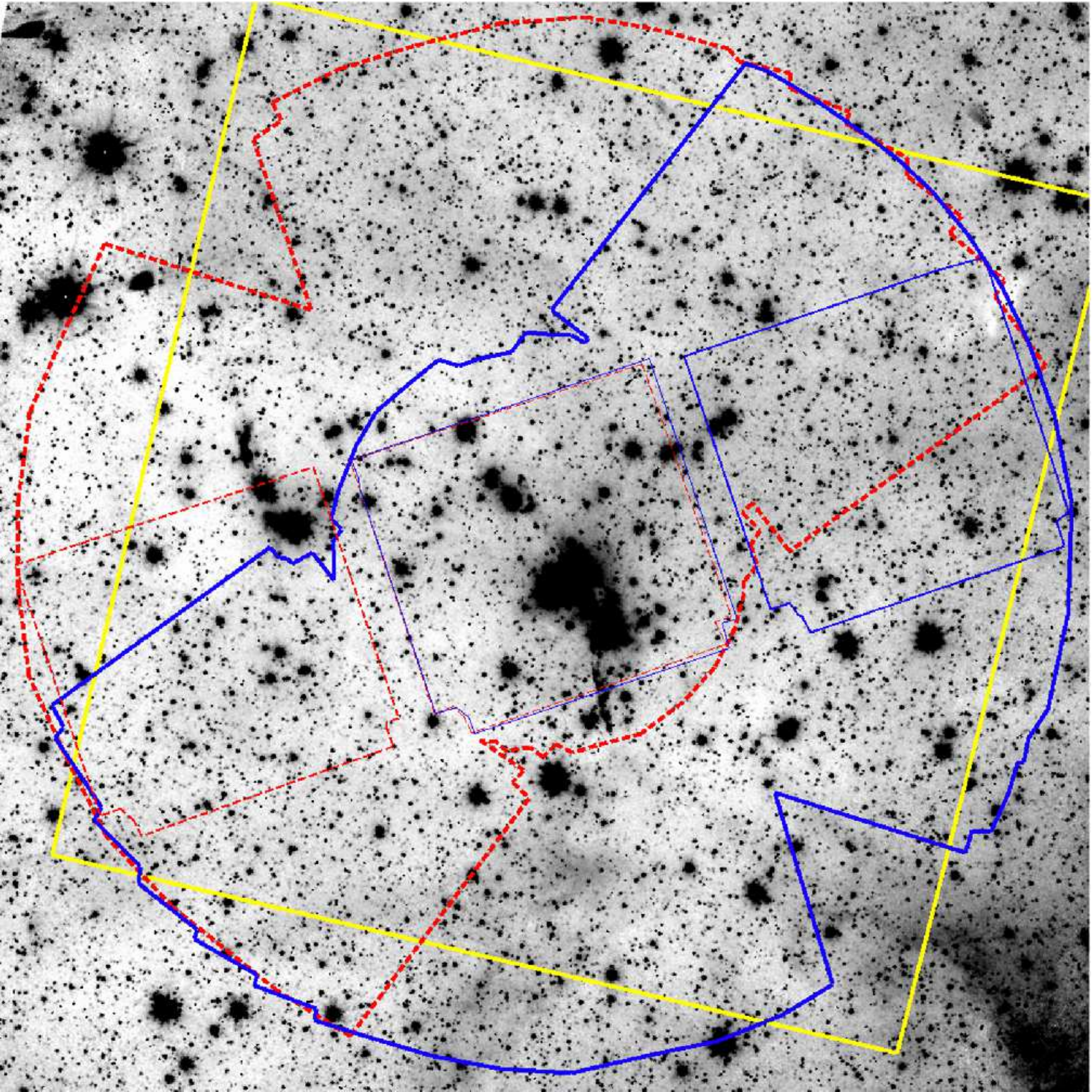}
\caption{As for Fig.~\ref{fig:n1333footprints}, but for our Ceph C
observations, with the north-south boundary extremes separated
by $\sim$20$\arcmin$. As for Fig.~\ref{fig:ggdfootprints}, a single
epoch of observation is also indicated. The field
rotation here is quite significant. }
\label{fig:cephcfootprints}
\end{figure}

\end{document}